\documentclass[12pt]{article}
\usepackage{graphicx}
\usepackage{amsmath,amsthm,amssymb, graphicx, multicol, array}
\usepackage{graphics,wrapfig,natbib,amssymb}
\usepackage[font=footnotesize,labelfont=bf]{caption}
\usepackage[pdfencoding=auto,psdextra,colorlinks=true,linkcolor=blue,citecolor=blue]{hyperref}
\begin{document}
\def\begeq{\begin{equation}}
\def\endeq{\end{equation}}
\def\begeqar{\begin{eqnarray}} 
\def\endeqar{\end{eqnarray}}
\def\half{\frac 12}
\def\non{\nonumber\\}
\def\bfx{{\bf r}}
\def\bfX{{\bf R}}
\def\bfr{{\bf r}}
\def\bfK{{\bf K}}
\def\bfz{{\bf z}}
\def\bfrhat{{\hat\bfr}}
\def\bfkhat{{\hat\bfk}}
\def\bfKhat{{\hat\bfK}}
\def\bfxhat{{\hat\bfx}}
\def\bfXhat{{\hat\bfX}}
\def\bfzhat{{\hat\bfz}}
\def\xhat{{\hat x}}
\def\Bfk{{\bf K}}
\def\bfL{{\bf L}}
\def\Bfkhat{{\hat\Bfk}}
\def\bfR{{R}}
\def\bfRhat{{\hat\bfR}}
\def\bfshat{{\hat\bfs}}
\def\bfs{{\bf s}}
\def\bfk{{\bf k}}
\def\ellb{{\bar\ell}}
\def\barL{{\bar L}}
\def\barm{{\bar m}}
\def\barbarm{{\bar{\bar m}}}
\def\ellbb{{\bar{\bar\ell}}}
\def\barbarL{{\bar{\bar L}}}
\def\Mbar{{\bar M}}
\def\Mbarbar{{\bar{\bar M}}}
\def\Lis{\Lambda_i \Lambda'_i \Lambda''_i}
\def\Mis{ M_i M'_i M''_i}
\def\mis{m_i m'_i m''_i}
\def\ellis{\ell_i \ell'_i \ell''_i}
\newcommand{\Mpch}{{\rm Mpc}/h}
\newcommand{\state} [1]{|L_{#1}m_{#1}\right>}
\newcommand{\cg}[2]{\left<L_{#1}m_{#1}L_{#2}m_{#2}|L_{#1#2}m_{#1#2}\right>}
\newcommand{\notprod}[2]{({#1},{#2})}
\newcommand{\dotprod}[2]{(\bfxhat_{#1}\cdot\bfxhat_{#2})}
\newcommand\crossprod[3]{\bfxhat_{#1}\cdot(\bfxhat_{#2}\times\bfxhat_{#3})}
\renewcommand{\thefigure}{\arabic{section}.\arabic{figure}}
\renewcommand{\thetable}{\arabic{section}.\arabic{table}}
\def\deltat{{\tilde\delta}}
\def\calN{{\cal N}}
\def\calY{{\cal P}}
\def\calZ{{\cal Z}}
\def\calC{{\cal C}}
\def\calR{{\cal R}}
\def\calD{{\cal D}}
\def\calK{{\cal K}}
\def\calJ{{\cal J}}
\def\calB{{\cal B}}
\def\calA{{\cal A}}
\def\calS{{\cal D}^{\rm I}}
\def\calQ{{\cal Q}}
\def\calG{{\cal G}}
\def\Rot{{\cal R}}
\def\calO{{\cal E}(\Lambda)}
\def\calOp{{\cal E}(\Lambda')}
\def\calOpp{{\cal E}(\Lambda'')}
\def\calP{{\cal S}}
\def\calQL{{\calQ^{\Lambda,\Lambda',\Lambda''}}}
\def\calQLL{{\calQ^{\Lambda,\Lambda',\Lambda''}}}
\def\calSL{{\calS_{\Lambda\Lambda'\Lambda''}}}
\def\snl{S(\Lambda)}
\def\snlp{S(\Lambda')}
\def\snlpp{S(\Lambda'')}

	\newcommand{\ket}[2]{|{#1}{#2}\right>}
	\newcommand{\ketx}[1]{|L_{#1}M_{#1}\right>}
	\newcommand{\ketxp}[1]{|L'_{#1}M'_{#1}\right>}
	\newcommand{\bra}[2]{\left<{#1}{#2}|}
	\newcommand{\brax}[1]{\left<L_{#1}M_{#1}|}
	\newcommand{\CG}[3]{\left<L_{#1}M_{#1}L_{#2}M_{#2}|L_{#3}M_{#3}\right>}

	\newcommand{\CGp}[3]{\left<L'_{#3}M'_{#3}|L'_{#1}M'_{#1}L'_{#2}M'_{#2}\right>}
	\newcommand{\ninej}[9]{\left\{\begin{array}{ccc}
									j_{#1}& j_{#2}& j_{#3}\\
									j_{#4}& j_{#5}& j_{#6} \\
									  j_{#7}& j_{#8}& j_{#9}
									  \end{array}\right\}}
	\newcommand{\nine}[9]{\left\{\begin{array}{ccc}
									{#1}& {#2}& {#3}\\
									{#4}& {#5}& {#6} \\
									  {#7}& {#8}& {#9}
									  \end{array}\right\}}
						  
	\def\sixcol{\setlength{\arraycolsep{50pt}}}
	\newcommand{\fifteen}[3]{\left\{\begin{array}{ccc}
									{#1}_1& {#2}_1& {#3}_1\\
									{#1}_2& {#2}_2& {#3}_2 \\
									  {#1}_3& {#2}_3& {#3}_3\\
									{#1}_{12}& {#2}_{12}& {#3}_{12} \\
									  {#1}_{123}& {#2}_{123}& {#3}_{123}								  \end{array}\right\}}
	\newcommand{\six}[6]{\left(\begin{array}{ccc}
									{#1}\!& {#2}\!& {#3}\!\\
									{#4}\!& {#5}\!& {#6}\! \\
											  \end{array}\right)}
											  
\newcommand{\sixj}[6]{\left\{\begin{array}{ccc}
									{#1}& {#2}& {#3}\\
									{#4}& {#5}& {#6} \\
											  \end{array}\right\}}
\newcommand{\ketone}[1]{|#1\right>}
\newcommand{\Ls}[1]{L_{#1}}
\newcommand{\Lsp}[1]{L'_{#1}}
\newcommand{\tildei}{{\tilde i}}

\newcommand{\deltaK}{\delta^{{\rm K}}}

\begin{Large}
\begin{center}
Isotropic N-Point Basis Functions and\\[0.2in]
Their Properties\\[0.2in]
R. N. Cahn and Z. Slepian\\[0.2in]
\today\\[0.2in]

\end{center}
\end{Large}

\section*{Abstract}
Isotropic functions of positions $\bfx_1,\bfx_2,
\ldots, \bfx_N$, i.e. functions invariant under simultaneous rotations of all the coordinates, are conveniently formed using spherical harmonics and Clebsch-Gordan coefficients.  An orthonormal basis of such functions provides a formalism suitable for analyzing isotropic distributions such as those that arise in cosmology, for instance in the clustering of galaxies as revealed by large-scale structure surveys. The algebraic properties of the basis functions are conveniently expressed in terms of 6-$j$ and 9-$j$ symbols. The calculation of relations among the basis functions is facilitated by ``Yutsis'' diagrams for the addition and recoupling of angular momenta.

\section{Introduction}
Analyses of angular behavior are ubiquitous in many branches of physics and astronomy. For instance in extracting physical parameters from measurements of the Cosmic Microwave Background (CMB), we compute the angular power spectrum, bispectrum \citep{Luo_1994, Spergel_1999}, and trispectrum \citep{Hu_2001} of temperature (and more recently polarization) anisotropies. This approach is also used on multiple concentric spherical shells at different redshifts for intensity mapping \citep{Liu_2016}. Similarly, in projected large-scale structure
surveys we often compute angular two- and three-point correlation
functions or their Fourier-space analogs the power spectrum and bispectrum (e.g. \citealt{Verde_2000}). More recently, methods have been developed for using ``local'' angular expansions (about each galaxy in a survey) to compute the full 
three-dimensional three-point correlation functions (\citealt{SE_3pt_alg, SE_3PCF_FT, SE_aniso_3pcf}). A similar expansion approach has been developed for the bispectrum \citep{Sugiyama_2018, Sugiyama_2020}. Angular behavior is also often considered to quantify dependence of clustering statistics on the line of sight in the presence of redshift-space distortions (RSD; e.g. \citealt{Hamilton_1998}). However in this work we will focus solely on analyses where the underlying signal is fully isotropic (forthcoming work treats the RSD; Li \& Slepian in prep.)

Spherical harmonics are the traditional means for expressing angular dependence.\footnote{As noted earlier, we focus here on isotropic signals; for spherical harmonic work in redshift space, see \citealt{Heavens_1995, Hamilton_1996, Tadros_1999, Szapudi_2004_wa, Castorina_2018_beyond_PP}, and for studies of large-scale deviations from statistical isotropy in the CMB using harmonics, see e.g. \citealt{Hajian_2003, Joshi_2010, Book_2012}. Also somewhat related to the present work is \cite{Dai_2012}, which proposes use of total angular momentum (TAM) basis functions for cosmological calculations; these TAM functions involve spherical harmonics and spherical Bessel functions, and are extended to describing vectors and tensors as well as the scalars to which the present work is restricted.}  It is often the case that the angular behavior of a multi-body distribution depends only on the relative orientations of the bodies. For the case of two 
positions, the spherical harmonic addition theorem 
provides the means for representing such behavior in terms of Legendre polynomials $P_{\ell}$. If $\bfxhat_1$ and $\bfxhat_2$ represent the directions to the points $\bfx_1$ and $\bfx_2$ then 
\begeq
P_\ell(\bfxhat_1\cdot\bfxhat_2)=\frac{4\pi}{2\ell+1}\sum_{m = -\ell}^{\ell} Y_{\ell m}(\bfxhat_1)
Y^*_{\ell m}(\bfxhat_2),
\label{eqn:harmonic_add_thm}
\endeq
where $Y_{\ell m}$ is a spherical harmonic. Here the limits on the sum over $m$ are explicit, but in the rest of this work, it will be implicit that $|m_i| \leq \ell_i$ any time we consider a set of angular momenta $\ell_i $. The Legendre polynomials then give the basis for an expansion of functions  invariant under simultaneous rotations of $\bfxhat_1$ and $\bfxhat_2$.  
Our purpose here is to provide a convenient generalization of this identity for more than two directions. That is, we construct a set of functions of the unit vectors $\bfxhat_i, i=1,\ldots, N$  that are invariant under simultaneous rotation of all the arguments and are adequate for expanding all functions sharing this invariance.

As briefly noted already, our immediate motivation comes from cosmology.
Correlations in density distributions play a central role in observational cosmology (e.g. \citealt{Peebles_1980}). The 
two-point correlation function and power spectrum of the density distribution are the primary examples. However, multi-point distributions, especially the three-point correlation function, contain valuable additional information (e.g. \citealt{Sefusatti_2005} Figure 3), reaching
signal-to-noise greater than the 
two-point correlation function on scales of about $5\;\Mpch$ for a Sloan Digital Sky Survey (SDSS)-like volume. 
The 2-Point Correlation Function (2PCF), by hypothesis, depends simply on the distance between the points.
For the 3-Point Correlation Function (3PCF) it suffices to consider two sides of the triangle and an expansion in Legendre polynomials of the dependence on the cosine of the included
angle. For the $N$-Point Correlation Functions with $N>3$, there are challenges that go beyond those of the 
3PCF in terms of both formalism and computability.  The formalism we develop here is particularly useful for $N$-Point Correlation Functions. Computability will be discussed in a separate paper (Slepian, Cahn \& Eisenstein in prep.).

One approach to the formalism is to use a Cartesian representation.
From the collection $\bfxhat_1,\bfxhat_2,\ldots, \bfxhat_N$ we can always form functions that are invariant under the simultaneous rotation of all the $\bfxhat_1,\bfxhat_2,\ldots, \bfxhat_N$ by taking multinomials created from the scalar and cross products
$\bfxhat_i\cdot\bfxhat_j$, $\bfxhat_i\cdot(\bfxhat_j\times\bfxhat_k).$
Multinomials with an odd number of cross products are parity-odd.   Those with an even number of cross products can be written in terms of dot-products alone and are parity even. Examples are given in Appendix \ref{ap:examples}. 

However, as we encounter polynomials of higher degree it becomes more convenient to work with functions built from spherical harmonics. Because rotations are generated by operators that have the quantum mechanical interpretation of angular momentum, we will use that language here.  We take advantage of the standard techniques for combining states of well-defined angular momentum to create functions that are invariant under rotations and that possess orthogonality properties analogous to those for Legendre polynomials and spherical harmonics. Because these functions, which we indicate by $\calY$, are isotropic, we anticipate that final results will be expressed in terms only of the angular momenta $L$ involved without reference to specific components along any axis, generically $L_z=m$. Moreover, rotationally-invariant functions are necessarily eigenstates of the total angular momentum operator $\bfL=\sum_i \bfL_i$ with eigenvalue $L=0$.

Our generalization of the addition theorem for spherical harmonics is a set of basis functions $\calY(\bfX)$, with $\bfX \equiv (\bfx_1,\bfx_2,\ldots,\bfx_N)$, invariant under simultaneous rotations $\Rot$ of all the coordinates:
\begeq
\calY(\Rot\bfX)=\calY(\bfX).
\endeq
 In carrying out computations with these functions we encounter products of these functions. We also encounter functions whose arguments are permuted. We show here how to evaluate these operations, relying on techniques developed long ago in studies of atomic and nuclear physics.  We find it convenient to use so-called Yutsis diagrams \citep{Yutsis_1962} to address complicated combinations of angular momenta. These are described in Appendix \ref{ap:yutsis}.
 
 We briefly discuss a few related works. During late stages of the preparation of this work, we became aware of \cite{Mitsou_2020}, which develops a general construction for $N$-point angular functions. Like our work, the starting point is averaging over rotations, and like our work, \cite{Mitsou_2020} use spherical harmonics. They define ``multilateral Wigner symbols'' (see their Figure 1) to represent the addition of arbitrary numbers of angular momenta. These symbols include intermediate sums which show how the different primary angular momenta (those on each spherical harmonic) are coupling. These intermediates form internal diagonals on their diagrams. \cite{Mitsou_2020} note there is an ambiguity as to which common origin point their diagonals share, and that different choices are related by the Wigner 6-$j$ symbols. Their work does not use Yutsis diagrams; our work takes advantage of this previously developed framework for handling angular momentum coupling. This framework allows us to efficiently obtain explicit expressions in terms of 3-$j$ symbols both for our basis functions and for reductions of products and reorderings. 
 
 Within the CMB literature, a number of authors have developed the formalism for describing angular correlations on a single spherical shell, most prominently \cite{Spergel_1999} for the angular bispectrum (harmonic-space analog of a 
 three-point correlation function with all three points on the same spherical shell) and \cite{Hu_2001, Okamoto_2002} for the angular trispectrum (harmonic-space analog of a 
 four-point correlation function with all four points on the same spherical shell). \cite{Kamion_2011, Regan_2010,  Abramo_2010, lacasa2014nongaussianity} discuss both bispectrum and trispectrum in the context of Primordial Non-Gaussianity (PNG; \citealt{Bartolo_2004} for a review), and these last two works also consider the problem of angular 
 N-point correlation functions respectively for $N=5$ and for arbitrary $N$; footnote 2 of \cite{Mitsou_2020} gives further useful discussion.

 \section{Overview of Method}
 To introduce our method, we show the relationship of the simplest $\calY$ to the addition theorem above.
 Spherical harmonics are eigenstates of the operators $\bfL^2$ and $\bfL_z$.  The standard Clebsch-Gordan coefficients \citep{DLMF_temp} enable us to 
 create functions of $\bfxhat_1$ and $\bfxhat_2$, acted on by $\bfL_1$ and $\bfL_2$, that are eigenstates of $\bfL^2=(\bfL_1+\bfL_2)^2$ and $\bfL_z=\bfL_{1z}+\bfL_{2z}$ with $L_z=m$. In particular, a rotationally-invariant function can be obtained with $L_1 = L_2 \equiv \Lambda$ as
  \begeqar
  \calY_{\Lambda,\Lambda}(\bfxhat_1,\bfxhat_2)&=&\sum_{m_1,m_2}\left<\Lambda m_1 \Lambda m_2|00\right>Y_{\Lambda m_1}(\bfxhat_1)Y_{\Lambda m_2}(\bfxhat_2)\non
  &=&\sum_m\frac {(-1)^{\Lambda -m}}{\sqrt{2\Lambda+1}}Y_{\Lambda m}(\bfxhat_1)Y_{\Lambda -m}(\bfxhat_2)\non
  &=&\sum_m\frac {(-1)^{\Lambda }}{\sqrt{2\Lambda+1}}Y_{\Lambda m}(\bfxhat_1)Y^*_{\Lambda m}(\bfxhat_2)\non
  &=&\frac{\sqrt{2\Lambda+1}}{4\pi} (-1)^{\Lambda} P_{\Lambda}(\bfxhat_1\cdot\bfxhat_2).
  \label{eq:nequal2}
\endeqar
To pass from first to the second equality, we substituted the value of the 
Clebsch-Gordan coefficient. Going from the second to the third equality, we used that $Y_{\ell m}^*(\hat{r}) =(-1)^m Y_{\ell -m}(\hat{r})$. We then used the spherical harmonic addition theorem (\ref{eqn:harmonic_add_thm}) to resum the 
spherical harmonics into a Legendre polynomial. 

Advancing now to three directions, the combination of the angular momenta associated with the first two directions, $\Lambda_1$ and $\Lambda_2$, will give a resulting total angular momentum $\Lambda_{12}$ between $|\Lambda_1-\Lambda_2|$ and $\Lambda_1+\Lambda_2$ due to the triangle rule on vector addition.  For the total angular momentum of the three vectors to equal zero, the value of the third angular momentum, $\Lambda_3$ must then equal $\Lambda_{12}$.

However, once we go beyond three vectors, there are choices to be made.  There may be more than one intermediate value $\Lambda_{12}$ that will yield a combination that is rotationally invariant (i.e. one that has total angular momentum zero).  If there are just four vectors, the combination of $\Lambda_{12}$ with $\Lambda_3$ is uniquely determined to equal $\Lambda_4$ since that is the only way the sum of all the angular momenta can be zero. Thus, we do not need to separately specify any intermediate angular momenta (i.e. the sum of two or more). 

If there are more than four vectors, we need to specify one or more additional, ``intermediate'' angular momenta.
Each intermediate is governed by a triangle inequality as is usual for angular momentum sums. There are multiple schemes by which one could choose to organize specifying the intermediate angular momenta. We set our canonical choice to be specifying intermediate momenta that are cumulative sums, i.e. $\Lambda_{12},\Lambda_{123},\Lambda_{1234},\ldots$, with the primary momenta indicated as $\Lambda_1, \Lambda_2,\ldots,\Lambda_N$. We indicate the full set of momenta, including both primary and intermediate, as $\Lambda=(\Lambda_1,\Lambda_2,(\Lambda_{12}),\Lambda_3,\ldots)$, 
Similarly, we designate the set of all $z$-components as $m = (m_1,m_2,m_3,\ldots)$.  Note that $m_{12}$ is uniquely determined to be $m_1+m_2$, etc.; the $z$ components satisfy simple scalar addition. 

We define the isotropic functions of order $N$ as the sum over $z$-components of a product of $N$ spherical harmonics weighted by a factor $\calC^\Lambda_m$:
\begeqar
\calY_\Lambda(\bfXhat)
&=&\sum_{m_1,m_2, \ldots }\calC^\Lambda_m \;Y_{\Lambda_1 m_1}(\bfxhat_1)\cdots Y_{\Lambda_N m_N}(\bfxhat_N),
\label{eq:calP_defn}
\endeqar
with $\bfXhat \equiv (\bfxhat_1, \ldots, \bfxhat_N)$ and the weight
\begeqar
\label{eq:calC1}
&\calC^\Lambda_m \equiv \calC^{(\Lambda_1 \Lambda_2 
\Lambda_{12},\ldots)}_{(m_1 m_2 m_3,\ldots)}
=\sum_{m_{12},m_{123}, \ldots }\left<\Lambda_1 m_1\Lambda_2 m_2|\Lambda_{12} m_{12}\right>\cdots\\
&\times \left<\Lambda_{12\ldots   N-2}m_{12\ldots N-2}\Lambda_{N-1} m_{N-1}|\Lambda_{12\ldots N-1} m_{12\ldots N-1}\right>\nonumber\\
&\times\left< \Lambda_{12\ldots N-1} m_{12\ldots N-1}\Lambda_N m_N|00\right>,\nonumber
\endeqar
This weight can be straightforwardly applied for $N\geq 4$; for $N=2$ and $N=3$ there are no intermediate momenta, and we discuss that case after the more general, $N\geq 4$ case.

$\calY_\Lambda(\bfXhat)$ is an eigenfunction of $\bfL^2$ with $L=0$ and also of $(\bfL_1+\bfL_2)^2$ with eigenvalue $\Lambda_{12}(\Lambda_{12}+1)$, and analogously for $(\bfL_1+\bfL_2+\bfL_3)^2$, etc.  We do not need to specify separately $\Lambda_{12\ldots N-1}$ since it must be equal to $\Lambda_N$: this follows from the last Clebsch-Gordan coefficient. The values of $m_{12},m_{123}\ldots$ are uniquely determined by the $m_i$ since the $z$-components satisfy a simple scalar sum rule. We note that in general, if we have $N>3$ vectors we need to specify $N-3$ intermediate values of angular momentum in order to determine uniquely a rotationally invariant function.  

Because the parity of $Y_{LM}(\bfxhat)$ is $(-1)^{L}$, the parity of $\calY_\Lambda$ is determined by the sum of the primary components $\sum_i \Lambda_i$. If this sum is even, $\calY_\Lambda$ is even under inversion of all the coordinates $\bfxhat_i\to -\bfxhat_i$.  Conversely, if this sum is odd, $\calY_\Lambda$ is odd.  Moreover, if the sum is even, $\calY_\Lambda$ is purely real while if the sum is odd, $\calY_\Lambda$ is purely imaginary. We now prove this last claim. 

First, our spherical harmonics (with the usual Condon-Shortley conventions) satisfy
\begeq
\label{eqn:ylm_conj}
Y_{LM}^*(\bfxhat)=(-1)^MY_{L,-M}(\bfxhat),
\endeq
and 
\begeq
\label{eqn:CG_prop}
\left<L_1,-M_1, L_2, -M_2|L,-M\right>=(-1)^{L_1+L_2-L}\left<L_1,M_1, L_2, M_2|L,M\right>.
\endeq
Inserting Eq. (\ref{eqn:CG_prop}) into Eq. (\ref{eq:calC1}) and noting that the final angular
momentum in Eq. (\ref{eq:calC1}) is zero, we find
\begeq 
\label{eqn:calC_S}
\calC^\Lambda_{-M}=\calO\calC^\Lambda_M
\endeq 
where because of its ubiquity we have defined
\begeq
\calO\equiv(-1)^{\sum_i\Lambda_i}
\label{eq:def-parity}
\endeq
which is $+1$ for parity-even $\Lambda$ and $-1$ for parity-odd $\Lambda$. We note that the sum above is taken only over what we term ``primary'' angular momenta, i.e. it does not include any intermediate angular momenta (which specify how primaries, or primaries and intermediates, couple to each other).

Thus
\begeqar
\calY^*_\Lambda(\bfXhat)&=&\sum_{m_1,m_2,\ldots }\calC^\Lambda_M \;Y^*_{\Lambda_1 m_1}(\bfxhat_1)\cdots Y^*_{\Lambda_N m_n}(\bfxhat_N)\\
&=&\sum_{m_1,m_2,\ldots }\calC^\Lambda_M \;Y_{\Lambda_1 -m_1}(\bfxhat_1)\cdots Y_{\Lambda_N -m_N}(\bfxhat_N)(-1)^{\sum_im_i}\nonumber
\endeqar
where we used Eq. (\ref{eqn:ylm_conj}) and the reality of $\calC^\Lambda_M$.  Now the sum of the $m_i$ is zero since the total $L$ is zero. So, using Eq. (\ref{eqn:calC_S}) we find
\begeqar
\calY^*_\Lambda(\bfXhat)
&=&\sum_{m_1,m_2,\ldots }\calO\calC^\Lambda_{-M}Y_{\Lambda_1 -m_1}(\bfxhat_1)\cdots Y_{\Lambda_N -m_N}(\bfxhat_N)\non
&=&\calO\calY_\Lambda(\bfXhat).
\label{eqn:Pconj}
\endeqar
To obtain the second line, we noticed that we can flip all of the signs on the $z$-components with no change since we are summing over all values positive and negative in any case. From Eq. (\ref{eqn:Pconj}) we thus see that if $\Lambda$ is even, and hence $\calO = +1$, then $\calY = \calY^*$ so $\calY$ must be purely real. If $\Lambda$ is odd, and hence $\calO = -1$, then $\calY^* = - \calY$ and $\calY$ must be purely imaginary.

It is convenient to re-express $\calC^\Lambda_M$ (Eq. \ref{eq:calC1}) in terms of 3-$j$ symbols
rather than Clebsch-Gordan coefficients using the relations in Appendix \ref{ap:symbols}. We note again that $\calO$ (Eq. \ref{eq:def-parity}) involves a sum over primary angular momenta only. We have
\begeqar 
\label{eqn:C_threej}
C^\Lambda_m&=&\calO\sqrt{2\Lambda_{12}+1}\times\cdots\times\sqrt{2\Lambda_{12\ldots N-2}+1}\non
&&\times\sum_{m_{12} \ldots }(-1)^\kappa
\six{\Lambda_1}{\Lambda_2}{\Lambda_{12}}{m_1}{m_2}{-m_{12}}
\six{\Lambda_{12}}{\Lambda_3}{\Lambda_{123}}{m_{12}}{m_3}{-m_{123}}\cdots\non
&&\qquad\times\six{\Lambda_{12\ldots N-2}}{\Lambda_{N-1}}{\Lambda_N}{m_{12\ldots N-2}}{m_{N-1}}{m_N}\label{eq:calC2}
\endeqar 
where
\begeq
\kappa=(\Lambda_{12}-m_{12}+\Lambda_{123}-m_{123}+\cdots+\Lambda_{12\ldots N-2}-m_{12\ldots N-2}).
\label{eq:caly3j2}
\endeq
The factor $(-1)^\kappa$, which has contributions from every intermediate $\Lambda$, is essential to the use of Yutsis diagrams, as described in Appendix \ref{ap:yutsis}. The quantities $m_{12}, m_{123},$ etc. appearing in $\kappa$ are uniquely determined by the $m_i$.  Thus the factor $(-1)^\kappa$ could be taken outside the summation with the understanding that $m_{12}=m_1+m_2$, etc. 

We also introduce more abstract notation for the isotropic states $\calY_{\Lambda}$:
\begeq
\calY_{\Lambda} \to \left.|\Lambda\right>=\sum_{m_1,m_2,\ldots}\calC^\Lambda_m|\Lambda_1m_1\left.\right>|\Lambda_2m_2\left.\right>\cdots|\Lambda_Nm_N\left.\right>
\label{eqn:abs_notation}
\endeq 
where $\bfL_i$ acts on the state $|\Lambda_i,m_i\left>\right.$ in the conventional fashion of rotation generators with $\bfL_i^2|\Lambda_i,m_i\left>\right.=\Lambda_i(\Lambda_i+1)|\Lambda_i,m_i\left>\right.$.

The cases $N=2$ (Eq. \ref{eq:nequal2}) and $N=3$ have no intermediate $\Lambda_{12 \ldots}$ values. Applying Eq. (\ref{eq:calC1}) with no intermediate momenta, we find:
\begeqar
\calC^{(\Lambda_1,\Lambda_2)}_{(m_1,m_2)}&=&\frac{(-1)^{\Lambda_1-m_1}}{\sqrt{2\Lambda_1+1}}\;\deltaK_{\Lambda_1,\Lambda_2}\deltaK_{m_1, -m_2},
\label{eq:14a}
\endeqar
and applying it with $\Lambda_{12} \equiv \Lambda_3$ we find
\begeqar
\calC^{(\Lambda_1,\Lambda_2,\Lambda_3)}_{(m_1,m_2,m_3)}&=&(-1)^{\Lambda_1+\Lambda_2+\Lambda_3}\six{\Lambda_1}{\Lambda_2}{\Lambda_3}{m_1}{m_2}{m_3}.\label{eq:14b}
\endeqar
$\deltaK$ is a Kronecker delta, unity when its arguments are equal and zero otherwise.  
The Cartesian versions of some simple $\calY_\Lambda$ functions are given in Appendix \ref{ap:examples}.
\section{Orthonormality and Completeness of $\calY$ Functions}
\label{sec:ortho}
The Clebsch-Gordan coefficients
\begeq
\left<L_1m_1L_2m_2|Lm\right>=\left<Lm|L_1m_1L_2m_2\right>
\endeq
satisfy the orthonormality conditions
\begeqar
\sum_{L,m}\left<L_1m_1L_2m_2|Lm\right>\left<Lm|L'_1m'_1L'_2m'_2\right>&=&\deltaK_{L_1L'_1}\deltaK_{L_2L'_2}\deltaK_{m_1m'_1}\deltaK_{m_2m'_2}\deltaK\{L_1,L_2,L\},\non
\sum_{m_1,m_2}\left<Lm|L_1m_1L_2m_2\right>\left<L_1m_1L_2m_2|L'm'\right>&=&\deltaK_{LL'}\deltaK_{mm'}\deltaK\{L_1,L_2,L\},
\endeqar
where $\deltaK\{L_1,L_2,L\}$ enforces the triangularity condition: it is unity if $|L_1-L_2|\leq L\leq L_1+L_2$ and is zero otherwise.
Writing $d\bfxhat$ for $d\Omega$,
the orthonormality of the spherical harmonics
\begeq
\int\ d\hat{\bf r}\, Y_{LM}(\hat{\bf r}) Y_{L'M'}^*(\hat{\bf r})=\deltaK_{LL'}\deltaK_{MM'},
\endeq
together with the orthonormality of the Clebsch-Gordan coefficients results in the orthonormality of the $\calY_\Lambda$: 
\begeq
\int  \,d\bfXhat\, \calY_{\Lambda}(\bfXhat)\calY^*_{\Lambda'}(\bfXhat)=\deltaK_{\Lambda_1\Lambda'_1}\deltaK_{\Lambda_2\Lambda'_2}\deltaK_{\Lambda_{12}\Lambda'_{12}}\cdots\label{eq:orthonormal}
\endeq
recall that $\bfXhat$ 
stands for $\bfxhat_1,\bfxhat_2,\ldots, \bfxhat_N$, and similarly
$d\bfXhat=d\bfxhat_1\,d\bfxhat_2\cdots d\bfxhat_N$.
The complex conjugation above is trivial since the functions are either purely real or purely imaginary.  

Clearly for functions of $N$ angular variables the set of all products 
$\prod_iY_{\Lambda_i m_i}(\bfxhat_i)$
is complete, enabling us to express any function of the variables as a linear combination of these functions. To expand just those functions that are rotationally invariant we need only the linear combinations of the products of spherical harmonics that are themselves rotationally invariant.

It follows that we can expand any isotropic function $F_{\rm{iso}}(\bfX)$ as
\begeqar
F_{\rm{iso}}(\bfX)&=&\sum_\Lambda \calZ_\Lambda(R)\calY_\Lambda(\bfXhat),
\endeqar
where  $R = (r_1, r_2,\ldots, r_N) = (|\bfx_1|,|\bfx_2|,\ldots, |\bfx_N|)$ and  the radial coefficients $\calZ_\Lambda(R)$ are obtained by orthogonality as 
\begeqar
\calZ_\Lambda(R)&=&\int d\bfXhat\;
F_{\rm{iso}}(\bfX)\calY^*_\Lambda(\bfXhat).
\endeqar
Again, the sum is over all possible $\Lambda$.
\section{Isotropy from Rotational Averaging}
\label{sec:isotropy}

Functions that are isotropic can be generated by averaging over the rotation group.  We indicate a single 3D rotation by $\Rot:\bfx_1,\ldots,\bfx_N\to \Rot\bfx_1,\ldots, \Rot\bfx_N$ and the invariant integration element by $d\Rot$ with the normalization
\begeq
\int d\Rot = 1.
\endeq
Given some function $F(\bfX)$, the rotational average
\begeq
{\cal F}(\bfX) \equiv \int d\Rot\; F(\Rot \bfX)
\endeq
is isotropic, as we will now show. 

For any fixed rotation $\Rot'$ we have $d(\Rot\Rot')=d\Rot$ because, when we average, it does not matter with what rotation we begin. Thus
\begeq
{\cal F}(\Rot' \bfX)=\int d\Rot\; F(\Rot(\Rot'\bfX))=\int d(\Rot\Rot')\;F(\Rot(\Rot'\bfX))={\cal F}(\bfX),
\endeq
demonstrating isotropy. Moreover, if we denote rotational averaging as an operator $\bf{\Pi}$, so that  ${\bf \Pi} (F(\bfX)) = {\cal F}(\bfX)$,  then
${\bf \Pi}$ is a projection operator because it satisfies ${\bf \Pi}^2 = {\bf \Pi}$:
\begeqar
{\bf \Pi}({\bf\Pi}({ F}(\bfX))&=&\int d\Rot' \; \int d\Rot\; F(\Rot(\Rot'\bfX))\non
&=&\int d\Rot'\;{\cal F}(\bfX)={\cal F}(\bfX)={\bf\Pi}(F(\bfX).
\endeqar
${\bf \Pi}$ projects out the $L=0$ states and can be represented symbolically by

\begeq
{\bf \Pi}\to\sum_\Lambda |\Lambda\big>\big<\Lambda|.
\endeq

To find the rotational average of a product of spherical harmonics we use the correspondence Eq. (\ref{eqn:abs_notation}).

Then the projection of the $L=0$ component is determined by the correspondences
\begeqar
\label{eqn:proj_prod}
\prod_i Y_{\Lambda_i m_i}(\bfxhat_i) &\to& \prod_i|\Lambda_i m_i\big >,\\
\int d\Rot\prod_i Y_{\Lambda_i m_i}(\Rot \bfxhat_i) &=& {\bf \Pi}\left\{ \prod_i Y_{\Lambda_i m_i}(\bfxhat_i)\right\} \to \sum_\Lambda |\Lambda\big>\big<\Lambda|\prod_i|\Lambda_i m_i\big>.\nonumber
\endeqar 
From Eq. (\ref{eqn:abs_notation}) we have that
\begeq
\big<\Lambda|\prod_i|\Lambda_i m_i\big >=\calC^\Lambda_m.
\label{eqn:calC_prod}
\endeq
Inserting Eq. (\ref{eqn:calC_prod}) in Eq. (\ref{eqn:proj_prod}) we see that
\begeq
\int d\Rot\prod_i Y_{\Lambda_i m_i}(\Rot \bfxhat_i)\to
\sum_\Lambda|\Lambda\big>\calC^\Lambda_m.
\label{eqn:implicit}
\endeq 
Finally using the correspondence Eq. (\ref{eqn:abs_notation}) in Eq. (\ref{eqn:implicit}) we obtain
\begeq 
\int d\Rot\prod_i Y_{\Lambda_i m_i}(\Rot \bfxhat_i)=\sum_\Lambda\calC^\Lambda_m\calY_\Lambda(\bfXhat)\label{eq:basic}
\endeq
where the sum over $\Lambda$ includes all those that can be made from the primary $\Lambda_i$.
This relation will be used repeatedly.

As an important application of this result consider the special case where $\bfxhat_i=\bfzhat$
for all $i$.  Then using

\begin{align}
  Y_{\ell m}( \bfzhat)=\deltaK_{ m 0}\sqrt{\frac{2\ell+1}{4\pi} }  
\end{align}
we find
\begeq
\calY_\Lambda(\bfzhat,\bfzhat,\ldots, \bfzhat)=(4\pi)^{-N/2}
\prod_{j=1}^N\sqrt{2\Lambda_j+1}\;\calC^\Lambda_0
\endeq
and
\begeqar
\int d\Rot \prod_{j=1}^N Y_{\Lambda_j M_j}(\Rot \bfzhat)
&=&(4\pi)^{-N/2} \prod_{j=1}^N\sqrt{2\Lambda_j+1}\;\sum_\Lambda\calC^\Lambda_M\calC^\Lambda_0\non
 &=& \frac{1}{4\pi}\int d\bfxhat \prod_{j=1}^N Y_{\Lambda_jM_j}(\bfxhat)
\endeqar 
 where the sum is over all $\Lambda$ consistent with the given $\Lambda_j$, $M$ stands for all the $M_j$, and the subscript $0$ stands for a sequence of $N$ zeros.  

We define 
\begeq
\calD^{\rm P}_\Lambda= \prod_{j=1}^N\sqrt{2\Lambda_j+1},
\label{eq:define-calD}
\endeq
where superscript ``${\rm P}$'' denotes that $\calD$ is evaluated at the $primary$ angular momenta $only$, rather than intermediate momenta resulting from couplings within a rotation-invariant function.
With this definition, for any unit vector $\bfxhat$ we find
\begeq
\int d\bfxhat\prod_{j=1}^N Y_{\Lambda_jM_j}(\bfxhat)
=(4\pi)^{1-N/2}\sum_\Lambda\calD^{\rm P}_\Lambda  \calC^\Lambda_0 \calC^\Lambda_M.
\label{eq:single_var}
\endeq 

\section{Contraction of $\calY_\Lambda$ functions}
We now consider integrating out two of the directions, $\bfxhat_i$ and $\bfxhat_j$ by setting $\bfxhat_i=\bfxhat_j=\bfxhat$ and then integrating
over $d^2\bfxhat$, as below:
\begeq
\int d^2\bfxhat\, \calY_\Lambda(\bfxhat_1,\ldots,\bfxhat_i=\bfxhat,\ldots,\bfxhat_j=\bfxhat,\ldots,\bfxhat_N).
\endeq
This integral will vanish unless $\Lambda_i=\Lambda_j$.  

For simplicity, suppose $i=1, j=2$.  Then we would have
\begeqar
&&\int d^2\bfxhat\, \sum_{m_1,m_2,\ldots}Y_{\Lambda_1 m_1}(\bfxhat)Y_{\Lambda_2 m_2}(\bfxhat) \left<\Lambda_1 m_1 \Lambda_2 m_2|\Lambda_{12}m_{12}\right>\left<\Lambda_{12}m_{12}\Lambda_3 m_3|\ldots.\right>\cdots.\non
&&\qquad =\sum_{m_1,m_2,\ldots}(-1)^{m_1}\deltaK_{\Lambda_1\Lambda_2}\deltaK_{m_1,-m_2}\left<\Lambda_1 m_1 \Lambda_2 m_2|\Lambda_{12}m_{12}\right>\left<\Lambda_{12}m_{12}\Lambda_3 m_3|\ldots.\right>\cdots.\non
\endeqar
But we can use
\begeq
\left<00|\Lambda m_1\Lambda  m_2\right>=\frac{(-1)^{\Lambda-m_1}\delta^K_{m_1,-m_2}}{\sqrt{2\Lambda+1}}
\endeq 
to write
\begeqar 
&&\int d^2\bfxhat\, \sum_{m_1,m_2,\ldots}Y_{\Lambda_1 m_1}(\bfxhat)Y_{\Lambda_2 m_2}(\bfxhat) \left<\Lambda_1 m_1 \Lambda_2 m_2|\Lambda_{12}m_{12}\right>\left<\Lambda_{12}m_{12}\Lambda_3 m_3|\ldots.\right>\cdots.\non
&&\qquad =\sqrt{2\Lambda_1+1}(-1)^{\Lambda_1}\deltaK_{\Lambda_1\Lambda_2}\sum_{m_1,m_2,\ldots}\left<00|\Lambda_1m_1\Lambda_2m_2\right>\left<\Lambda_1 m_1 \Lambda_2 m_2|\Lambda_{12}m_{12}\right>\non
&&\qquad\qquad\qquad\times\left<\Lambda_{12}m_{12}\Lambda_3 m_3|\ldots.\right>\ldots.\non
&&\qquad=\sqrt{2\Lambda_1+1}(-1)^{\Lambda_1}\deltaK_{\Lambda_1\Lambda_2}\deltaK_{\Lambda_{12},0}\deltaK_{m_{12},0}\left<\Lambda_{12}m_{12}\Lambda_3 m_3|\ldots.\right>\cdots.
\endeqar
This relation shows that the contracted $\calY$ is non-zero only if $\Lambda_1=\Lambda_2$ and $\Lambda_{12}=0$, conditions as anticipated as a consequence of rotational averaging.  Thus if initially $\Lambda=(\Lambda_1,\Lambda_2=\Lambda_1,\Lambda_{12}=0,\Lambda_3,\Lambda_{123}=\Lambda_3,\Lambda_4,\ldots)$ then defining $\Lambda'=(\Lambda_3,\Lambda_4,\Lambda_{1234}\ldots)$ we have
\begeq
\int d^2\bfxhat\, \calY_\Lambda(\bfxhat,\bfxhat,\bfxhat_3,\ldots,\bfxhat_N)=\sqrt{2\Lambda_1+1}(-1)^{\Lambda_1}\calY_{\Lambda'}.
\endeq
In the event that the two variables to be contracted are not $\bfxhat_1$ and $\bfxhat_2$, we can use the techniques of \S\ref{sec:reordering} to move the involved variables to the first two positions.  The resulting sum of $\calY$ can then be treated as described above.

As a simple example consider
\begeqar
\int d\bfxhat\; \calY_{11(0)11}(\bfxhat,\bfxhat,\bfxhat_3,\bfxhat_4)&=&\int d\bfxhat\; \frac 3{(4\pi)^2}(\bfxhat\cdot\bfxhat)(\bfxhat_3\cdot\bfxhat_4)=\frac 3{4\pi}(\bfxhat_3\cdot\bfxhat_4)\non 
&=&-\sqrt 3 \calY_{11}(\bfxhat_3,\bfxhat_4).
\endeqar 

\section{Reduction of Products of $\calY_\Lambda$ Functions}
\label{sec:products}
\setcounter{figure}{0}  
The completeness of the $\calY_\Lambda$ functions means that it is always possible to express a product of $\calY_\Lambda$ functions as a sum of $\calY_\Lambda$ functions:
\begeq
\calY_\Lambda(\bfXhat)\calY_{\Lambda'}(\bfXhat)=\sum_{\Lambda''}\calOp \calG^{\Lambda\Lambda'\Lambda''}\calY_{\Lambda''}(\bfXhat).
\label{eq:Gdef}
\endeq
We recall that $\calO$ is defined in Eq. (\ref{eq:def-parity}), and is $+1$ for even-parity $\Lambda''$ and $-1$ for odd-parity $\Lambda''$. We have chosen $\calG$ for ``Gaunt'' because the coefficient $\calG$ is a generalized Gaunt integral. Here $\Lambda$ stands for $\Lambda_1,\Lambda_2,\ldots,\Lambda_{12}, \Lambda_{123}\ldots, \Lambda_{N}$ etc. and similarly for $\Lambda'$ and ${\Lambda''}$, while $\bfXhat=(\bfxhat_1,\ldots,\bfxhat_N)$. We have inserted the factor $\calOp$ so that $\calG^{\Lambda\Lambda'\Lambda''}$ will be completely symmetric in its arguments. Using orthonormality and the parity relation $\calY_{\Lambda''}^*(\bfXhat)=\calOpp \calY_{\Lambda''}(\bfXhat)$ (see Eq. \ref{eq:def-parity}), we find
\begeqar
\label{eq:calG_integral}
\calG^{\Lambda\Lambda'\Lambda''}&=&\int d\bfXhat\,\calY_\Lambda(\bfXhat)\calY_{\Lambda'}(\bfXhat)\calY_{\Lambda''}(\bfXhat)\\
&=&\sum_{MM'M''}\calC^{\Lambda}_M\,\calC^{\Lambda'}_{M'}\,\calC^{{\Lambda''}}_{M''}\,\int d\bfXhat\,\, \left[Y_{\Lambda_1M_1}(\bfxhat_1)\cdots Y_{\Lambda_{N}M_{N}}(\bfxhat_{N})\right]\nonumber\\
&&\;\;\times\, \left[Y_{\Lambda'_1M'_1}(\bfxhat_1)\cdots Y_{\Lambda'_{N}M'_{N}}(\bfxhat_{N})\right]
 \left[ Y_{{\Lambda''}_1M''_1}(\bfxhat_1)\cdots Y_{{\Lambda''}_{N}M''_{N}}(\bfxhat_{N})\right].\nonumber
\endeqar
Here $M$ stands for the complete expression  $(M_1,M_2,M_{12})\ldots,
M_{N}$, as in Eqs. (\ref{eq:calC1}) and (\ref{eq:calC2}). 

The angular integrals are given by Eq. (\ref{eq:single_var}) with $N=3$. This is because we are integrating across each angular variable $\bfxhat_i$ separately, and each appears three times. We write out the $i^{th}$ such integral below to show explicitly the values of the components $\Lambda_i$ of $\Lambda$ and similarly for $M$; there will be $N$ copies of this integral as $i$ runs from $1$ to $N$.
\begin{align}
&\int d\bfxhat_i\; Y_{\Lambda_i M_i}(\bfxhat_i) Y_{\Lambda'_i M'_i}(\bfxhat_i) Y_{{\Lambda''}_i M''_i}(\bfxhat_i) \nonumber\\
 &= (4\pi)^{-1/2}\calD^{\rm P}_{(\Lambda_i \Lambda_i' \Lambda_i'')}\calC^{(\Lambda_i \Lambda_i' \Lambda_i'')}_0 \calC^{(\Lambda_i \Lambda_i' \Lambda_i'')}_{(M_i M_i' M_i'')};
 \label{eq:Ylm_triple}
\end{align}
$\calD^{\rm P}$ is defined in Eq. (\ref{eq:define-calD}) and $\calC$ in Eq. (\ref{eq:calC1}). 
Altogether

\begin{align}
\calG^{\Lambda \Lambda' \Lambda''} = (4\pi)^{-N/2} 
\left[\prod_i\calD^{\rm P}_{(\Lambda_i \Lambda_i' \Lambda_i'')}\calC^{(\Lambda_i \Lambda_i' \Lambda_i'')}_0\right]\calQ^{\Lambda \Lambda' \Lambda''}
\label{eq:AQ}
\end{align}
where we have defined
\begeq
\calQ^{\Lambda\Lambda'\Lambda''} \equiv \sum_{M_i,M'_i,M''_i}\left[\prod_{i=1}^N\calC^{(\Lambda_i\Lambda'_i\Lambda''_i)}_{(M_i,M'_iM''_i)}\right]
\calC^\Lambda_M \calC^{\Lambda'}_{M'} \calC^{\Lambda''}_{M''}
\label{eq:define_calQ}
\endeq 
with $M=(M_1,M_2,\ldots, M_N)$ and similarly for $M'$ and $M''$. The $N$-fold product of $\calD^P$ in Eq. (\ref{eq:AQ}) and the $N$-fold product of $\calC$ within $\calQ$ come from Eq. (\ref{eq:Ylm_triple}); the factor of $\calC^\Lambda_M \calC^{\Lambda'}_{M'} \calC^{\Lambda''}_{M''}$ within $\calQ$ comes from the pre-factor on the angular integral in Eq. (\ref{eq:calG_integral}) and originally 
stems from the pre-factors of the three $\calY$ involved in this calculation.
\subsection{Evaluation of $Q^{\Lambda\Lambda'\Lambda''}$}

Parity requires that among $\Lambda,\Lambda',\Lambda''$ either two or none are parity-odd so 
\begeq 
\prod_{i=1}^N\calC^{(\Lis)}_{(\Mis)}=\prod_{i=1}^N\calC^{(\Lis)}_{(-M_i -M'_i -M''_i)}
\endeq 
and using Eq. (\ref{eqn:C_threej}) for the $\calC$ we can write
\begin{align}
\label{eq:mainresult}
&\calQ^{\Lambda\Lambda'{\Lambda''}}=\sqrt{(2\Lambda_{12}+1)\cdots(2\Lambda_{12\ldots N-2}+1)}\\
&\times\sqrt{(2\Lambda'_{12}+1)\cdots(2\Lambda'_{12\ldots N-2}+1)} \sqrt{(2{\Lambda''}_{12}+1)\cdots(2{\Lambda''}_{12\ldots N-2}+1)}\non
&\times\sum_{M_iM_{12},M_{123}\ldots}(-1)^{\kappa_\Lambda}\six{\Lambda_1}{\Lambda_2}{\Lambda_{12}}{M_1}{M_2}{-M_{12}}\ldots\six{\Lambda_{12\ldots N-2}}{\Lambda_{N-1}}{\Lambda_{N}}{M_{12\ldots N-2}}{M_{N-1}}{M_{N}}\non
&\times\sum_{M'_i,M'_{12},M'_{123}\ldots}(-1)^{\kappa_{\Lambda'}}\six{\Lambda'_1}{\Lambda'_2}{\Lambda'_{12}}{M'_1}{M'_2}{-M'_{12}}\ldots\six{\Lambda'_{12\ldots N-2}}{\Lambda'_{N-1}}{\Lambda'_{N}}{M'_{12\ldots N-2}}{M'_{N-1}}{M'_{N}}\non
&\times\sum_{M''_i,M''_{12},M''_{123}\ldots}(-1)^{\kappa_{\Lambda''}}\six{{\Lambda''}_1}{{\Lambda''}_2}{{\Lambda''}_{12}}{M''_1}{M''_2}{-M''_{12}}\ldots\six{{\Lambda''}_{12\ldots N-2}}{{\Lambda''}_{N-1}}{{\Lambda''}_{N}}{M''_{12\ldots N-2}}{M''_{N-1}}{M''_{N}}\non
&\times \six{\Lambda_1}{\Lambda'_1}{{\Lambda''}_1}{-M_1}{-M'_1}{-M''_1}\ldots\six{\Lambda_N}{\Lambda'_N}{{\Lambda''}_N}{-M_N}{-M'_N}{-M''_N}.\nonumber
\end{align}
The pre-factors and the three lines of sums stem from the individual $\calC$, while the last line of 3-$j$ symbols coupling unprimed, primed, and double-primed momenta stems from the angular integrals over each $\bfxhat_i$. Here the factors of $(-1)$ are determined by Eq. (\ref{eq:caly3j2})
\begeq
\kappa_\Lambda=\Lambda_{12}-M_{12}+\ldots+\Lambda_{12\ldots N-2}-M_{12\ldots N-2}
\label{eq:kappa}
\endeq 
and similarly for $\kappa_{\Lambda'}$ and $\kappa_{\Lambda''}$.  Any possible factor $(-1)^{\sum_i(\Lambda_i+\Lambda'_i+\Lambda''_i)}$ can be ignored because, of the set $\calY_\Lambda, \calY_{\Lambda'}$ and $\calY_{\Lambda''}$, either none or two of them are odd parity.  

We separate the result into two factors:
\begeq 
\calQ^{\Lambda\Lambda'\Lambda''}=\calS_{\Lambda\Lambda'\Lambda''}
\calP^{\Lambda\Lambda'\Lambda''}
\label{eq:QTP}
\endeq 
with
\begeqar
\calS_{\Lambda\Lambda'\Lambda''}&\equiv &\sqrt{(2\Lambda_{12}+1)\ldots(2\Lambda_{12\ldots N-2}+1)}\sqrt{(2\Lambda'_{12}+1)\ldots(2\Lambda'_{12\ldots N-2}+1)}\non 
&&\qquad\qquad \times\sqrt{(2{\Lambda''}_{12}+1)\ldots(2{\Lambda''}_{12\ldots N-2}+1)}.
\label{eq:T}
\endeqar
$\calS$ is the analog of ${\cal D}^{\rm P}$ as defined in Eq. (\ref{eq:define-calD}) but is a product over only the $intermediate$
angular momenta, which come from coupling within each rotation-invariant function; hence the superscript ``${\rm I}$''. $\calP^{\Lambda\Lambda'\Lambda''}$, for ``sums'', is given by the remaining four lines of Eq. (\ref{eq:mainresult}) (the three sums and then the final line's 3-$j$ symbols). We note that only $\calQ$ with intermediate momenta require a factor ${\cal D}^{\rm I}$; otherwise this factor 
is unity.

In the expression for $\calP$ every $M$-type index occurs twice with, once with a positive sign and once with a negative sign. Each occurrence has the same $\Lambda$-type value. The intermediate quantities occur in $\kappa$ always as differences such as $\Lambda_{12}-M_{12}$.  Analogous differences like $\Lambda_1-M_1$ can be added in to the power because sums such as $M_1+\ldots +M_N$ vanish and because the sum $\sum_i(\Lambda_i+\Lambda'_i+{\Lambda''}_i)$ is necessarily even by parity arguments.  As a consequence, for every entry column with $j$ over $m$ in a 3-$j$ symbol there is a factor $(-1)^{j-m}$.  Thus $\calP^{\Lambda\Lambda'\Lambda''}$ fulfills all the requirements described in Appendix \ref{ap:yutsis} to be expressed as a Yutsis diagram.  
 \subsection{ $\calG^{\Lambda\Lambda'\Lambda''}$ for $N=2$} 
 Eq. (\ref{eq:AQ}) is the generalization of the Gaunt integral, which evaluates the integral of the product of three Legendre polynomials between $-1$ and $1$.  Explicitly, for $N=2$, with $\Lambda=(\ell,\ell), \Lambda'=(\ell',\ell'),\Lambda''=(\ell'',\ell'')$, we use Eq. (\ref{eq:AQ}) to obtain $\calG$. First, we have from Eq. (\ref{eq:define-calD}) that
 \begeq 
 \calD^{\rm P}_{(\ell \ell' \ell'')} \mathcal{C}_0^{(\ell \ell' \ell'')}=\sqrt{(2\ell+1)(2\ell'+1)(2\ell''+1)}\six\ell{\ell'}{\ell''}000 .
 \endeq
 The 3-$j$ symbol requires that $\ell+\ell'+\ell''$ is even, a consequence of parity. We next find that $\calQ$ is
 \begeqar
 &\calQ^{\Lambda\Lambda'\Lambda''}=\left[(2\ell+1)(2\ell'+1)(2\ell''+1)\right]^{-1/2}\nonumber\\
 &\times\sum_{m_1,m'_1,m''_1}\six\ell{\ell'}{\ell''}{m_1}{m'_1}{m''_1}\six\ell{\ell'}{\ell''}{-m_1}{-m'_1}{-m''_1}\nonumber\\
 &=\left[(2\ell+1)(2\ell'+1)(2\ell''+1)\right]^{-1/2},
 \endeqar
 where we applied the orthogonality relation for 3-$j$ symbols Eq. (\ref{eq:3j_orth_two}) to perform the sums in the middle line.
 Combining $\calD$ and $\calQ$ as dictated by Eq. (\ref{eq:AQ}) for $\calG$ we obtain
 \begeqar
 \calG^{\Lambda\Lambda'\Lambda''}&=&\frac{\sqrt{(2\ell+1)(2\ell'+1)(2\ell''+1)}}{4\pi}\six\ell{\ell'}{\ell''}000^2,
 \endeqar
 in agreement (up to pre-factors) with the Gaunt integral
 \begeq 
 \int_{-1}^1 d\mu\; P_\ell(\mu)P_{\ell'}(\mu)P_{\ell''}(\mu)= 2
 \six{\ell}{\ell'}{\ell''}000^2.
 \label{eq:G_for_two}
\endeq

We can demonstrate this with a calculation in Cartesian form for $N=2$ using the examples in Appendix \ref{ap:examples}.

\begeqar
\calY_{00}(\bfxhat_1,\bfxhat_2)&=&\frac1{4\pi}\non
\calY_{11}(\bfxhat_1,\bfxhat_2)&=-&\sqrt{\frac{3}{(4\pi)^2}}\;\left(\bfxhat_1\cdot\bfxhat_2\right)\non
\calY_{22}(\bfxhat_1,\bfxhat_2)&=&\frac 32\sqrt{\frac{5}{(4\pi)^2}}\left[(\bfxhat_1\cdot\bfxhat_2)^2-\frac 13\right]
\endeqar
from which we find directly that
\begin{align}
&\calY_{11}(\bfxhat_1,\bfxhat_2)\calY_{11}(\bfxhat_1,\bfxhat_2)=\frac 3{(4\pi)^2}(\bfxhat_1\cdot\bfxhat_2)^2\nonumber\\
&=\frac 1{4\pi}\left[\frac 2{\sqrt 5}\calY_{22}(\bfxhat_1,\bfxhat_2)+\calY_{00}(\bfxhat_1,\bfxhat_2)\right].
\label{eq:explicit_N_two}
\end{align}
Alternatively, inserting the appropriate values of the angular momenta into Eq. (\ref{eq:G_for_two}) for $\calG$ for the $N=2$ functions, and then using that
\begeq
\six 112000^2=\frac 2{15};\quad \six 110000^2=\frac 13,
\endeq
we find agreement with the coefficients obtained by explicit calculation in Eq. (\ref{eq:explicit_N_two}). 

 \subsection{Products of $\calY_\Lambda$ for $N=3$} 
 We indicate the components of $\Lambda$ by $\Lambda_1,\Lambda_2,\Lambda_3$, and similarly for $\Lambda'$ and $\Lambda''$. For $N=3$ there is no independent $\Lambda_{12}$, but rather $\Lambda_{12}=\Lambda_3$. We recall that for $\calG^{\Lambda \Lambda' \Lambda''}$ (Eq. \ref{eq:AQ}), we need the $\calD^{\rm P}_{\Lambda_i \Lambda'_i \Lambda''_i}$ and also $\calQ^{\Lambda \Lambda' \Lambda''}$. We focus first on $\calQ^{\Lambda \Lambda' \Lambda''}$, which is split into $\calSL$ and $\calP^{\Lambda,\Lambda',\Lambda''}$ via Eq. (\ref{eq:QTP}). Since there is no independent $\Lambda_{12}$, $\calSL=1$ and we have from Eq. (\ref{eq:QTP}) that
 \begeqar
 \label{eq:N3}
 &&\calQ^{\Lambda \Lambda' \Lambda''}= \calP^{\Lambda \Lambda'\Lambda''}=\non
 &&\sum_{M,M',M''}\six{\Lambda_1}{\Lambda_2}{\Lambda_3}{M_1}{M_2}{M_3}\six{\Lambda'_1}{\Lambda'_2}{\Lambda'_3}{M'_1}{M'_2}{M'_3}\six{{\Lambda''}_1}{{\Lambda''}_2}{{\Lambda''}_3}{M''_1}{M''_2}{M''_3}\\
&&\times \six{\Lambda_1}{\Lambda'_1}{{\Lambda''}_1}{-M_1}{-M'_1}{-M''_1}\six{\Lambda_2}{\Lambda'_2}{{\Lambda''}_2}{-M_2}{-M'_2}{-M''_2}\six{\Lambda_3}{\Lambda'_3}{{\Lambda''}_3}{-M_3}{-M'_3}{-M''_3}.\nonumber
\endeqar
This expression satisfies the conditions for a Yutsis diagram, as described more fully in Appendix \ref{ap:yutsis} and summarized below:

\begin{enumerate}
    \item Each 3-$j$ symbol is represented by a vertex joining three lines.
    \item A line connects vertices with elements of the form
    $(j,m)$ and $(j,-m)$.
    \item At each vertex assign $+$ if the order, left to right, of the $j_i$ in the 3-$j$ symbol is counterclockwise in the diagram.  Otherwise assign $-$.
    \end{enumerate}

 \begin{figure}
\begin{center}
\includegraphics[width=3.5 in]{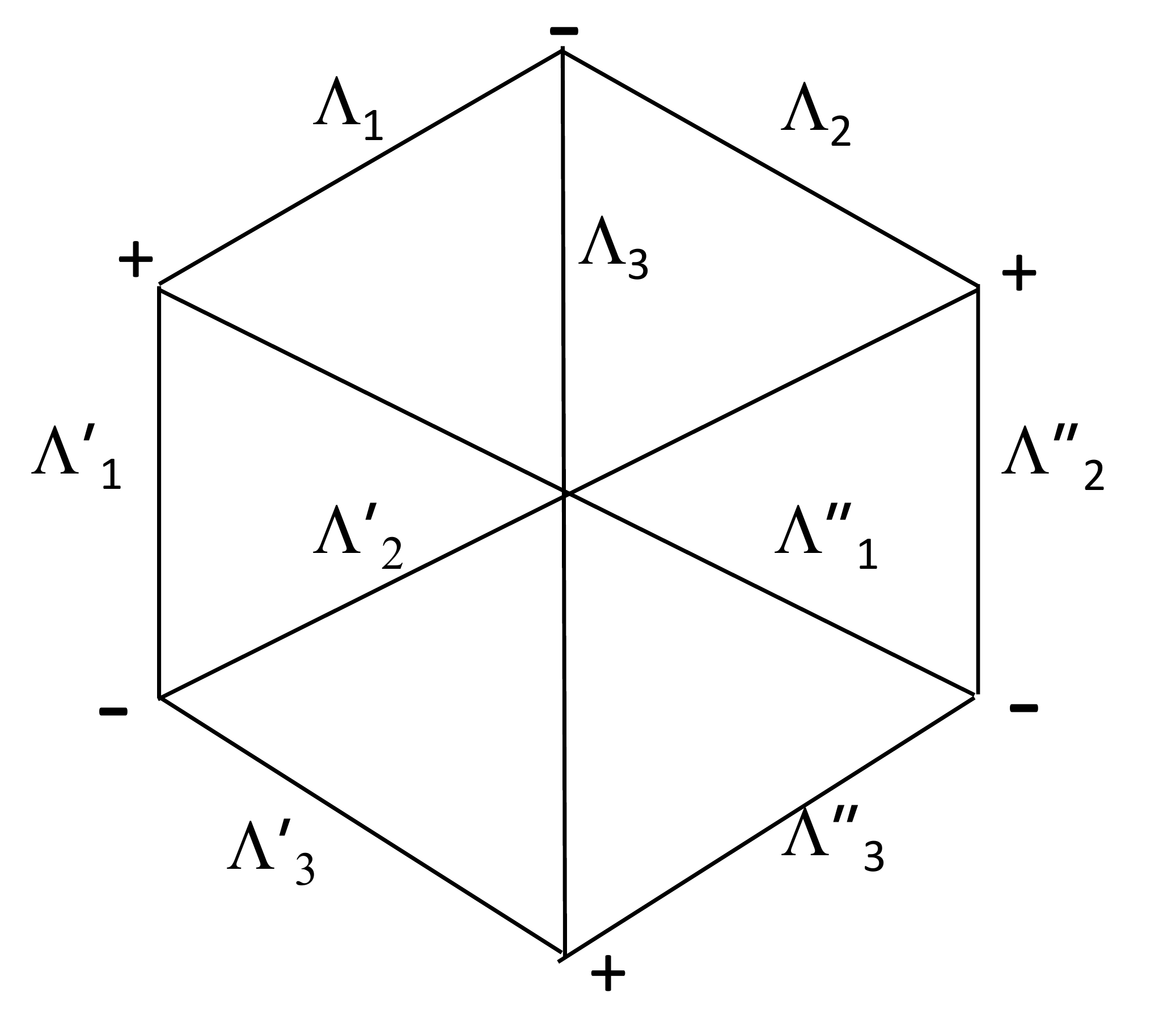}
\caption{Yutsis diagram representing Eq. (\ref{eq:N3}). We note that to determine signs at the vertices for some pairs (e.g. $\Lambda_1', \Lambda_3'$ one needs to read the 3-$j$ in a cyclic (ring-like) fashion. For instance, with the ordering $\Lambda_1' \Lambda_2' \Lambda_3'$, one should read cyclically as $\Lambda_1' \Lambda_2' \Lambda_3' \Lambda_1' \ldots$ so that $\Lambda_3'$ in fact comes before $\Lambda_1'$. This ordering is obtained by moving clockwise on the Yutsis diagram above, so we assign a $-$ sign at the vertex between $\Lambda_3'$ and $\Lambda_1'$.} 
\label{fig:5pt}
\end{center}
\end{figure}

We find the representation of Eq. (\ref{eq:N3}) shown in Fig. \ref{fig:5pt}. Comparing with Fig. \ref{fig:9j} or Eq. (\ref{eq:9pt})
we identify the 9-$j$ symbol 
\begeq \calQ^{\Lambda \Lambda' \Lambda''}= \calP^{\Lambda \Lambda' \Lambda''}=\nine{\Lambda_1}{\Lambda'_1}{{\Lambda''}_1}{\Lambda_2}{\Lambda'_2}{{\Lambda''}_2}{\Lambda_3}{\Lambda'_3}{{\Lambda''}_3}.
\endeq 
Inserting this result into Eq. (\ref{eq:AQ}), as well as computing $\calD^{\rm P}$ and inserting it also, we find
\begeqar
\calG^{\Lambda\Lambda'\Lambda''}&=(4\pi)^{-3/2}\left[\prod_{i=1}^3\sqrt{(2\Lambda_i+1)(2\Lambda'_i+1)(2\Lambda''_i+1)} \six{\Lambda_i}{\Lambda'_i}{{\Lambda''}_i}{0}{0}{0}\right]\nonumber\\
&\times \nine{\Lambda_1}{\Lambda'_1}{{\Lambda''}_1}{\Lambda_2}{\Lambda'_2}{{\Lambda''}_2}{\Lambda_3}{\Lambda'_3}{{\Lambda''}_3}.
\label{eq:calG_Nthree}
\endeqar

As an example, let us consider $\calY_{110}(\bfxhat_1,\bfxhat_2,\bfxhat_3)\calY_{011}(\bfxhat_1,\bfxhat_2,\bfxhat_3)$, where we indicate $\Lambda$ in the subscripts by $\Lambda_1,\Lambda_2,\Lambda_3$. We will use:
\begeqar
\calY_{110}(\bfxhat_1,\bfxhat_2,\bfxhat_3)&=&-\sqrt{\frac{3}{(4\pi)^3}}\left(\bfxhat_1\cdot\bfxhat_2\right),\non
\calY_{011}(\bfxhat_1,\bfxhat_2,\bfxhat_3)&=&-\sqrt{\frac{3}{(4\pi)^3}}\left(\bfxhat_2\cdot\bfxhat_3\right),\non
\calY_{121}(\bfxhat_1,\bfxhat_2,\bfxhat_3)&=&\sqrt{\frac{27}{2(4\pi)^3}}\left[(\bfxhat_1\cdot\bfxhat_2)\,(\bfxhat_3\cdot\bfxhat_2)-\frac 13\bfxhat_1\cdot\bfxhat_3\right].
\label{eq:list_to_use}
\endeqar
We note that the second line comes from simply interchanging arguments $3 \leftrightarrow 1$ in the first line; $\calY_{121}$ similarly comes from interchanging $2 \leftrightarrow 3$ in $\calY_{112}$ as given in Appendix \ref{ap:examples}.
We now may determine the product
\begeqar
&&\calY_{110}(\bfxhat_1,\bfxhat_2,\bfxhat_3)\calY_{011}(\bfxhat_1,\bfxhat_2,\bfxhat_3)=\frac{3}{(4\pi)^3}(\bfxhat_1\cdot\bfxhat_2)\,(\bfxhat_2\cdot\bfxhat_3)\non
&&\qquad=\frac 1{(4\pi)^{3/2}}\left[\sqrt{\frac{2}3}\calY_{121}(\bfxhat_1,\bfxhat_2,\bfxhat_3)-\sqrt{\frac13} \calY_{101}(\bfxhat_1,\bfxhat_2,\bfxhat_3)\right].
\label{eq:explicit_N3_result}
\endeqar
We note that $\calY_{101}$ is obtained by interchanging arguments as $1 \leftrightarrow 2$ in $\calY_{011}$ of Eq. (\ref{eq:list_to_use}).

We can compare Eq. (\ref{eq:explicit_N3_result}) to our general result Eq. (\ref{eq:Gdef}) using the values
\begeqar
\calG^{(110) (011)(121)}&=&\frac{27\sqrt{5}}{(4\pi)^{3/2}}
\nine110011121\six{1}10{0}{0}{0}\six 011{0}{0}{0}\six1210{0}{0}\non
&=&\frac{27\sqrt 5}{(4\pi)^{3/2}}\frac 19\left(-\frac 1{\sqrt 3}\right)^2\sqrt{\frac 2{15}}=\frac{1}{(4\pi)^{3/2}}\;\sqrt\frac{2}{3} .
\label{eq:coeff_1}
\endeqar
and
\begeqar
\calG^{(110) (011)(101)}&=&\frac{27}{(4\pi)^{3/2}}
\nine110011101
\six{1}10{0}{0}{0}\six 011{0}{0}{0}\six1010{0}{0}\non
&=&\frac{27}{(4\pi)^{3/2}}\frac 19\left(-\frac 1{\sqrt 3}\right)^3=\frac{1}{(4\pi)^{3/2}}\left(-\sqrt{\frac{1}{3}}\right).
\label{eq:coeff_2}
\endeqar
We have used \textsc{sympy.physics.wigner} to evaluate the 9-$j$ and 3-$j$ symbols above; these can also be computed using a program by A. Stone available through a web interface.\footnote{\url{http://www-stone.ch.cam.ac.uk/wigner.shtml}} We find agreement between the coefficients as given by Eqs. (\ref{eq:coeff_1}) and (\ref{eq:coeff_2}) and as given by Eq. (\ref{eq:explicit_N3_result}).

An example involving parity-odd functions is
\begeqar
&&\calY_{111}(\bfxhat_1,\bfxhat_2,\bfxhat_3)\calY_{111}(\bfxhat_1,\bfxhat_2,\bfxhat_3)=\frac 1{(4\pi)^{3/2}}\Big[
\frac{\sqrt{14}}5\calY_{222}(\bfxhat_1,\bfxhat_2,\bfxhat_3)+\non
&&\qquad +\frac 1{\sqrt 5}\left[\calY_{220}(\bfxhat_1,\bfxhat_2,\bfxhat_3)+\calY_{202}(\bfxhat_1,\bfxhat_2,\bfxhat_3)+\calY_{022}(\bfxhat_1,\bfxhat_2,\bfxhat_3)\right]\non
&&\qquad\qquad-\calY_{000}(\bfxhat_1,\bfxhat_2,\bfxhat_3)\Big].
\endeqar
The functions used above are listed in Appendix \ref{ap:examples}. We inserted in Eq. (\ref{eq:calG_Nthree}) $(\Lambda_1, \Lambda_2, \Lambda_3) = (1, 1, 1)$, $(\Lambda'_1, \Lambda'_2, \Lambda'_3) = (1, 1, 1)$, and $(\Lambda''_1, \Lambda''_2, \Lambda''_3) = (2, 2, 2)$ to obtain the $\calY_{222}$ coefficient, $(\Lambda''_1, \Lambda''_2, \Lambda''_3) = (2, 2, 0)$ to obtain the $\calY_{220}$ coefficient, $(\Lambda''_1, \Lambda''_2, \Lambda''_3) = (2, 0, 2)$ to obtain the $\calY_{202}$ coefficient, etc. We note that the middle three coefficients must be the same by inspection because of the interchange symmetry of the left-hand side.

\subsection{Products of $\calY_\Lambda$ for $N=4$} 
\begin{figure}
\begin{center}
\includegraphics[width=5.5 in]{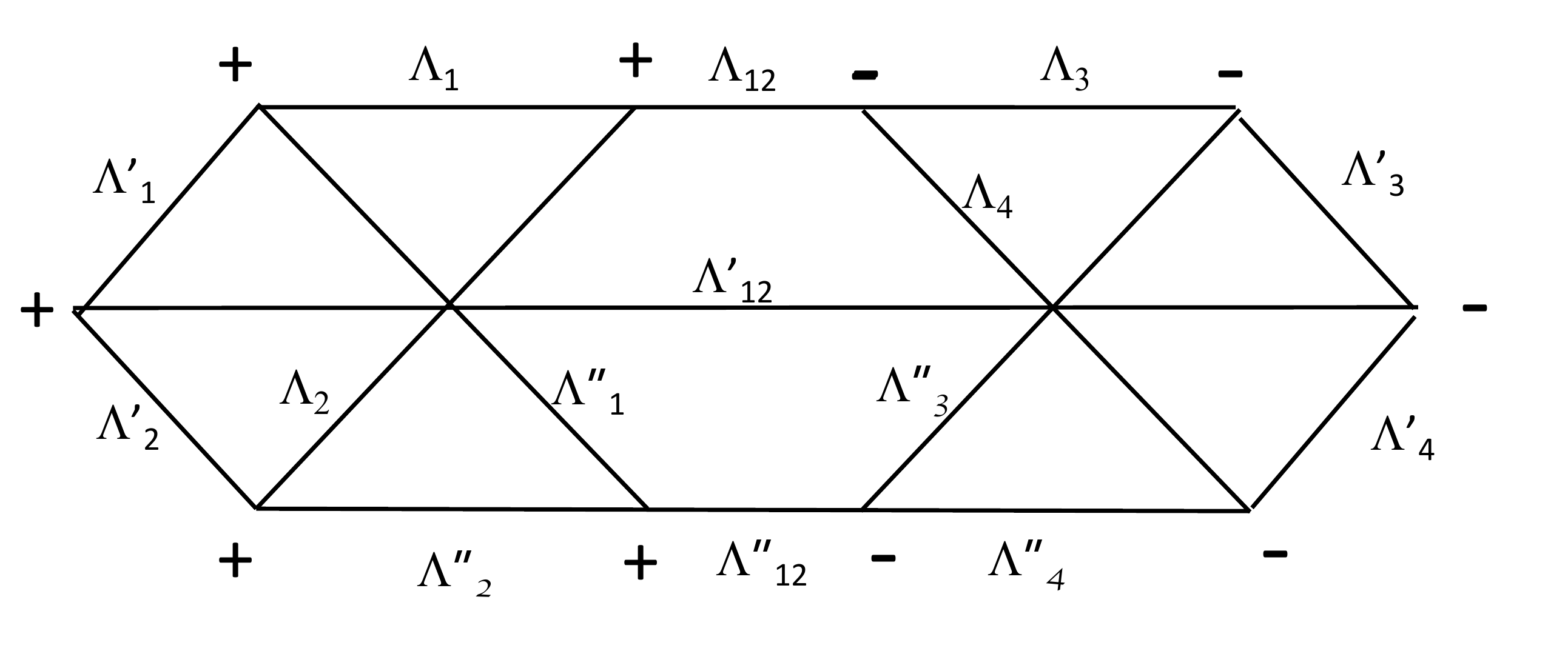}
\caption{Yutsis diagram describing the coefficient of $\calY_{\Lambda''}$ in the product  of the four-point functions $\calY_\Lambda$ and $\calY_{\Lambda'}$.} \label{fig:Neq4}
\end{center}
\end{figure}

\begin{figure}
\begin{center}
\includegraphics[width=5.5 in]{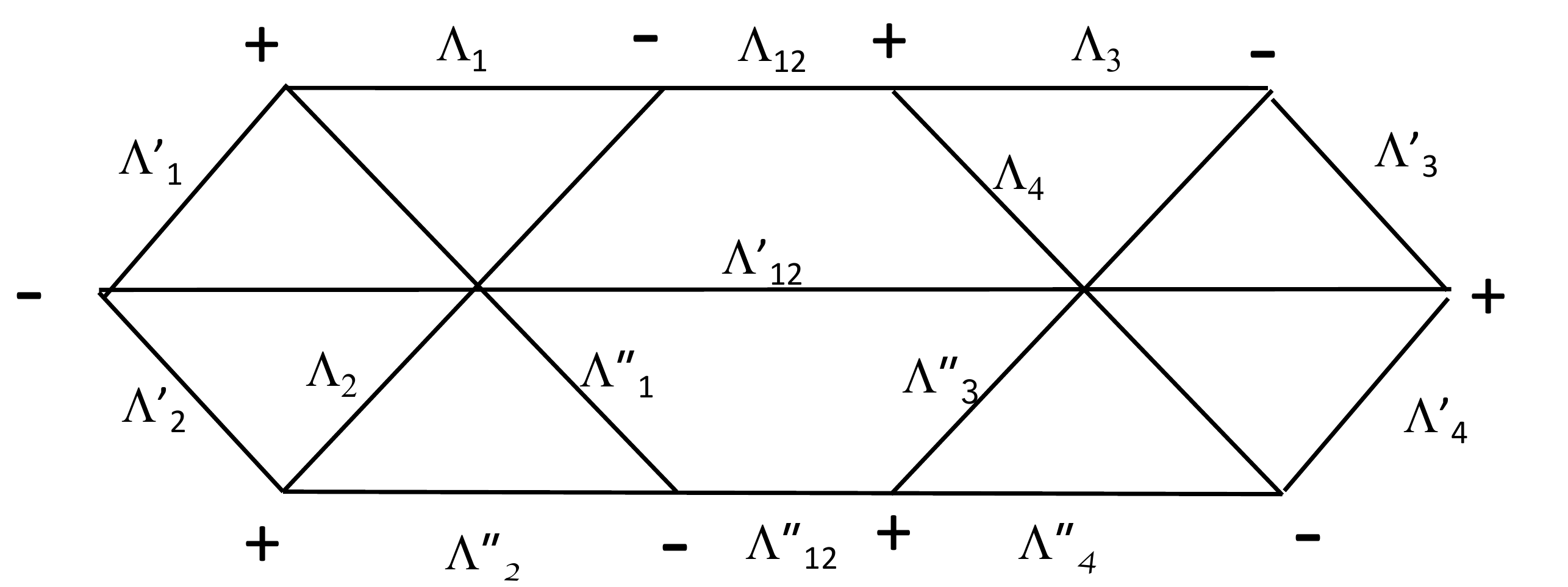}
\caption{Yutsis diagram  describing the coefficient of $\calY_{\Lambda''}$ in the product  of the four-point functions $\calY_\Lambda$ and $\calY_{\Lambda'}$, but with the factors associated with some vertices changed to create the canonical alternating signs at the vertices.} 
\label{fig:5pt-flipped}
\end{center}
\end{figure}

For $N=4$ there are intermediate angular momenta: $\Lambda_{12},\; \Lambda'_{12},$ and $\Lambda''_{12}$. We recall that the general expansion coefficients are given by Eq. (\ref{eq:mainresult}), and the phases required for the Yutsis diagram are supplied by the factor
$
(-1)^{\kappa_\Lambda+\kappa_{\Lambda'}+\kappa_{\Lambda''}}
$
in this result, where $\kappa$ is itself defined in Eq. (\ref{eq:kappa}). To construct the Yutsis diagram it suffices to examine the upper portions of the 3-$j$ symbols dictated by Eq. (\ref{eq:mainresult}) since all the various $m_i$ are simply summed over. The upper portion written out schematically is:
\begeqar
&&(\Lambda_1,\Lambda_2,\Lambda_{12});(\Lambda'_1,\Lambda'_2,\Lambda'_{12});({\Lambda''}_1,{\Lambda''}_2,{\Lambda''}_{12});\non
&&(\Lambda_{12},\Lambda_3,\Lambda_4);(\Lambda'_{12},\Lambda'_3,\Lambda'_4)({\Lambda''}_{12},{\Lambda''}_3,{\Lambda''}_4);\non
&&(\Lambda_1,\Lambda'_1,{\Lambda''}_1);(\Lambda_2,\Lambda'_2,{\Lambda''}_2);(\Lambda_3,\Lambda'_3,{\Lambda''}_3);(\Lambda_4,\Lambda'_4,{\Lambda''}_4).
\endeqar
From this we derive the Yutsis diagram in Fig. \ref{fig:Neq4}.
The plus and minus signs are determined by the order of the $j$ values in the various 3-$j$ symbols.  To reach the canonical forms we change some of the signs to produce Fig. \ref{fig:5pt-flipped}. At each vertex where the sign is reversed a factor $(-1)^{j_a+j_b+j_c}$ is induced, where $j_a,j_b$ and $j_c$ are the angular momenta in the associated 3-$j$ symbol. In our set of flips, the overall factor is $-1$ raised to the sum of the $\Lambda_i$, the $\Lambda'_i$, the ${\Lambda''}_i$, and $2(\Lambda_{12} + \Lambda'_{12} + {\Lambda''}_{12})$. This sum is even, so the overall factor is unity.

\begin{figure}
\begin{center}
\includegraphics[width=5.5 in]{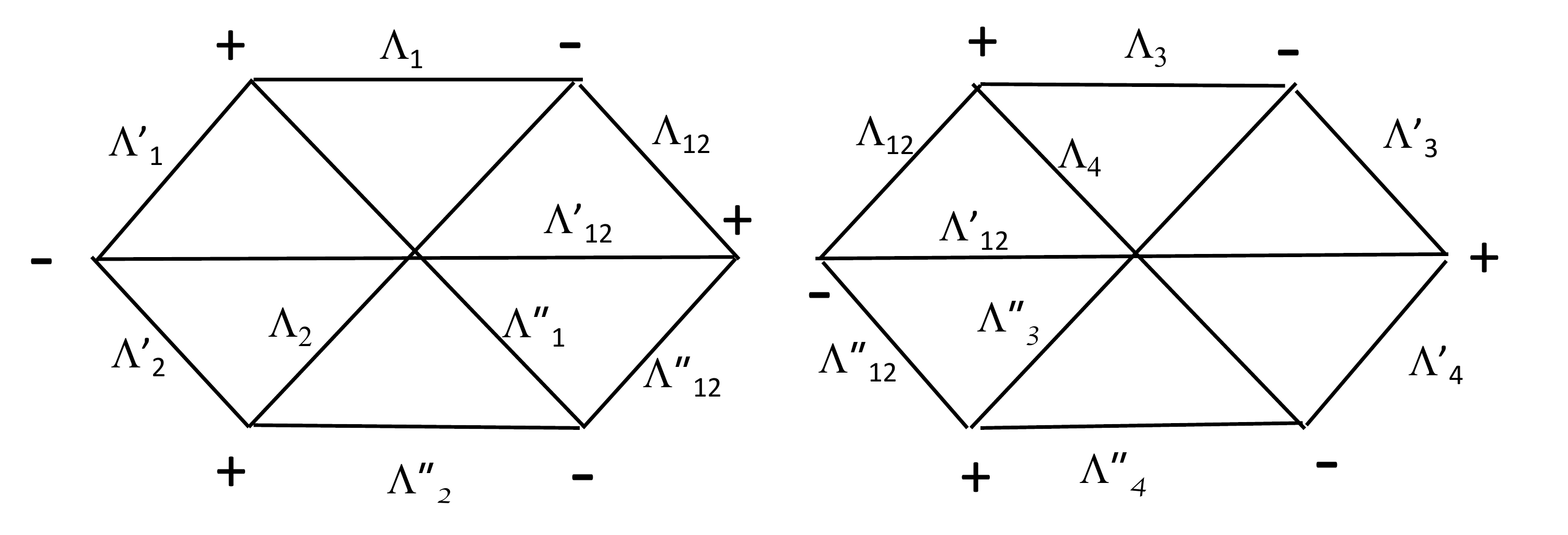}
\caption{Yutsis diagram showing the result of cutting the three lines in Fig. \ref{fig:5pt-flipped} corresponding to $\Lambda_{12},\Lambda'_{12},$ and $\Lambda''_{12}$. } 
\label{fig:Neq4-split}
\end{center}
\end{figure}

\cite{Yutsis_1962} show that any diagram that can be separated into two disconnected pieces by cutting just three lines is a product of two simpler factors.  This is a consequence of their theorem (see Appendix \ref{ap:theorem}) that a product $F_{(j_1,m_1), (j_2,m_2), (j_3,m_3)}$ of 3-$j$ symbols with three unsummed  $j,m$ columns, $(j_1,m_1), (j_2,m_2), (j_3,m_3)$, is necessarily proportional to a 3-$j$ symbol
\begeq
F_{(j_1,m_1), (j_2,m_2), (j_3,m_3)}=\six{j_1}{j_2}{j_3}{m_1}{m_2}{m_3}F.
\endeq
The factored diagram stemming from Fig. \ref{fig:5pt-flipped} is shown in Fig. \ref{fig:Neq4-split}.

To compute $\calG^{\Lambda \Lambda' \Lambda''}$ using Eq. (\ref{eq:AQ}), we need $\calQ^{\Lambda \Lambda' \Lambda''}$, which in turn requires $\calP^{\Lambda \Lambda' \Lambda''}$ by Eq. (\ref{eq:QTP}).
Comparing the two diagrams in Fig. \ref{fig:Neq4-split} to that for the canonical 9-j symbol, given in Fig. \ref{fig:9j}, we find that
\begeqar
\label{eq:4pt_calS}
\calP^{\Lambda\Lambda'\Lambda''}&=&\nine{\Lambda_1}{\Lambda_{12}}{\Lambda_2}{\Lambda'_1}{\Lambda'_{12}}{\Lambda'_2}{{\Lambda''}_1}{\Lambda''_{12}}{{\Lambda''}_2}
\nine{\Lambda_3}{\Lambda_{12}}{\Lambda_4}{\Lambda'_3}{\Lambda'_{12}}{\Lambda'_4}{{\Lambda''}_3}{{\Lambda''}_{12}}{{\Lambda''}_4}\non
&=&\nine{\Lambda_1}{\Lambda_{2}}{\Lambda_{12}}{\Lambda'_1}{\Lambda'_{2}}{\Lambda'_{12}}{{\Lambda''}_1}{{\Lambda''}_{2}}{{\Lambda''}_{12}}
\nine{\Lambda_{12}}{\Lambda_3}{\Lambda_4}{\Lambda'_{12}}{\Lambda'_3}{\Lambda'_4}{{\Lambda''}_{12}}{{\Lambda''}_3}{{\Lambda''}_4}.
\endeqar
We note that we chose $j_1 = \Lambda_1$, $j_2 = \Lambda_{12}$ when comparing the left-hand Yutsis diagram of Fig. \ref{fig:Neq4-split} to Fig. \ref{fig:9j}. For the right-hand Yutsis diagram of Fig. \ref{fig:Neq4-split}, we chose $j_1 =\Lambda_3$, $j_2 =\Lambda_{12} $ 
when comparing to Fig. \ref{fig:9j}; this was motivated by wanting to preserve mirror symmetry regarding the role of $\Lambda_{12}$ between the two halves of Fig. \ref{fig:Neq4-split}. We note that the signs in Fig. \ref{fig:Neq4-split} flip under this symmetry, as is indeed seen in the diagrams. We note also that the rearrangement of the columns going from the first to the second line above leaves the overall sign unchanged. The coefficient we seek is, using Eqs. (\ref{eq:AQ}), (\ref{eq:QTP}), and  (\ref{eq:T}):
\begin{align}
\calG^{\Lambda\Lambda'\Lambda''}&=
\sqrt{(2\Lambda_{12}+1)(2\Lambda'_{12}+1)(2{\Lambda''}_{12}+1)}\non
&\times(4\pi)^{-2} \prod_{i=1}^4\sqrt{(2\Lambda_i+1)(2\Lambda'_i+1)(2{\Lambda''}_i+1)}\six{\Lambda_i}{\Lambda'_i}{{\Lambda''}_i}000\non
&\times\nine{\Lambda_1}{\Lambda_{2}}{\Lambda_{12}}{\Lambda'_1}{\Lambda'_{2}}{\Lambda'_{12}}{{\Lambda''}_1}{{\Lambda''}_{2}}{{\Lambda''}_{12}}
\nine{\Lambda_{12}}{\Lambda_3}{\Lambda_4}{\Lambda'_{12}}{\Lambda'_3}{\Lambda'_4}{{\Lambda''}_{12}}{{\Lambda''}_3}{{\Lambda''}_4}.
\label{eq:fiveptprod}
\end{align}

As an example, we calculate $\calY_{11(1)11}\calY_{10(1)01}$. By applying the triangle rule on angular momentum addition, we see that the possible values for ${\Lambda''}_i$ are ${\Lambda''}_1=0,1,2$, ${\Lambda''}_2=1$, ${\Lambda''}_3=1$, 
${\Lambda''}_4=0,1,2$ and ${\Lambda''}_{12}=0,1,2$. Algebraic manipulation of the Cartesian forms in Eq. (\ref{eq:cartesian}) shows
\begeqar
&&\calY_{11(1)11}(\bfxhat_1,\bfxhat_2,\bfxhat_3,\bfxhat_4)\calY_{10(1)01}(\bfxhat_1,\bfxhat_2,\bfxhat_3,\bfxhat_4)\non
&&\quad=\frac 1{(4\pi)^2}\bigg[
\frac3{2\sqrt 5}\calY_{21(2)12}(\bfxhat_1,\bfxhat_2,\bfxhat_3,\bfxhat_4)-\frac{1}{2\sqrt3}\calY_{21(1)12}(\bfxhat_1,\bfxhat_2,\bfxhat_3,\bfxhat_4)\non &&\qquad\qquad-\frac{1}{\sqrt 6}\left[\calY_{21(1)10}(\bfxhat_1,\bfxhat_2,\bfxhat_3,\bfxhat_4)+\calY_{01(1)12}(\bfxhat_1,\bfxhat_2,\bfxhat_3,\bfxhat_4)\right]\non
&&\qquad\qquad\qquad -\frac{1}{\sqrt 3}\calY_{01(1)10}(\bfxhat_1,\bfxhat_2,\bfxhat_3,\bfxhat_4)\bigg].
\endeqar
The same result is obtained from Eq. (\ref{eq:fiveptprod}), which can be evaluated conveniently with the \textsc{python} package \textsc{sympy.physics.wigner}. We note that mapping $\bfxhat_4\to \bfxhat_1$ converts $\calY_{21(1)10}$ to $\calY_{01(1)12}$, and given that the Universe is insensitive to our choice of labels, this implies that these two functions' coefficients must be the same in our expansion above.

As a second example we consider a product with a parity-odd factor. Using Eq. (\ref{eq:fiveptprod})
we obtain:
\begeqar
&&\calY_{11(1)10}(\bfxhat_1,\bfxhat_2,\bfxhat_3,\bfxhat_4)\calY_{10(1)01}(\bfxhat_1,\bfxhat_2,\bfxhat_3,\bfxhat_4)\non 
&&\qquad=\frac 1{(4\pi)^2}\bigg[
\sqrt{\frac 12}\calY_{21(2)11}(\bfxhat_1,\bfxhat_2,\bfxhat_3,\bfxhat_4)-\sqrt{\frac 16}\calY_{21(1)11}(\bfxhat_1,\bfxhat_2,\bfxhat_3,\bfxhat_4)\non
&&\qquad\qquad-\sqrt{\frac13}\calY_{01(1)11}(\bfxhat_1,\bfxhat_2,\bfxhat_3,\bfxhat_4) \bigg].
\endeqar

\subsection{Products of $\calY_\Lambda$ for $N=5$} 
\begin{figure}
\begin{center}
\includegraphics[width=5.5 in]{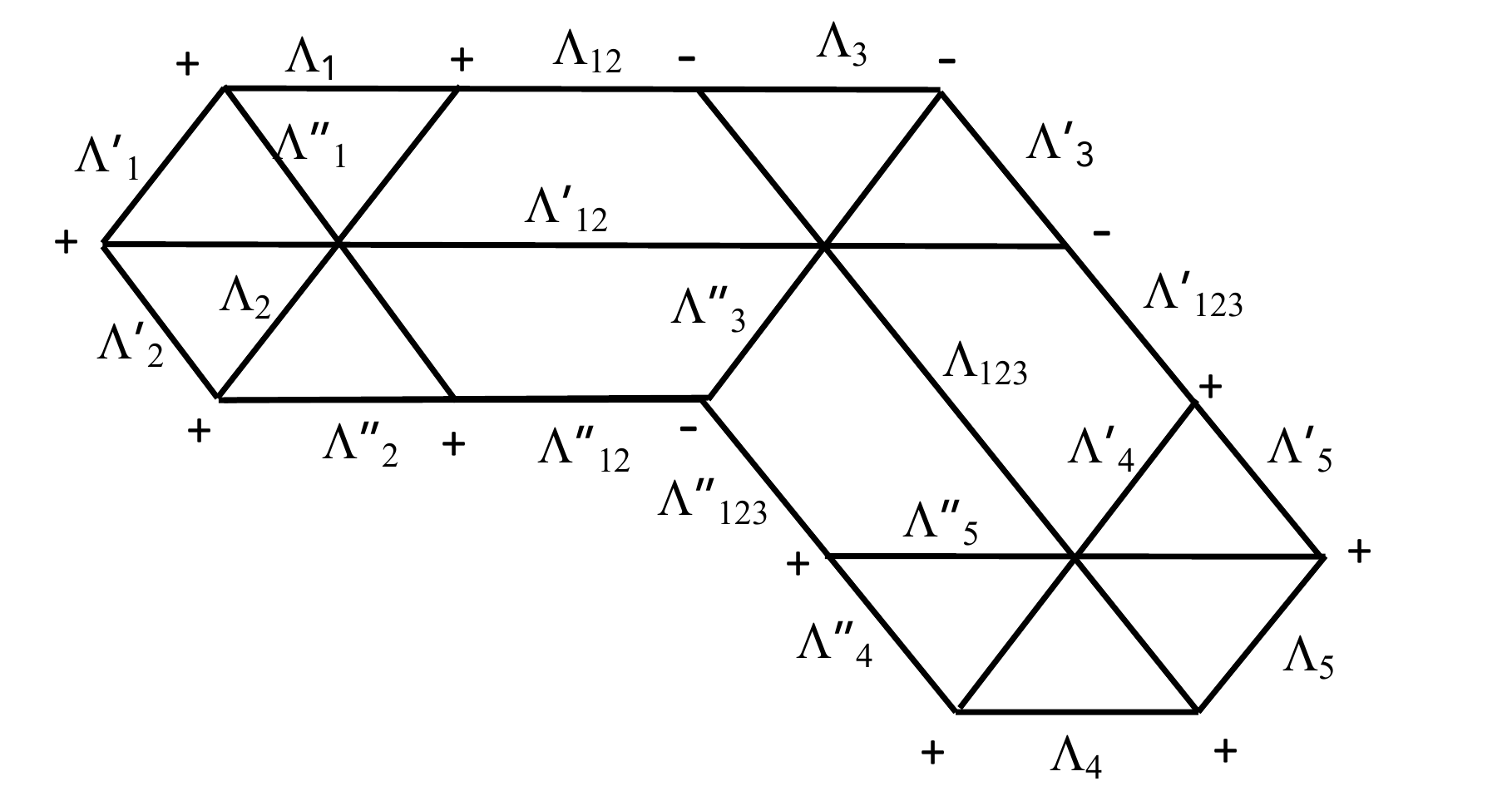}
\caption{Yutsis diagram describing the coefficient of $\calY_{\Lambda''}$ in the product  of the five-point functions $\calY_\Lambda$ and $\calY_{\Lambda'}$; the diagram is derived from Eq. (\ref{eq:5pt_split_analysis}). We will cut it across the set of three lines representing the intermediate momenta $\Lambda_{12}, \Lambda'_{12}, \Lambda_{12}''$ and also across the set of three lines representing $\Lambda_{123}, \Lambda'_{123}, \Lambda_{123}''$ to obtain Fig. \ref{fig:Neq5-split}.} 
\label{fig:Neq5}
\end{center}
\end{figure}

\begin{figure}
\begin{center}
\includegraphics[width=5.5 in]{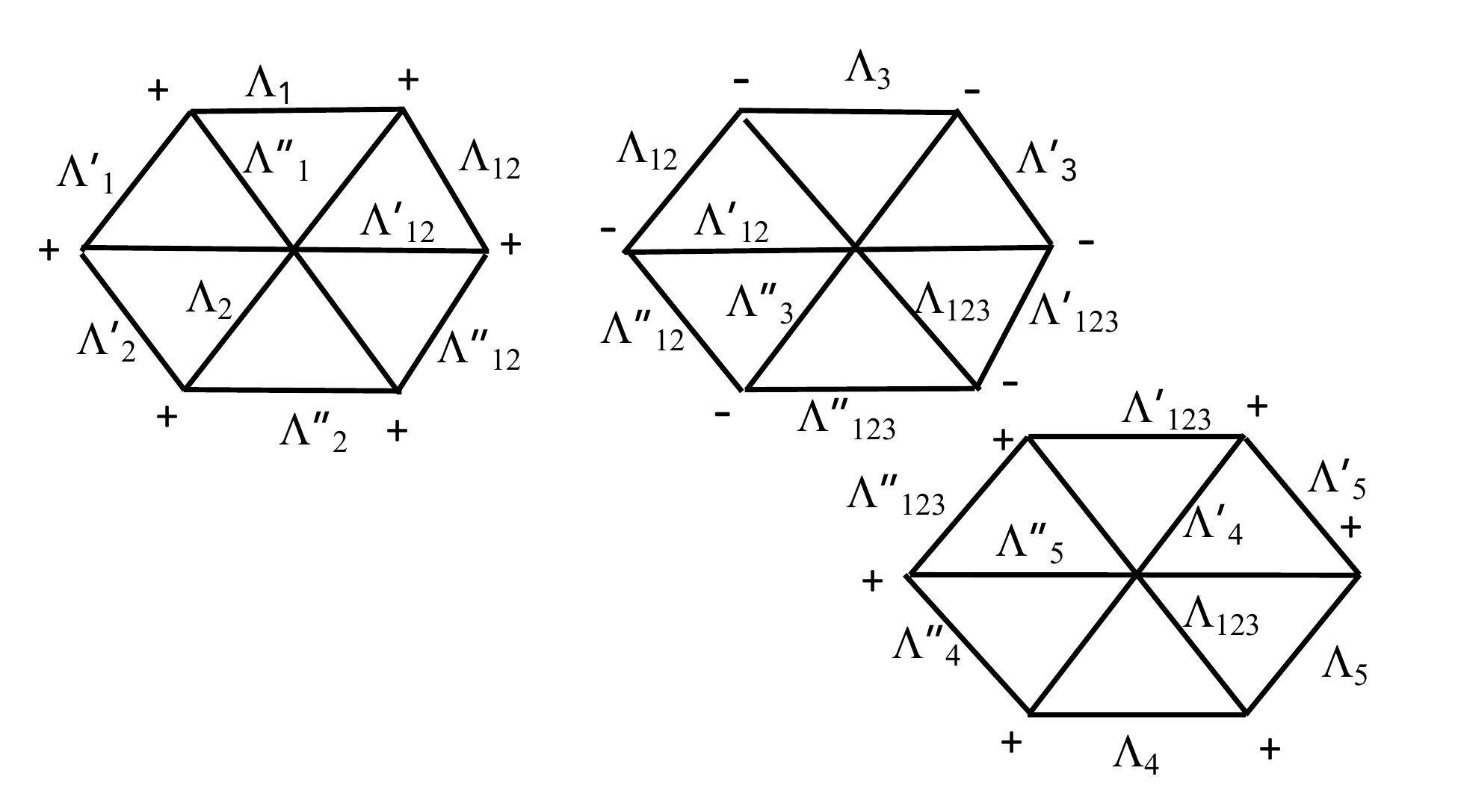}
\caption{Yutsis diagram  Fig. \ref{fig:Neq5} factored by cutting that diagram at the lines representing the intermediate momenta, as detailed in the caption for Fig. \ref{fig:Neq5}.} 
\label{fig:Neq5-split}
\end{center}
\end{figure}

For $N=5$ there are two sets of intermediate angular momenta: $\Lambda_{12}, \Lambda'_{12},{\Lambda''}_{12}$ and $\Lambda_{123}, \Lambda'_{123},{\Lambda''}_{123}$.  The phases required for the Yutsis diagram are supplied by the factor $(-1)^{\kappa_\Lambda+\kappa_{\Lambda'}+\kappa_{\Lambda''}}$, itself defined in Eq. (\ref{eq:kappa}). To construct the Yutsis diagram it suffices to examine the upper portions of the 3-$j$ symbols in $\mathcal{G}$ (Eq. \ref{eq:AQ}) and specifically its factor $\mathcal{Q}$ as defined in Eq. (\ref{eq:mainresult}):
\begeqar
\label{eq:5pt_split_analysis}
&&(\Lambda_1,\Lambda_2,\Lambda_{12});(\Lambda'_1,\Lambda'_2,\Lambda'_{12});({\Lambda''}_1,{\Lambda''}_2,{\Lambda''}_{12});\\
&&(\Lambda_{12},\Lambda_3,\Lambda_{123});(\Lambda'_{12},\Lambda'_3,\Lambda'_{123});({\Lambda''}_{12},{\Lambda''}_3,{\Lambda''}_{123});\non
&&(\Lambda_{123},\Lambda_4,\Lambda_5);(\Lambda'_{123},\Lambda'_4,\Lambda'_5);({\Lambda''}_{123},{\Lambda''}_4,{\Lambda''}_5);\non
&&(\Lambda_1,\Lambda'_1,{\Lambda''}_1);(\Lambda_2,\Lambda'_2,{\Lambda''}_2);(\Lambda_3,\Lambda'_3,{\Lambda''}_3);(\Lambda_4,\Lambda'_4,{\Lambda''}_4);(\Lambda_5,\Lambda'_5,{\Lambda''}_5).\nonumber
\endeqar
From this we derive the diagram in Fig. \ref{fig:Neq5}.
This diagram can be divided into three separate pieces by cutting through three lines in two different places. We cut through the lines representing  $\Lambda_{12},\Lambda'_{12}$ and $\Lambda''_{12}$ and also the lines representing $\Lambda_{132},\Lambda'_{123},\Lambda''_{123}$. Notably, these are the lines representing all of the intermediate momenta. The result is shown in Fig. \ref{fig:Neq5-split}.

To bring the three separate diagrams to canonical form for 9-$j$ symbols introduces a factor
$(-1)^{\sum_{i=1}^5(\Lambda_i+\Lambda'_i+{\Lambda''}_i)}
$,
which is simply $+1$ by consideration of parity.
\begin{figure}
\begin{center}
\includegraphics[width=5.5 in]{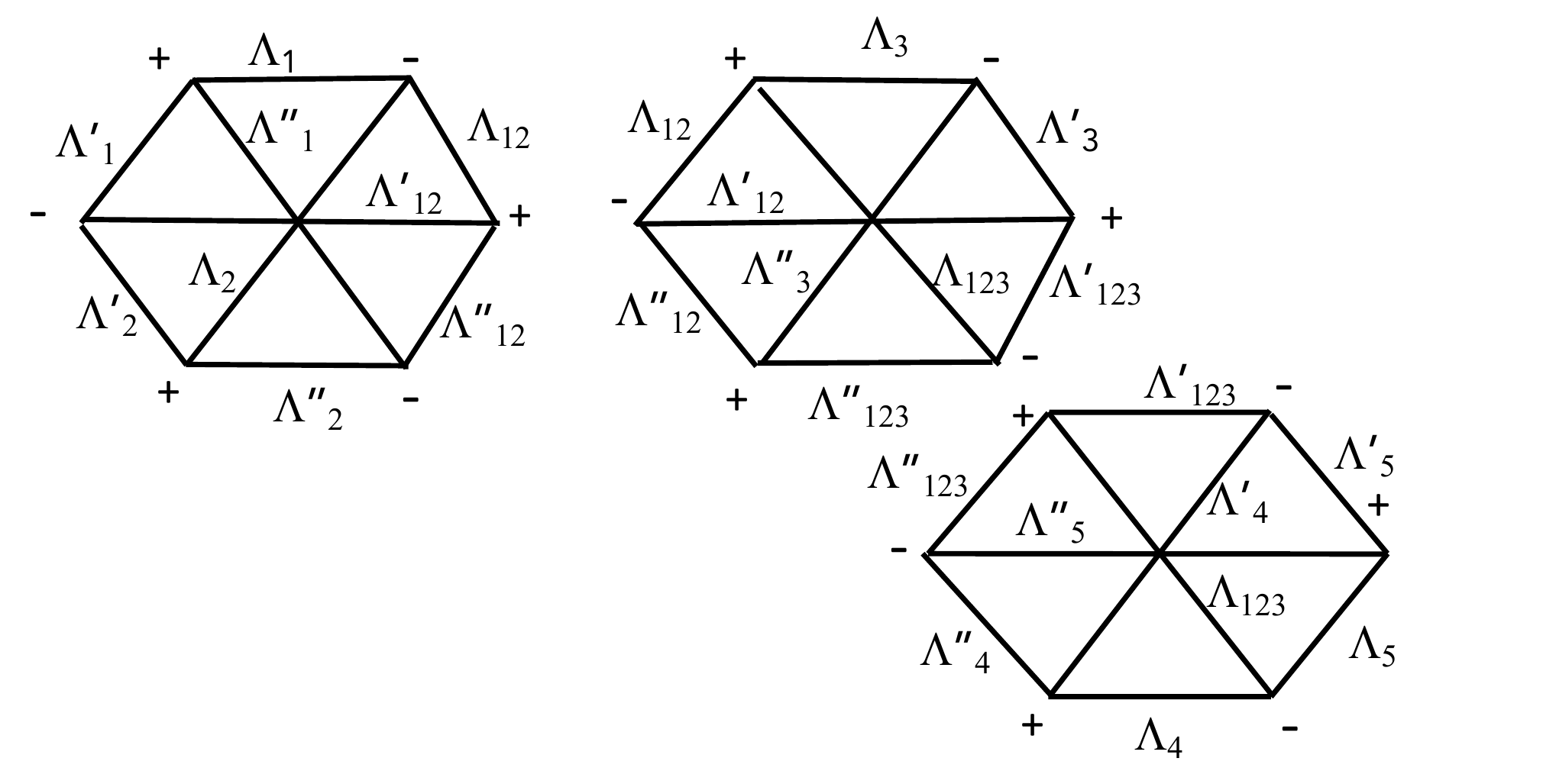}
\caption{The Yutsis diagram for $N=5$ (Fig. \ref{fig:Neq5}) after splitting (Fig. \ref{fig:Neq5-split}) and now with canonical signs at the vertices (compare with Fig. \ref{fig:Neq5-split}).} \label{fig:5pt-fixed}
\label{fig:Neq5-split-fixed}
\end{center}
\end{figure}

Comparing with Fig. \ref{fig:9j} showing the canonical diagram for a 9-$j$ symbol we find that the three diagrams give directly
\begin{align}
&\calP^{\Lambda\Lambda'\Lambda''}\\
&=\nine{\Lambda_1}{\Lambda_{12}}{\Lambda_2}{\Lambda'_1}{\Lambda'_{12}}{\Lambda'_2}{{\Lambda''}_1}{{\Lambda''}_{12}}{{\Lambda''}_2}
\nine{\Lambda_3}{\Lambda_{12}}{\Lambda_{123}}{\Lambda'_3}{\Lambda'_{12}}{\Lambda'_{123}}{{\Lambda''}_3}{{\Lambda''}_{12}}{{\Lambda''}_{123}}
\nine{\Lambda_5}{\Lambda_{4}}{\Lambda_{123}}{\Lambda'_5}{\Lambda'_{4}}{\Lambda'_{123}}{{\Lambda''}_5}{{\Lambda''}_{4}}{{\Lambda''}_{123}}\non
&=\nine{\Lambda_1}{\Lambda_{2}}{\Lambda_{12}}{\Lambda'_1}{\Lambda'_{2}}{\Lambda'_{12}}{{\Lambda''}_1}{{\Lambda''}_{2}}{{\Lambda''}_{12}}
\nine{\Lambda_{12}}{\Lambda_3}{\Lambda_{123}}{\Lambda'_{12}}{\Lambda'_3}{\Lambda'_{123}}{{\Lambda''}_{12}}{{\Lambda''}_3}{{\Lambda''}_{123}}
\nine{\Lambda_{123}}{\Lambda_{4}}{\Lambda_{5}}{\Lambda'_{123}}{\Lambda'_{4}}{\Lambda'_{5}}{{\Lambda''}_{123}}{{\Lambda''}_{4}}{{\Lambda''}_{5}}.\nonumber
\end{align}
We recall that $\calP^{\Lambda\Lambda'\Lambda''}$ is one of the two factors in $\mathcal{Q}$, which in turn enters $\mathcal{G}$, the overall expansion coefficient for turning a product of two isotropic functions into a sum over single ones. We recall further that $\calP^{\Lambda\Lambda'\Lambda''}$ is defined through Eqs. (\ref{eq:mainresult} ) and (\ref{eq:QTP}). When comparing Fig. \ref{fig:Neq5-split-fixed} with Fig. \ref{fig:9j} we identify for the first diagram $j_1 = \Lambda_1,j_2 = \Lambda_{12}$, etc. For the second diagram, we identify $j_1 = \Lambda_3, j_2 = \Lambda_{12}$, etc. For the third diagram, we identify $j_1 = \Lambda_5, j_2 = \Lambda_4$, etc. These choices give the 9-$j$ symbols in the first line above. We have made them with the same reasoning about mirror symmetry about the points where the diagrams separated as discussed below Eq. (\ref{eq:4pt_calS}). We can then use the symmetries of the 9-$j$ symbol (described in Appendix \ref{ap:symbols}, \S\ref{subsec:9j}) 
to interchange two columns in each of the 9-$j$ symbols.  Because parity requires $\calO\calOp\calOpp=1$, this does not change the overall sign. 

The full coefficient we seek is then
\begeqar
\calG_{\Lambda\Lambda'}^{\Lambda''}&=&
\sqrt{(2\Lambda_{12}+1)(2\Lambda'_{12}+1)(2{\Lambda''}_{12}+1)(2\Lambda_{123}+1)(2\Lambda'_{123}+1)(2{\Lambda''}_{123}+1)}\non
&&\times(4\pi)^{-5/2}\left[\prod_{i=1}^5\sqrt{(2\Lambda_i+1)(2\Lambda'_i+1)(2{\Lambda''}_i+1)}\six{\Lambda_i}{\Lambda'_i}{{\Lambda''}_i}000\right]\\
&&\quad\times\nine{\Lambda_1}{\Lambda_{2}}{\Lambda_{12}}{\Lambda'_1}{\Lambda'_{2}}{\Lambda'_{12}}{{\Lambda''}_1}{{\Lambda''}_{2}}{{\Lambda''}_{12}}
\nine{\Lambda_{12}}{\Lambda_3}{\Lambda_{123}}{\Lambda'_{12}}{\Lambda'_3}{\Lambda'_{123}}{{\Lambda''}_{12}}{{\Lambda''}_3}{{\Lambda''}_{123}}
\nine{\Lambda_{123}}{\Lambda_{4}}{\Lambda_{5}}{\Lambda'_{123}}{\Lambda'_{4}}{\Lambda'_{5}}{{\Lambda''}_{123}}{{\Lambda''}_{4}}{{\Lambda''}_{5}}.\nonumber
\endeqar
\clearpage
\section{Reordering of arguments in $\calY_\Lambda$}
\label{sec:reordering}
\setcounter{figure}{0} 
The $\calY_\Lambda(\bfxhat_1,\ldots, \bfxhat_N)$ provide an orthonormal basis for expanding isotropic functions of $\bfxhat_i$, $i=1,\ldots, N$. The definition of the canonical order $i=1,\ldots, N$ must be supplied, for example by an ordering of the $r_i \equiv |\bfx_i|$.  A computation in terms of the $\calY_\Lambda$ can, however, lead to an expression in terms of a $\calY_\Lambda$
whose arguments are not in the canonical order. 
Including these additional isotropic functions would result in an over-complete basis, which would prevent identification of physically significant quantities.  So motivated, we consider a reordering of $\bfxhat_1,\ldots, \bfxhat_N$ as $\bfxhat_{F1},\ldots, \bfxhat_{FN}$ and our $\calY_{\Lambda}$ evaluated at those arguments, as
$\calY_{\Lambda}(\bfxhat_{F1},\ldots, , \bfxhat_{FN})$.  Since the canonically-ordered $\calY_{\Lambda}$ are a complete  basis, the non-canonically ordered function must be expressible in terms of a sum of canonically ordered ones, each weighted by a coefficient $\calB^F_{\Lambda,\Lambda'}$:
\begeq
\calY_{\Lambda}(\bfxhat_{F1},\ldots,\bfxhat_{FN})=\sum_{\Lambda'}\calB^F_{\Lambda,\Lambda'}\calY_{\Lambda'}(\bfxhat_1,\ldots,\bfxhat_N).
\endeq
Using Eq. (\ref{eq:calC1}), orthogonality then determines that
\begeqar
\calB^F_{\Lambda,\Lambda'}&=&\int d\bfXhat\,\calY_{\Lambda}(\bfxhat_{F1},\ldots,\bfxhat_{FN})\calY^*_{\Lambda'}(\bfxhat_1,\ldots,\bfxhat_N)\non
&=&\int d\bfXhat\,\sum_{M_i,M'_j}\calC^{\Lambda}_{M}\prod_{i=1}^N Y_{\Lambda_iM_i}(\bfxhat_{Fi})\;\calC^{\Lambda'}_{M'}\prod_{j=1}^N Y^*_{\Lambda'_jM'_j}(\bfxhat_j)
\label{eqn:calB_result}
\endeqar 
where $d\bfXhat\,=d\bfxhat\,_1 \cdots d\bfxhat\,_N$ and we used the definition Eq. (\ref{eq:calP_defn}) of $\calY$.

Now, $F$ takes $i$ to some value within the original set $1,\ldots, N$. Hence $\Lambda_1, M_1$ now goes with an argument $\hat{\bf r}_{F1}$, etc. We integrate a spherical harmonic of this argument against the conjugate harmonic with the same argument, so from the transformation of e.g. $\hat{\bf r}_1$ we have (pulling out from Eq. (\ref{eqn:calB_result}) the $j = F1$ piece)
\begin{align}
\int d\hat{\bf r}_{F1}\; Y_{\Lambda_1 M_1}(\hat{\bf r}_{F1}) Y^*_{\Lambda'_{F1} M'_{F1}}(\hat{\bf r}_{F1}), 
\end{align}
and similarly for the transformations of the other $\hat{\bf r}_i$.

By orthogonality of the spherical harmonics, we thus see that the integral in Eq. (\ref{eqn:calB_result}) will vanish unless $\Lambda_{1}=\Lambda'_{F1}$, and the same argument applies to $\Lambda_{2} = \Lambda'_{F2}$, etc. Thus overall we conclude that $\calB^F_{\Lambda,\Lambda'}$ must include a factor of $
\deltaK_{\Lambda_1 \Lambda'_{F1}}\cdots\deltaK_{\Lambda_N \Lambda'_{FN}}
$.

The notation simplifies if we use the inverse permutation $G=F^{-1}$. 
When we use $G$, the Kronecker delta factor derived above becomes $\prod_{i=1}^N \deltaK_{\Lambda'_i \Lambda_{Gi}}$. We note that the Kronecker deltas act only on the primary momenta, not the intermediate ones, so primed intermediate momenta remain in the calculation. Performing all of the angular integrals in Eq. (\ref{eqn:calB_result}) and then applying the resulting Kronecker deltas to the $primary$ primed momenta, we obtain
\begeq 
\calB^{F=G^{-1}}_{\Lambda,\Lambda'}=\sum_{M_i} \calC^{(\Lambda_{1},\Lambda_{2},\Lambda_{12},\ldots\Lambda_N)}_{(M_{1},M_{2},\ldots M_N)} \calC^{(\Lambda_{G1},\Lambda_{G2},\Lambda'_{12},\ldots\Lambda_{GN})}_{(M_{G1},M_{G2},\ldots M_{GN})}  \prod_{i=1}^N \deltaK_{\Lambda'_i \Lambda_{Gi}}.
\label{eq:calB}
\endeq 
We note that the permutation $F$ or $G$ does not appear explicitly in the intermediate momenta $\Lambda_{12},\Lambda'_{12}$, etc. Their values are specified initially and are not altered by the permutation. Though we have applied the Kronecker delta factor to the second $\calC$, we still need to include it on the righthand side to specify the values of the $\Lambda'_i$.

Now, given $\Lambda=\Lambda_1,\Lambda_2,\Lambda_{12},\ldots, \Lambda_N$ and the permutation $F$, our task is to find the coefficients $\calB^F_{\Lambda,\Lambda'}$, for the (fixed) values of the intermediate $\Lambda'_{12}$, etc.

Re-expressing Eq. (\ref{eq:calB}) using 3-$j$ symbols as in Eq. (\ref{eq:calC2}) we find
\begeqar 
\calB^{F=G^{-1}}_{\Lambda,\Lambda'}&=&\calOp\sum_{M_i}\calC^{(\Lambda_{1},\Lambda_{2},\Lambda_{12},\ldots\Lambda_N)}_{(M_{1},M_{2},\ldots M_N)} \calC^{(\Lambda_{G1},\Lambda_{G2},\Lambda'_{12},\ldots\Lambda_{GN})}_{(-M_{G1},-M_{G2},\ldots -M_{GN})} \prod_{i=1}^N\deltaK_{\Lambda'_i\Lambda_{Gi}}\non 
&=&\calOp\prod_{i=1}^N\deltaK_{\Lambda'_i\Lambda_{Gi}}\sum_{M_i,M_{12}\ldots M'_{12}\ldots}(-1)^{(\Lambda_{12}-M_{12}+\ldots)}(-1)^{ (\Lambda'_{12}-M'_{12}+\ldots)}\non
&&\qquad\times \sqrt{2\Lambda_{12}+1}\sqrt{2\Lambda'_{12}+1}\cdots\sqrt{2\Lambda_{12\ldots N-2}+1}\sqrt{2\Lambda'_{12\ldots N-2}+1}\non
&&\qquad\times\six{\Lambda_1}{\Lambda_2}{\Lambda_{12}}{M_1}{M_2}{-M_{12}}\six{\Lambda_{12}}{\Lambda_3}{\Lambda_{123}}{M_{12}}{M_3}{-M_{123}}\ldots\non 
&&\qquad\qquad\qquad\qquad\qquad\qquad\times\six{\Lambda_{12\ldots N-2}}{\Lambda_{N-1}}{\Lambda_{N}}{M_{12\ldots N-2}}{M_{N-1}}{M_{N}}\\ 
&&\qquad\times\six{\Lambda_{G1}}{\Lambda_{G2}}{\Lambda'_{12}}{-M_{G1}}{-M_{G2}}{M'_{12}}\six{\Lambda'_{12}}{\Lambda_{G3}}{\Lambda'_{123}}{-M'_{12}}{-M_{G3}}{M'_{123}}\ldots\non 
&&\qquad\qquad\qquad\qquad\qquad\qquad\times \six{\Lambda'_{12\ldots N-2}}{\Lambda_{G(N-1)}}{\Lambda'_{GN}}{-M'_{12\ldots N-2}}{-M_{G(N-1)}}{-M_{GN}}.\nonumber
\label{eq:full_reorder1}
\endeqar 
The factors $\calOp=\calO$ and $(-1)^{(\Lambda_{12}-M_{12}+\ldots \Lambda'_{12}-M'_{12}+\ldots)}$ are just what is required to make the product of the 3-$j$ symbols describable with Yutsis diagrams.   

We now discuss some very simple limits to gain intuition about the behavior of $\calB$. First, consider the case where the permutation $F$ leaves alone ``1" and ``2", i.e. $G1 = 1$ and $G2 = 2$. We may then use the orthogonality relation on the 3-$j$ symbols (\ref{eqn:3j_orth}), which we rewrite here with the relevant angular momenta filled in for clarity:
\begeqar
&&\sum_{M_1,M_2}\six{\Lambda_1}{\Lambda_2}{\Lambda_{12}}{M_1}{M_2}{-M_{12}}\six{\Lambda_1}{\Lambda_2}{\Lambda'_{12}}{-M_1}{-M_2}{M'_{12}}\\
&&\qquad=\frac{\deltaK_{\Lambda_{12}\Lambda'_{12}}\deltaK_{M_{12}M'_{12}} }{2\Lambda_{12}+1}(-1)^{\Lambda_1+\Lambda_2+\Lambda_{12}}.
\nonumber
\endeqar 
Inserting this in Eq. (\ref{eq:full_reorder1}) we find
\begeqar 
\label{eq:full_reorder2}
&&\calB^{F=G^{-1}}_{\Lambda,\Lambda'} = \calOp\deltaK_{\Lambda_{12},\Lambda'_{12}}\prod_{i=1}^N\deltaK_{\Lambda'_i\Lambda_{Gi}}\sum_{M_{12},M_3\ldots,\atop M_{123}\ldots M'_{123}\ldots}(-1)^{\Lambda_1+\Lambda_2+\Lambda_{12}}\\
&&\times (-1)^{(\Lambda_{123}-M_{123}+ \Lambda'_{123}-M'_{123}+\ldots)}\times \sqrt{2\Lambda_{123}+1}\sqrt{2\Lambda'_{123}+1}\nonumber\\
&&\times\six{\Lambda_{12}}{\Lambda_3}{\Lambda_{123}}{M_{12}}{M_3}{-M_{123}}\cdots \six{\Lambda_{12\ldots N-2}}{\Lambda_{N-1}}{\Lambda_{N}}{M_{12\ldots N-2}}{M_{N-1}}{M_{N}} \non 
&&\times\six{\Lambda_{12}}{\Lambda_{G3}}{\Lambda'_{123}}{-M_{12}}{-M_{G3}}{M'_{123}}\cdots\six{\Lambda'_{12\ldots N-2}}{\Lambda_{G(N-1)}}{\Lambda'_{GN}}{-M'_{12\ldots N-2}}{-M_{G(N-1)}}{-M_{GN}}.\nonumber
\endeqar 
Now specializing further to the case where $F$ is the identity permutation, so that $Gi = i$, we recover
\begeq
\calB^F_{\Lambda\Lambda'}=\deltaK_{\Lambda_{12}\Lambda'_{12}}\deltaK_{\Lambda_{123}\Lambda'_{123}}\cdots\prod_{i=1}^N\deltaK_{\Lambda_i\Lambda'_{Fi}}
\label{eq:identity_perm}
\endeq

Returning now to the consideration of a general permutation, $F$, we see that the cases $N=2$ and $N=3$ are trivial since they have no intermediate value $\Lambda_{12}$. For $N=2$, from Eq. (\ref{eq:nequal2}) we have simply
\begeq
\calB^F_{\Lambda_1,\Lambda'_1}=\deltaK_{\Lambda_1\Lambda'_1}.
\endeq
\begeq
\calB^{F=G^{-1}}_{\Lambda_1,\Lambda_2\Lambda_3,\Lambda_{G1},\Lambda_{G2},\Lambda_{G3}} =P(F)\prod_{i=1}^N\deltaK_{\Lambda'_i\Lambda_{Gi}}
\label{eq:perm_for_N3}
\endeq 
where $P(F)=P(G)$ 
is $+1$ unless $both$ the permutation $F$ is odd $and$ $\sum_i\Lambda_i$ is odd 
in which case $P(F)=-1$. This behavior is seen, for example, in the explicit Cartesian representations of $\calY_{111}$ and $\calY_{112}$ in Appendix \ref{ap:examples}, \S\ref{subsec:A_N3}. For instance, one can explicitly find that
\begin{align}
\calY_{110}(\hat{\bf r}_1, \hat{\bf r}_3, \hat{\bf r}_2) = \calY_{101}(\hat{\bf r}_1, \hat{\bf r}_2, \hat{\bf r}_3)    \end{align}
and this agrees with Eq. (\ref{eq:perm_for_N3}). To use the latter, we note that $\Lambda_{G1} = \Lambda_1 = 1, \Lambda_{G2} = \Lambda_3 = 0$, and $\Lambda_{G3} = \Lambda_2 = 1$, and the permutation is odd but the overall parity of the isotropic function involved is even. It is necessary that one has both an odd permutation and and an odd sum to get a minus sign. For instance, the identity permutation is even, but one could have an odd $\sum_i \Lambda_i$, yet clearly there should be no negative sign.

Now we consider a case where the permutation and $\sum_i \Lambda_i$ are both odd, and by explicit computation find that
\begin{align}
\calY_{111}(\hat{\bf r}_1, \hat{\bf r}_3, \hat{\bf r}_2) = - \calY_{111} (\hat{\bf r}_1, \hat{\bf r}_2, \hat{\bf r}_3)
\end{align}

This is in agreement with our result from Eq. (\ref{eq:perm_for_N3}).

\subsection{Reordering for $N=4$}

Applying the general formula Eq. (\ref{eq:full_reorder1}) to the case $N=4$ we obtain
\begeqar
\label{eq:CF4p}
\calB^F_{\Lambda,\Lambda'}
&=&\calOp\prod_{i=1}^4\deltaK_{\Lambda'_i\Lambda_{Gi}}(-1)^{\Lambda_{12}-M_{12}+\Lambda'_{12}-M'_{12}}\sqrt{2\Lambda_{12}+1}\sqrt{2\Lambda'_{12}+1}\non
&&\times\quad\sum_{M_i,M'_{12},M_{12}}
\six{\Lambda_{1}}{\Lambda_{2}}{\Lambda_{12}}{M_{1}}{M_{2}}{-M_{12}}
\six{\Lambda_{12}}{\Lambda_{3}}{\Lambda_{4}}{M_{12}}{M_{3}}{M_{4}}\\
&&\quad\times\six{\Lambda_{G1}}{\Lambda_{G2}}{\Lambda'_{12}}{-M_{G1}}{-M_{G2}}{M'_{12}}
\six{\Lambda'_{12}}{\Lambda_{G3}}{\Lambda_{G4}}{-M'_{12}}{-M_{G3}}{-M_{G4}}.\nonumber
\endeqar
Recall that the $\Lambda_{Gi}$ are simply a permutation of the $\Lambda_i$ and similarly the $M_{Gi}$ are the same permutation of the $M_i$.  Thus the sum over the $M_{i}$ implies a sum over the $M_{Gi}$.
\subsubsection{Identity and related permutations}
\label{subsubsec:identity_perm_N4}
Consider first the case where $F$ is simply the identity mapping, so $F = G^{-1} = I$.  As shown in Eq. (\ref{eq:identity_perm}) this results in the factor $\deltaK_{\Lambda_{12} \Lambda'_{12}}. $ Interchanging the values of ${G1}$ and ${G2}$ in the 3-$j$ symbol in the third line of Eq. (\ref{eq:CF4p}) will introduce a factor $(-1)^{\Lambda_{G1}+\Lambda_{G2}+\Lambda'_{12}}$ (via the 3-$j$ symbol identity Eq. \ref{eq:3j_perms}). Likewise, the same result applies 
$mutatis$ $mutandis$ when we interchange the values of $G3$ and $G4$. These considerations enable us to deduce the result for eight of the 24 permutations of the $\Lambda_i$; these results are displayed in Table \ref{tab:cat0}.

\begin{table}[ht]
Prefactor: $\deltaK_{\Lambda_{12}\Lambda'_{12}}\prod_{i=1}^4\deltaK_{\Lambda'_i\Lambda_{Gi}}$
\def\arraystretch{1.5}
\begin{center}
\begin{tabular}{lll}\hline\hline
$F$&$G=F^{-1}$&phase\\ \hline
$(1234)\to(1234)$&$(1234)\to(1234)$&$1$\\
$(1234)\to(1243)$&$(1234)\to(1243)$&$(-1)^{\Lambda_3+\Lambda_4+\Lambda_{12}}$\\
$(1234)\to(2134)$&$(1234)\to(2134)$&$(-1)^{\Lambda_1+\Lambda_2+\Lambda_{12}}$\\
$(1234)\to(2143)$&$(1234)\to(2143)$&$(-1)^{\Lambda_1+\Lambda_2+\Lambda_3+\Lambda_4}$\\
$(1234)\to(3412)$&$(1234)\to(3412)$&$1$\\
$(1234)\to(3421)$&$(1234)\to(4312)$&$(-1)^{\Lambda_3+\Lambda_4+\Lambda_{12}}$\\
$(1234)\to(4312)$&$(1234)\to(3421)$&$(-1)^{\Lambda_1+\Lambda_2+\Lambda_{12}}$\\
$(1234)\to(4321)$&$(1234)\to(4321)$&$(-1)^{\Lambda_1+\Lambda_2+\Lambda_3+\Lambda_4}$\\

\hline\hline
\end{tabular}
\caption{Factor converting the $N=4$ isotropic functions to standard ordering of angular arguments for eight  permutations as discussed in \S\ref{subsubsec:identity_perm_N4}. The value of $\calB^F_{\Lambda\Lambda'}$ is given by the product of the prefactor given above the table and the phase of the corresponding permutation $F$. The phase stems from the properties of the 3-$j$ symbols under permutations.  To find the inverse, $G$ of $F$ one simply reverses the action of $F$.  If $F:(1234)\to (abcd)$ then $G:(abcd)\to (1234)$.  Rearranging $abcd$ in numerical order and $(1234)$ in the same way provides the output of $G$ on $(1234)$.  Thus if $F:(1234)\to (3124)$ then $G(1234)\to(2314)$.
We observe that moving a whole pair relative to another pair, e.g. as in the fifth row, gives no phase. However, flipping $within$ a pair gives a phase equal to $(-1)$ raised to the sum of the angular momenta associated with that pair, plus the intermediate momenta, e.g. as in the second row. This rule also captures the fourth row and the last row, as one can interpret those rows as having a factor of $(-1)^{2 \Lambda_{12}} = 1$, one from each of the two ``within-pair'' flips. 
\label{tab:cat0}
}
\end{center}
\end{table}

\subsubsection{General Case for $N=4$}
We can represent Eq. (\ref{eq:CF4p}) with Yutsis diagrams. There are three distinct classes,
which can be understood by examining the indices $G1,G2, G3$, and  $G4$.  These four 
are a permutation of $(1,2,3,4)$.  The value $1$ must occur  either in the pair $G1,G2$
or $G3,G4$.  If the $1$ occurs in a pair with $2$, then the result is found in Table \ref{tab:cat0}. In these eight permutations where $1$ and $2$ are paired, the first two elements, $G1,G2$ is either $(1,2)$ (or $(2,1)$)  or $3,4$ (or $(4,3)$). The intermediate $\Lambda'_{12}$ must then be $\Lambda_{12}$. These results are thus obtained without needing to construct Yutsis diagrams, which will be required for the other 16 cases. To summarize this result, for the $N=4$ case, the intermediate angular momentum represents coupling of the first and second primaries together, and that must be equal to the coupling of the third and fourth together. In view of this equality, as long as we preserved the two pairings that each couple together into the intermediate, we were able to switch the ordering of these two pairs (so that the third and fourth momenta become the first and second).

\begin{figure}

\begin{center}
\includegraphics[width=3. in]{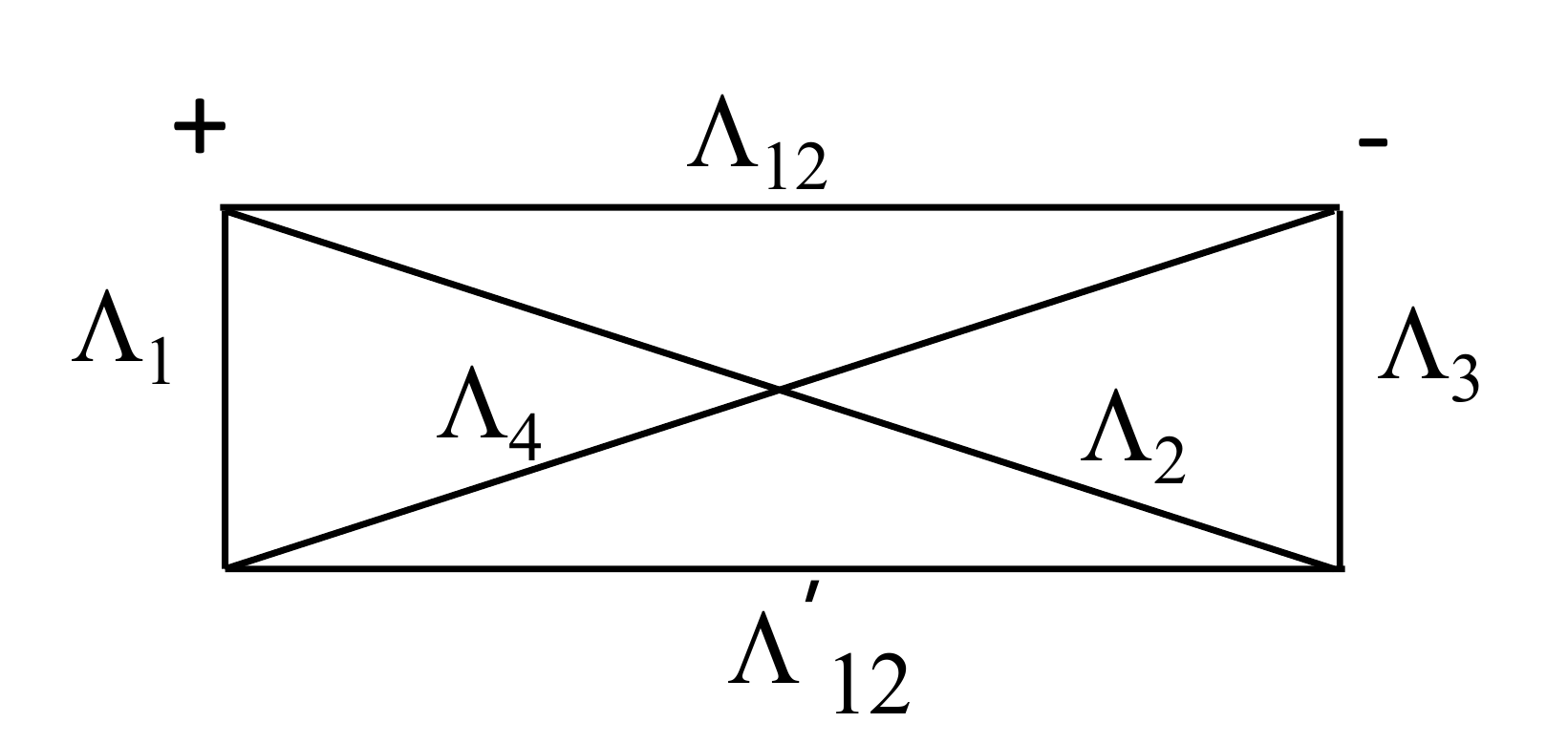}
\caption{The Yutsis diagram reordering arguments of the $N=4$ isotropic function when
$\Lambda_1$ shares a vertex with $\Lambda_4$; this stems from Eq. (\ref{eq:CF4p}) with particular choices of the $Gi$. The three lines coming from the upper left and upper right vertices are given by the top rows of the first two 3-$j$ symbols in Eq. (\ref{eq:CF4p}). The signs at these vertices are determined by the ordering of angular momenta in these first two 3-$j$ symbols. Our sign convention for the Yutsis diagrams is defined just below Eq. (\ref{eq:N3}). The lines coming from, and the signs at, the lower left and right vertices are determined by the specific values of $G1, G2, G3,$ and $G4$. For instance, if $G1=1$ and $G2 = 4$ the lower left vertex should be marked with a $-$, reflecting that we must read the lines emanating from that vertex clockwise if we read them in the order dictated by the first 3-$j$ symbol in the last line of Eq. (\ref{eq:CF4p}). If $G3=3$ and $G4=2$ the lower right vertex receives a $+$ mark, reflecting that we must read the lines emanating from that vertex counterclockwise if we read them in the order dictated by the second 3-$j$ symbol in the last line of Eq. (\ref{eq:CF4p}). Comparing with Fig. \ref{fig:6j}, the value of the diagram here, with $G1=1$, $G2=4$, $G3=3$, and $G4=2$, is given by a 6-$j$ symbol with no additional phase as shown in Table \ref{tab:cat1}. 
Changing the order of $(G1,G2)$ introduces a phase $(-1)^{\Lambda_{G1}+\Lambda_{G2}+\Lambda'_{12}}$ and similarly for changing the order of $(G3,G4)$. See Table \ref{tab:cat1}.
 }
\label{fig:N=4cat1}
\end{center}
\end{figure}

\begin{figure}
\begin{center}
\includegraphics[width=3. in]{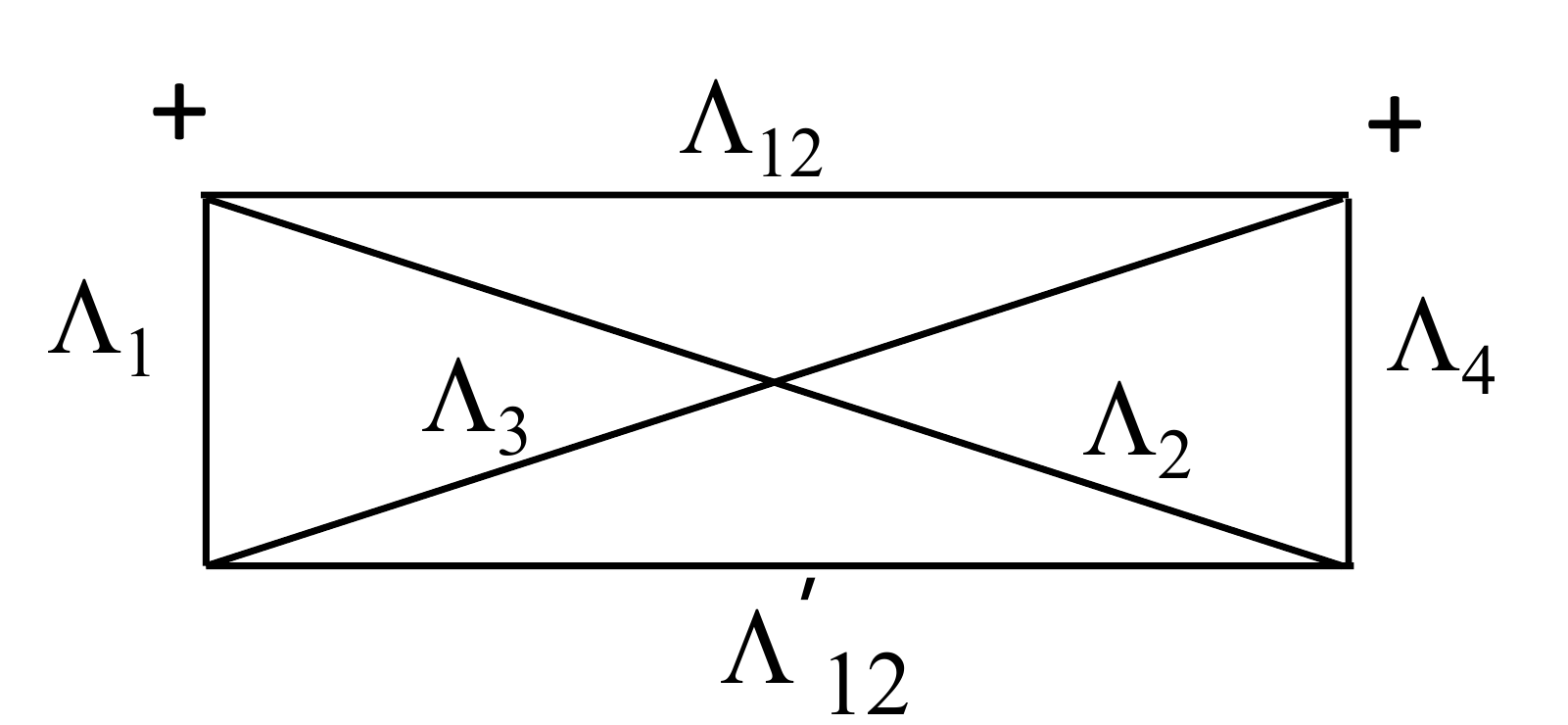}
\caption{The Yutsis diagram reordering arguments of the $N=4$ isotropic function when $\Lambda_1$ shares a vertex with $\Lambda_3$, that is, they occur in the same 3-$j$ symbol in the last line of Eq. (\ref{eq:CF4p}). This diagram is the same as that in Fig. \ref{fig:N=4cat1} save for switching 
$\Lambda_3$ with $\Lambda_4$. The orientations at the upper vertices are determined by the ordering of the angular momenta in the first two 3-$j$ symbols in Eq. (\ref{eq:CF4p}). The signs at the lower vertices are determined by the $Gi$, appearing in the 3-$j$ symbols in the last line of Eq. (\ref{eq:CF4p}). For instance, if $G1=1$ and $G2=3$, the lower lefthand vertex should be marked with a $-$. This sign is because reading in the order the momenta are listed in the relevant 3-$j$ symbol means going clockwise around the vertex. For the analogous reason, if $G3=4$ and $G4=2$, the lower righthand vertex receives a $+$ mark. Comparing this Figure with Fig. \ref{fig:6j}, the value of the diagram is given by a 6-$j$ symbol, with the addition of the phase $(-1)^{\Lambda_3+\Lambda_4 +\Lambda_{12}}$, required to bring the upper right vertex to its canonical value. See Table \ref{tab:cat2}.} 
\label{fig:N=4cat2} 
\end{center}
\end{figure}

For the remaining 16 permutations, where 1 and 2 are not paired with each other, $\Lambda'_{12}$ and $\Lambda_{12}$ are not necessarily equal. If the $1$ occurs in a pair with $4$, the resulting Yutsis diagram is shown in Fig. \ref{fig:N=4cat1}.  The phase is given in Table \ref{tab:cat1}.  Otherwise, $1$ occurs with $3$ and the Yutsis diagram is shown in Fig. \ref{fig:N=4cat2} and the phase is given in Table \ref{tab:cat2}.
In this way we have determined $\calB^F_{\Lambda,\Lambda'}$ for the 16 remaining permutations.

We now display two specific simple examples of reordering for $N=4$. We note that the lefthand side in each expression below is $not$ in canonical order, while the functions on the righthand side are.
\begeqar
\calY_{11(0)11}(\bfxhat_1,\bfxhat_3,\bfxhat_2,\bfxhat_4)&=& \frac 3{(4\pi)^2}\dotprod 13 \dotprod 24\non 
&=&\frac{\sqrt 5}3 \calY_{11(2)11}(\bfxhat_1,\bfxhat_2,\bfxhat_3,\bfxhat_4)+\frac 1{\sqrt 3}\calY_{11(1)11}(\bfxhat_1,\bfxhat_2,\bfxhat_3,\bfxhat_4)\non
&&\qquad\qquad+\frac 1{ 3}\calY_{11(0)11}(\bfxhat_1,\bfxhat_2,\bfxhat_3,\bfxhat_4).
\endeqar 

\begeqar
\calY_{11(2)11}(\bfxhat_1,\bfxhat_3,\bfxhat_2,\bfxhat_4)&=& \frac{9}{2\sqrt 5(4\pi)^2}\bigg[\dotprod 12\dotprod 34+\dotprod14\dotprod 23\non &&\qquad\qquad\qquad\qquad\qquad\qquad -\frac 23\dotprod 13\dotprod 24\bigg]\non
&=&\frac 16 \calY_{11(2)11}(\bfxhat_1,\bfxhat_2,\bfxhat_3,\bfxhat_4)-\frac {\sqrt 5}{2\sqrt 3}\calY_{11(1)11}(\bfxhat_1,\bfxhat_2,\bfxhat_3,\bfxhat_4)\non
&&\qquad\qquad+\frac {\sqrt 5}{ 3}\calY_{11(0)11}(\bfxhat_1,\bfxhat_2,\bfxhat_3,\bfxhat_4).
\endeqar 
These relations can be determined directly from the Cartesian representations or by use of Table \ref{tab:cat2}.

\def\arraystretch{1.5}
\begin{table}
Prefactor: $\sixj{\Lambda_1}{\Lambda_2}{\Lambda_{12}}{\Lambda_3}{\Lambda_4}{\Lambda'_{12}}\sqrt{(2\Lambda_{12}+1)(2\Lambda'_{12}+1)}\prod_{i=1}^4\deltaK_{\Lambda'_i\Lambda_{Gi}}$
\begin{center}
\begin{tabular}{lll}\hline\hline
$F$&$G=F^{-1}$&phase\\ \hline
$(1234)\to(1342)$&$(1234)\to(1423)$&$(-1)^{\Lambda_2+\Lambda_3+\Lambda'_{12}}$\\
$(1234)\to(1432)$&$(1234)\to(1432)$&$1$\\
$(1234)\to(2341)$&$(1234)\to(4123)$&$(-1)^{\Lambda_1+\Lambda_2+\Lambda_3+\Lambda_4}$\\
$(1234)\to(2431)$&$(1234)\to(4132)$&$(-1)^{\Lambda_1+\Lambda_4+\Lambda'_{12}}$\\
$(1234)\to(3124)$&$(1234)\to(2314)$&$(-1)^{\Lambda_2+\Lambda_3+\Lambda'_{12}}$\\
$(1234)\to(3214)$&$(1234)\to(3214)$&$1$\\
$(1234)\to(4123)$&$(1234)\to(2341)$&$(-1)^{\Lambda_1+\Lambda_2+\Lambda_3+\Lambda_4}$\\
$(1234)\to(4213)$&$(1234)\to(3241)$&$(-1)^{\Lambda_1+\Lambda_4+\Lambda'_{12}}$\\
\hline\hline
\end{tabular}
\caption{Factor converting the $N=4$ 
isotropic functions to the standard ordering of angular arguments for eight permutations, all of which are proportional to one 6-$j$ symbol. 
The value of $\calB^F_{\Lambda\Lambda'}$ is given by the product of the prefactor and the phase of the corresponding permutation $F$. 
The diagram corresponding to this Table is given in Fig. \ref{fig:N=4cat1}.
\label{tab:cat1}}
\end{center}
\end{table}
\begin{table}
Prefactor: $\sixj{\Lambda_1}{\Lambda_2}{\Lambda_{12}}{\Lambda_4}{\Lambda_3}{\Lambda'_{12}}\sqrt{(2\Lambda_{12}+1)(2\Lambda'_{12}+1)}\prod_{i=1}^4\deltaK_{\Lambda'_i\Lambda_{Gi}}$
\begin{center}
\begin{tabular}{lll}\hline\hline
$F$&$G=F^{-1}$&phase\\ \hline
$(1234)\to(1324)$&$(1234)\to(1324)$&$(-1)^{\Lambda_2+\Lambda_3+\Lambda_{12}+\Lambda'_{12}}$\\
$(1234)\to(1423)$&$(1234)\to(1342)$&$(-1)^{\Lambda_3+\Lambda_4+\Lambda_{12}}$\\
$(1234)\to(2314)$&$(1234)\to(3124)$&$(-1)^{\Lambda_1+\Lambda_2+\Lambda_{12}}$\\
$(1234)\to(2413)$&$(1234)\to(3142)$&$(-1)^{\Lambda_1+\Lambda_4+\Lambda_{12}+\Lambda'_{12}}$\\
$(1234)\to(3142)$&$(1234)\to(2413)$&$(-1)^{\Lambda_2+\Lambda_3+\Lambda_{12}+\Lambda'_{12}}$\\
$(1234)\to(3241)$&$(1234)\to(4213)$&$(-1)^{\Lambda_3+\Lambda_4+\Lambda_{12}}$\\
$(1234)\to(4132)$&$(1234)\to(2431)$&$(-1)^{\Lambda_1+\Lambda_2+\Lambda_{12}}$\\
$(1234)\to(4231)$&$(1234)\to(4231)$&$(-1)^{\Lambda_1+\Lambda_4+\Lambda_{12}+\Lambda'_{12}}$\\
\hline\hline
\end{tabular}
\caption{Factor converting the $N=4$ isotropic functions to  the standard ordering of angular arguments for eight permutations,  all of which are proportional to a second 6-$j$ symbol 
The value of $\calB^F_{\Lambda\Lambda'}$ is given by the product of the prefactor and the phase of the corresponding permutation $F$. The diagram corresponding to this Table is given in Fig. \ref{fig:N=4cat2}.}
\label{tab:cat2}
\end{center}
\end{table}

\subsection{Reordering for $N=5$}
By analogy with the $N=4$ case, we have for $N=5$ that 
\begeqar
&&\calB^{F=G^{-1}}_{\Lambda,\Lambda'}=\int d\bfXhat\,\calY_{\Lambda'}(\bfxhat_1,\ldots, \bfxhat_5)\calY^*_{\Lambda}(\bfxhat_{F1},\ldots,\bfxhat_{F5})\non
&&=\calOp\sqrt{2\Lambda_{12}+1}\sqrt{2\Lambda'_{12}+1}
\sqrt{2\Lambda_{123}+1}\sqrt{2\Lambda'_{123}+1}\prod_{i=1}^5\deltaK_{\Lambda'_i\Lambda_{Gi}}\non
&&\times\sum_{M_i, M_{12}, M'_{12}\ldots}(-1)^{\Lambda_{12}-M_{12}+\Lambda_{123}-M_{123}+\Lambda'_{12}-M'_{12}+\Lambda'_{123}-M'_{123}}\non 
&&\quad\times
\six{\Lambda_1}{\Lambda_2}{\Lambda_{12}}{M_1}{M_2}{-M_{12}}
\six{\Lambda_{12}}{\Lambda_3}{\Lambda_{123}}{M_{12}}{M_3}{-M_{123}}\six{\Lambda_{123}}{\Lambda_4}{\Lambda_{5}}{M_{123}}{M_4}{M_{5}}\non
&&\quad\times\six{\Lambda_{G1}}{\Lambda_{G2}}{\Lambda'_{12}}{-M_{G1}}{-M_{G2}}{-M'_{12}}
\six{\Lambda'_{12}}{\Lambda_{G3}}{\Lambda'_{123}}{M'_{12}}{-M_{G3}}{-M_{123}}\non
&&\qquad\qquad\qquad \times\six{\Lambda'_{123}}{\Lambda_{G4}}{\Lambda_{G5}}{M'_{123}}{-M_{G4}}{-M_{G5}}.
\label{eq:CF5}
\endeqar

\begin{figure}
\begin{center}
\includegraphics[width=3. in]{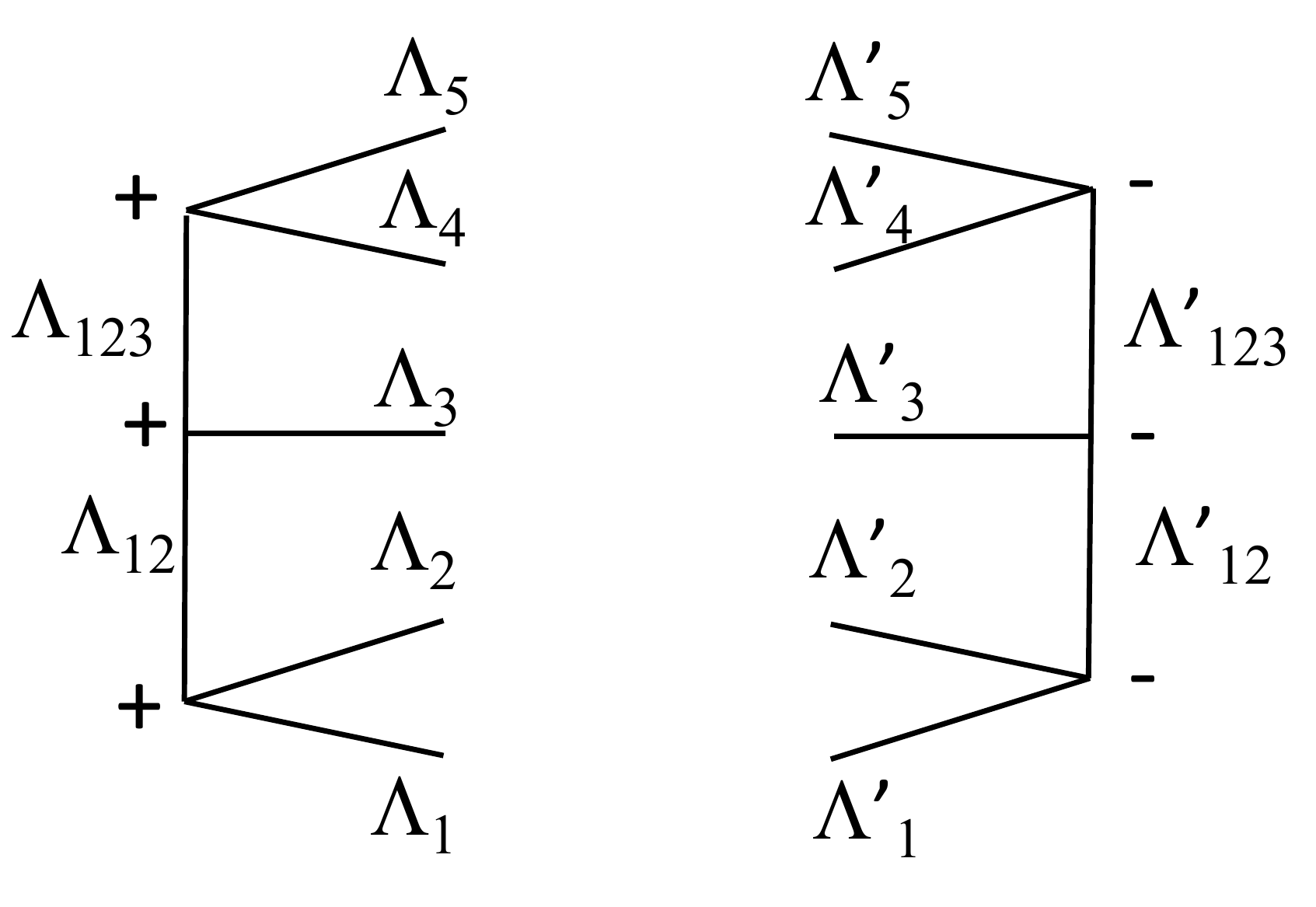}
\caption{The Yutsis diagram reordering the arguments of an $N=5$ isotropic function. The permutation is not yet specified.} \label{fig:yutsisN5}
\end{center}
\end{figure}

Again, the factor $\calO$ and the power of $(-1)$ supply what is needed for a treatment with Yutsis diagrams. The situation is displayed in Fig. \ref{fig:yutsisN5}.  

The values of $\calB^F_{\Lambda\Lambda'}$ fall into several distinct groups defined by the nature of the Yutsis diagram associated with the permutation $F$, leading to results that include a single 6-$j$ symbol, a product of two 6-$j$ symbols, a 9-$j$ symbol, or no such symbol. 

\begin{figure}
\begin{center}
\includegraphics[width=3. in]{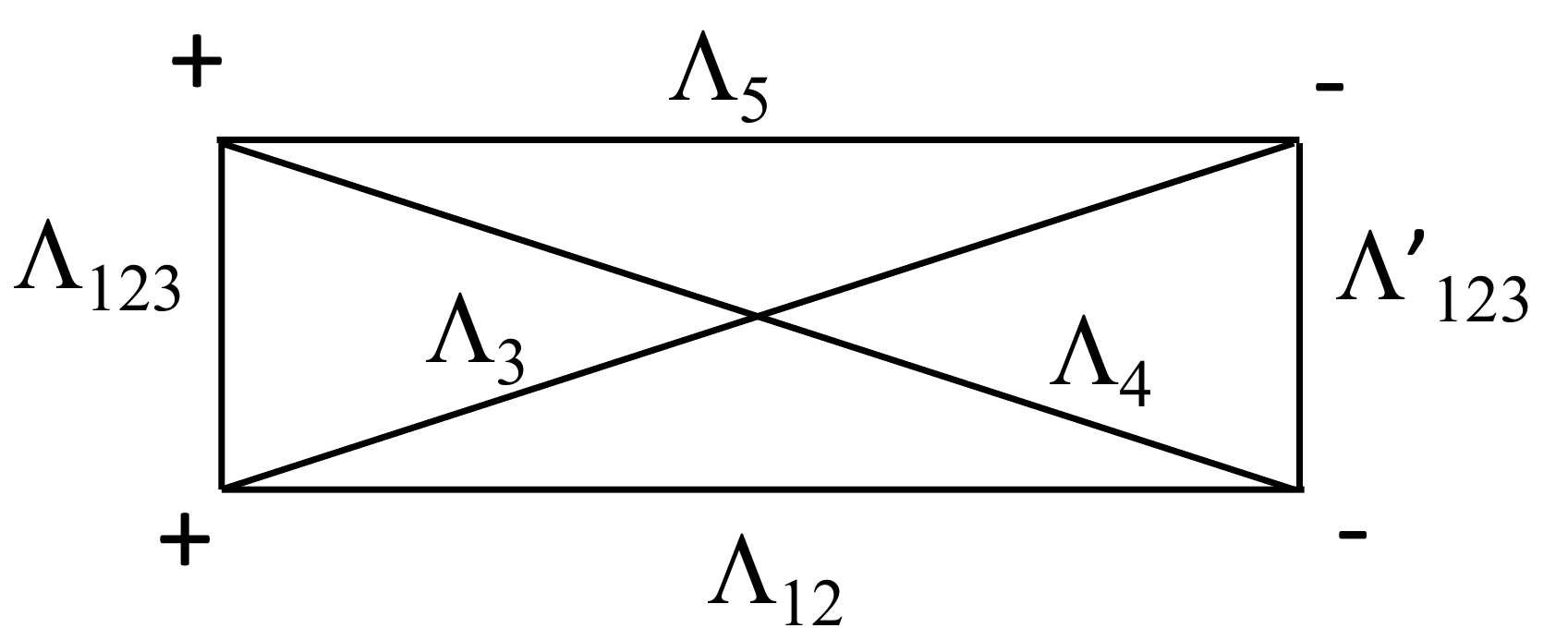}
\caption{The Yutsis diagram reordering arguments of the $N=5$ isotropic function with the permutation $G$ that maps $(12345)\to (12435)$.} 
\label{fig:yutsisN5-joined}
\end{center}
\end{figure}

Let us begin with a simple circumstance where $F$ is such that $G1=1,\; G2=2$.  Then we can do the sum over $M_1$ and $M_2$ as before.  This will leave us with 
\begeqar
\label{eq:CF5p}
\calB^{F=G^{-1}}_{\Lambda,\Lambda'}
&=&\sqrt{2\Lambda_{123}+1}\sqrt{2\Lambda'_{123}+1}\deltaK_{\Lambda_{12}\Lambda'_{12}} \sum_{M_i,M_{12},M'_{12}}(-1)^{\Lambda_{123}-M_{123}+\Lambda'_{123}-M'_{123}}\non
&&\quad\times
\six{\Lambda_{12}}{\Lambda_3}{\Lambda_{123}}{M_{12}}{M_3}{-m_{123}}\six{\Lambda_{123}}{\Lambda_4}{\Lambda_{5}}{M_{123}}{M_4}{M_{5}}\\
&&\quad\times
\six{\Lambda_{12}}{\Lambda_{G3}}{\Lambda'_{123}}{-m_{12}}{-M_{G3}}{-M_{123}}\six{\Lambda'_{123}}{\Lambda_{G4}}{\Lambda_{G5}}{M'_{123}}{-M_{G4}}{-M_{G5}}\prod_{i=1}^5\deltaK_{\Lambda'_i\Lambda_{Gi}}.\nonumber
\endeqar
In general, the condition $G1=1,\; G2=2$ yields a 6-$j$ symbol. For example, if $G(3)=4, G(4)=3$, and $G(5)=5$, so that $G$ maps $(12345)\to (12435)$, we will have the Yutsis diagram in Fig. \ref{fig:yutsisN5-joined}.
The 6-$j$ symbol is
\begeq
\sixj{\Lambda_{12}}{\Lambda_3}{\Lambda_{123}}{\Lambda_5}{\Lambda_4}{\Lambda'_{123}}
\endeq
and there is a factor $(-1)^{\Lambda_3+\Lambda_4+\Lambda_{123}+\Lambda'_{123}}$ arising from returning the vertex signs to canonical form.  
\subsubsection{Permutations requiring no 6-$j$ or 9-$j$ symbol}
\label{subsubsec:no_symbol}
If, however, we choose a permutation like $G$: $(12345)\to (21345)$ we see that both the sum over $M_1$ and $M_2$ and the sum over $M_4$ and $M_5$ are trivial reductions. In particular, they are equivalent to the identity permutation up to an overall sign which is determined by the factors associated with the changes like $(12)\to (21)$.  Clearly there are four such permutations in which each pair $(12)$ and $(45)$ transforms into itself. In addition, there are four more permutations like $G:(12345)\to (45312)$ in which $(1,2)$ and $(4,5)$ transform into each other.  Thus there are eight permutations for which 
\begeq
\calB^F_{\Lambda,\Lambda'}=\pm \deltaK_{\Lambda_{12}\Lambda'_{12}}\deltaK_{\Lambda_{123}\Lambda'_{123}}\prod \deltaK_{\Lambda_i,\Lambda_{Gi}}.
\endeq
\subsubsection{Permutations resulting in one 6-$j$ symbol}
\label{subsubsec:6j_symbol}
In contrast the number of permutations of the sort $G$: $(12345)\to (12435)$ in which $(12)$ transforms separately and $(345)$ transforms so that ``3'' is no longer in its original position is $2\times 4=8$.  There are eight more of the form $G:(12345)\to(34512)$, where the $(12)$ pair are at the end.  Similarly, there are 16 where we use $(45)$ in place of $(12)$.  Altogether, there are 32 permutations giving rise to a 6-$j$ symbol.
\subsubsection{Permutations yielding a 9-$j$ symbol}
\label{subsubsec:9j_symbol}
\begin{figure}
\begin{center}
\includegraphics[width=3. in]{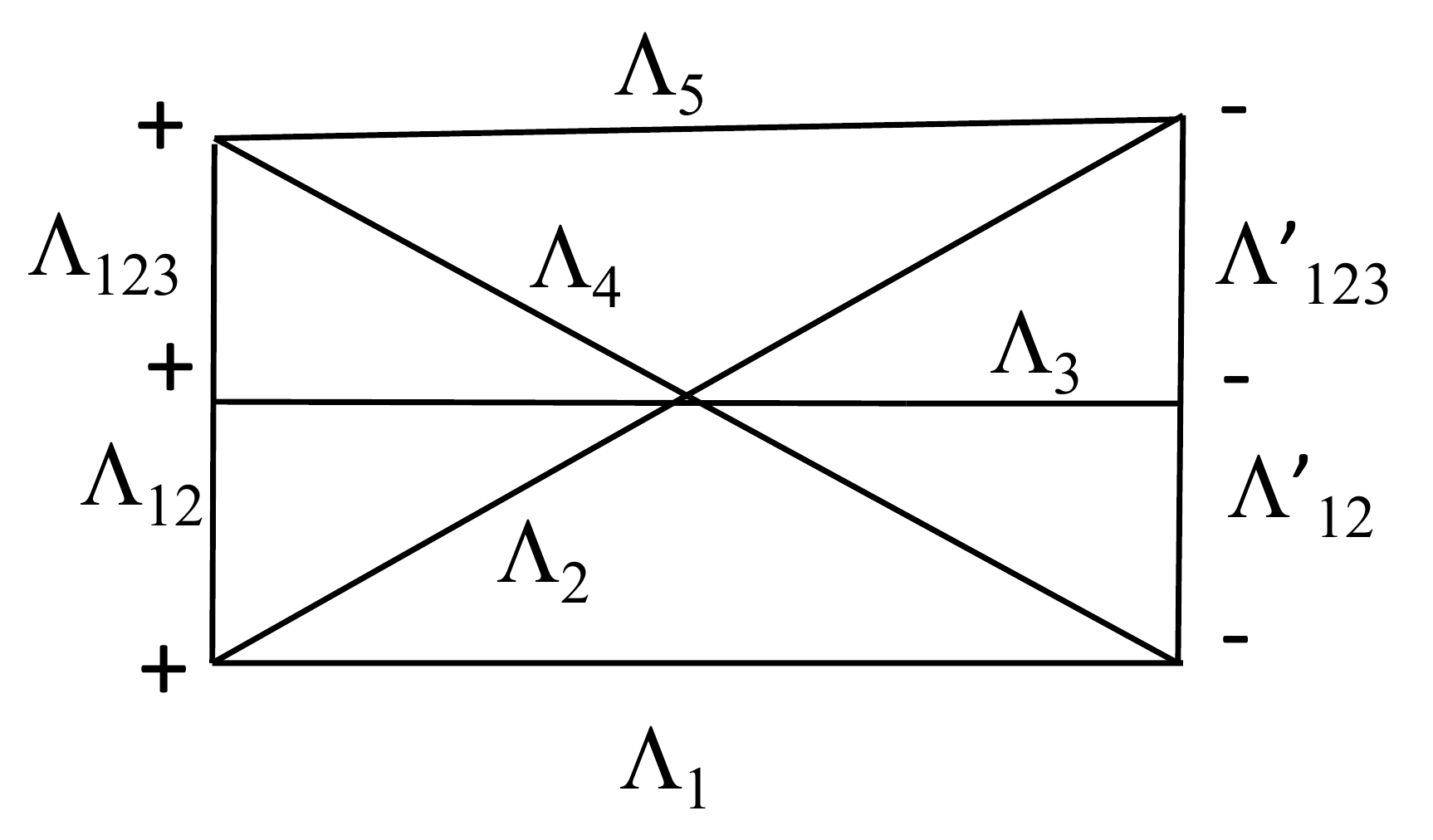}
\caption{The Yutsis diagram reordering arguments of $N=5$ isotropic function with permutation $G:(12345)\to (14325)$.  It is clear by comparison with Fig. \ref{fig:9j} that this is a 9-$j$ symbol.} \label{fig:14325}
\end{center}
\end{figure}

Now consider permutations that lead to no simple sums, i.e. neither pair $(12)$ or $(45)$ is left intact. In one class, $3$ is left in its original position, for example, $G$: $(12345)\to (14325)$. The Yutsis diagram is shown in Fig. \ref{fig:14325}.
Up to a sign and the factor
\begeq
\sqrt{(2\Lambda_{12}+1)(2\Lambda'_{12}+1)(2\Lambda_{123}+1)(2\Lambda'_{123}+1)}
\endeq
the result is
\begeq
\nine{\Lambda_{12}}{\Lambda_{1}}{\Lambda_{2}}{\Lambda_{123}}{\Lambda_{4}}{\Lambda_{5}}{\Lambda_{3}}{\Lambda'_{12}}{\Lambda'_{123}}.
\endeq
There are 16 permutations in this class.  
\subsubsection{Permutations yielding two 6-$j$ symbols}
\label{subsubsec:two_symbols}
Next consider permutations of the form $G$: $(12345)\to (31xxx)$ that have not been considered so far. The third element cannot be $2$ since this would leave the last two elements to be $(45)$ and these have already been counted. This leaves four possibilities. But in place of 1, we could use $2,4$ or $5$, each of which would itself provide four possibilities. Thus there are 16 new permutations beginning with $3$. But then we could equally well put the $3$ anywhere else but in its original position, so there are 64 permutations in this category.  

\begin{figure}
\begin{center}
\includegraphics[width=3. in]{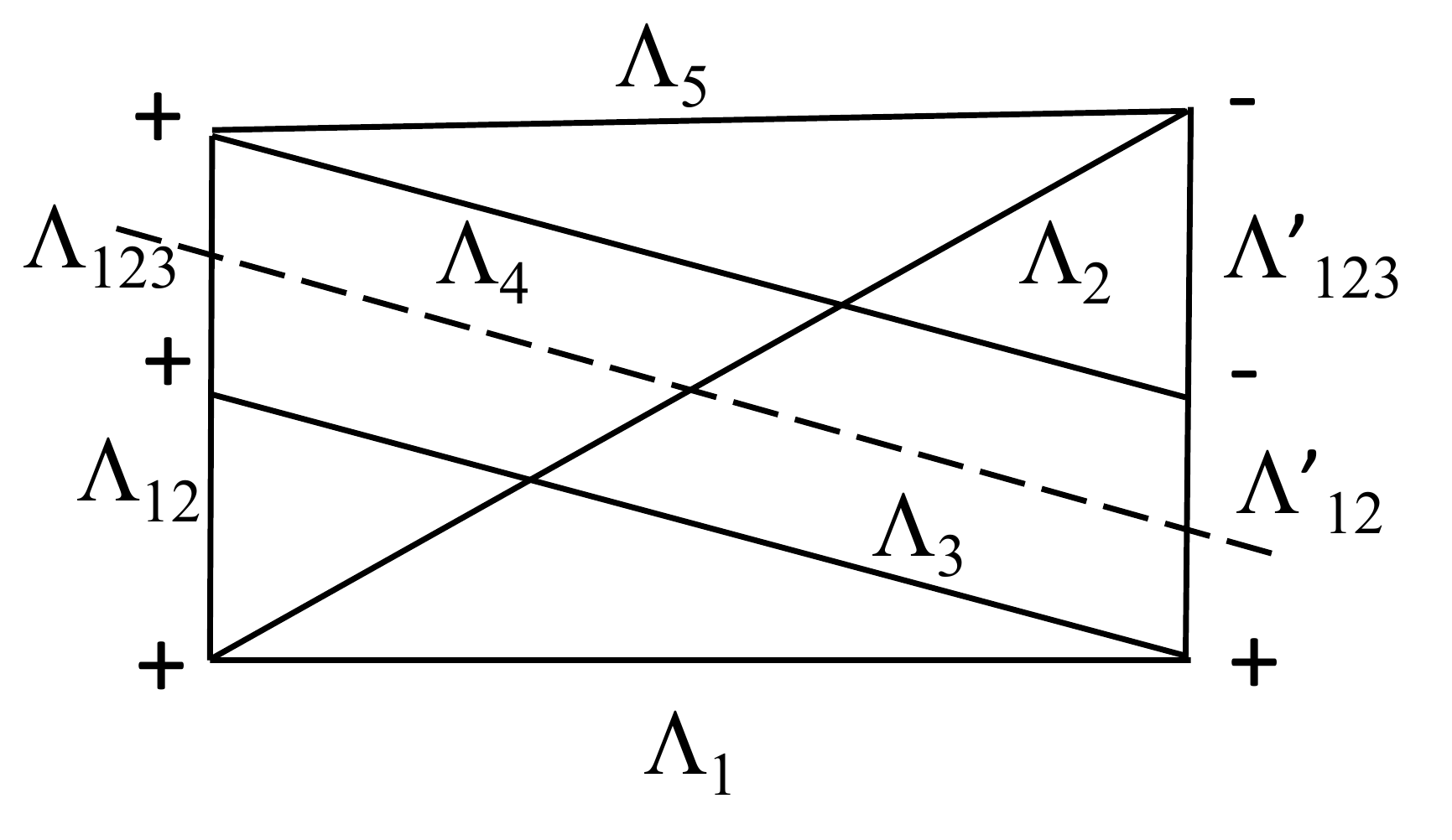}
\caption{The Yutsis diagram reordering arguments of the $N=5$ isotropic functions. The permutation is $G:(12345)\to (31425)$. The dotted line shows how the figure can be split into two separate pieces while crossing only three lines. As a result, this diagram will represent the product of two 6-$j$ symbols.} \label{fig:31425}
\end{center}
\end{figure}

\begin{figure}
\begin{center}
\includegraphics[width=3. in]{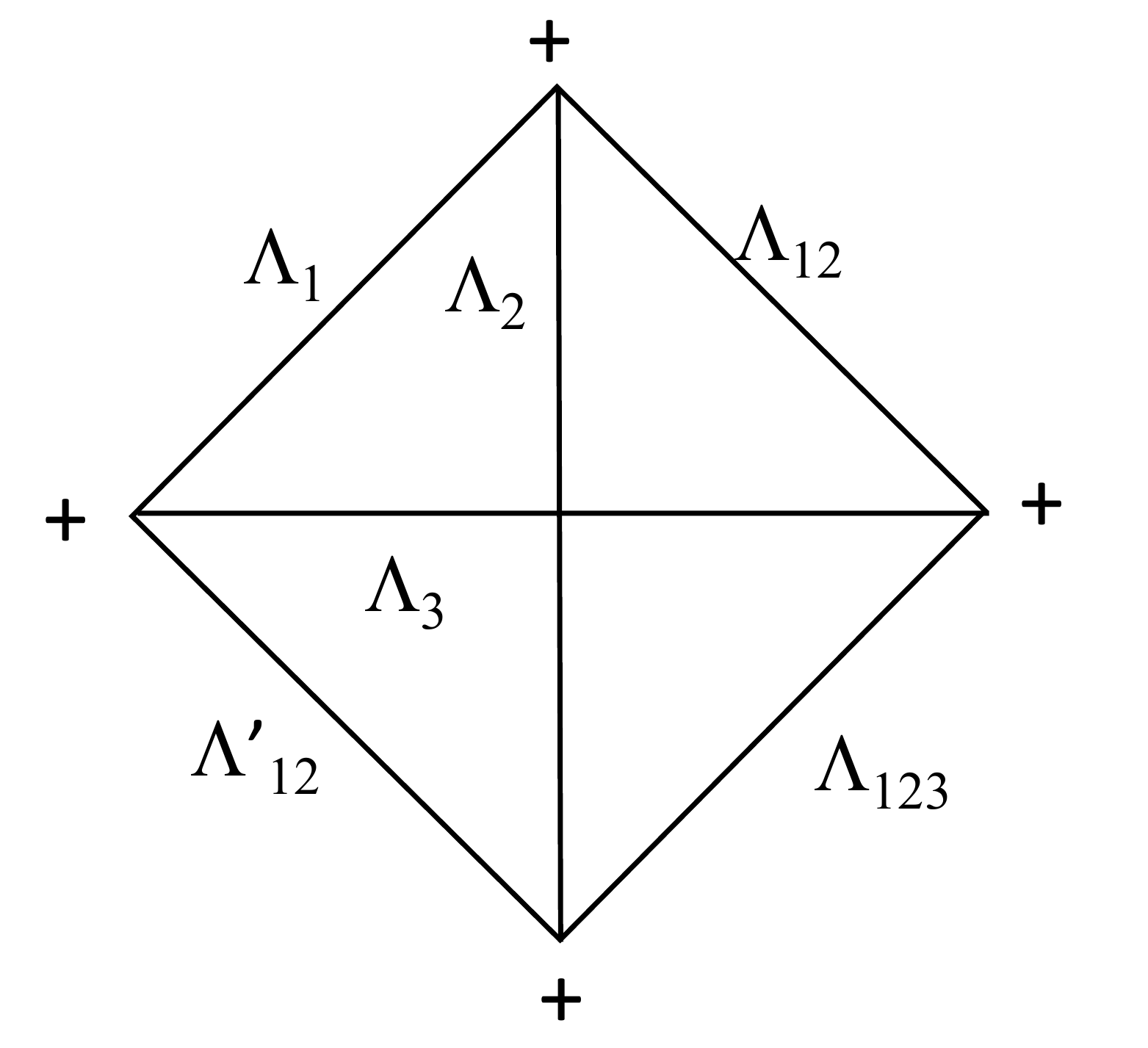}\includegraphics[width=3.3 in]{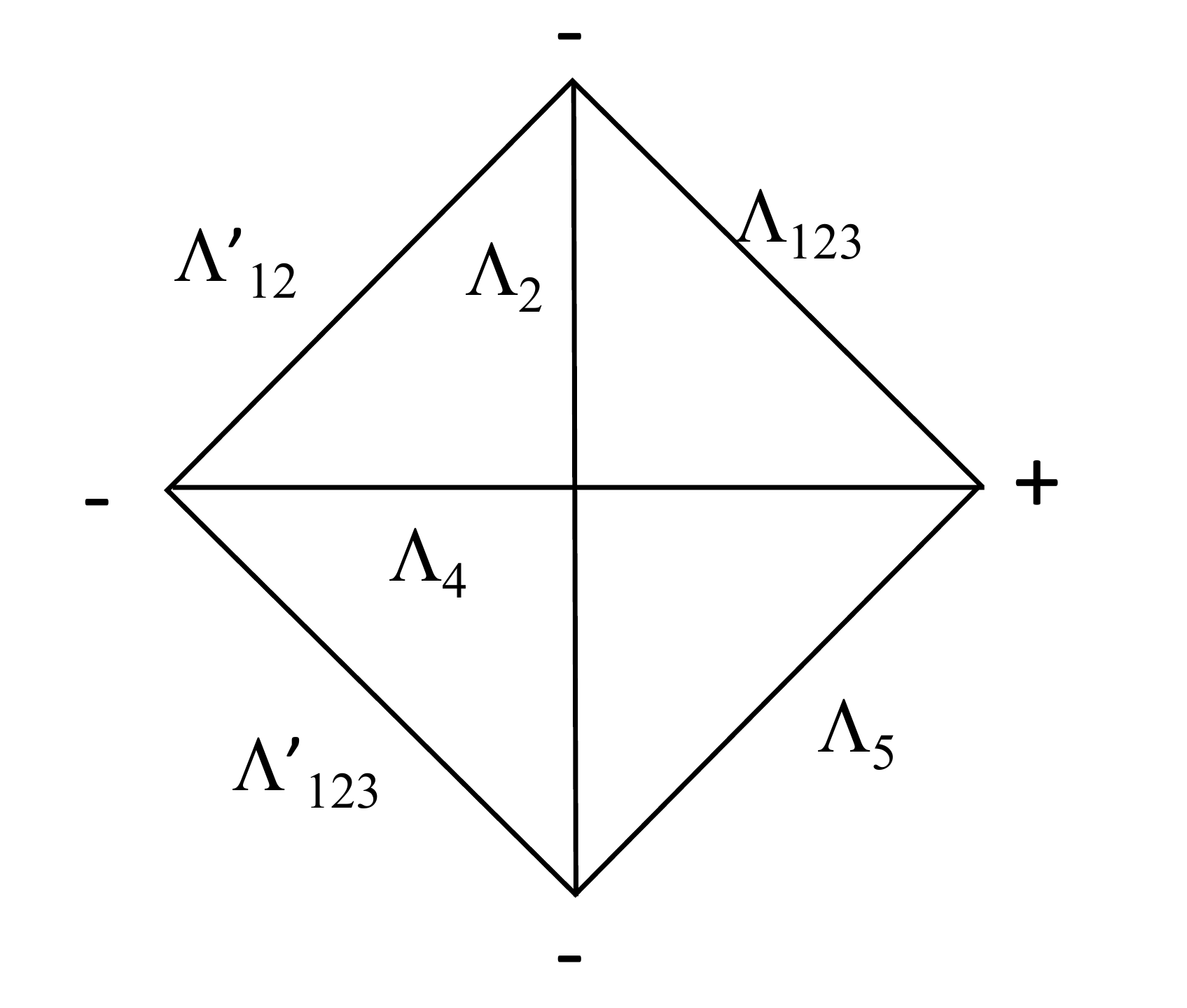}
\caption{The two diagrams obtained by splitting Fig. \ref{fig:31425}.  The signs do not have the canonical values, which would require alternating signs along the periphery.} \label{fig:31425by2}
\end{center}
\end{figure}

An example is $G$: $(12345)\to (31425)$, whose diagram in shown in Fig. \ref{fig:31425}.  
Applying the factorization theorem described in Appendix \ref{ap:theorem}, which shows that a diagram that can be split into two separate pieces by a single cut crossing only three lines,  we turn the diagram in Fig. \ref{fig:31425} into the two diagrams shown in Fig. \ref{fig:31425by2}.

The diagrams in Fig. \ref{fig:31425by2} have the value
\begeq
\sixj{\Lambda_1}{\Lambda_2}{\Lambda_{12}}{\Lambda_{123}}{\Lambda_3}{\Lambda'_{12}}
\sixj{\Lambda'_{12}}{\Lambda_2}{\Lambda_{123}}{\Lambda_{5}}{\Lambda_4}{\Lambda'_{123}}(-1)^{
\Lambda_1+\Lambda_3+\Lambda_4+\Lambda'_{123}}
\endeq
where the phase arises from restoring the canonical signs at the vertices.

Altogether we have eight permutations leading to values of $\calB^F_{\Lambda,\Lambda'}$ that have no 6-$j$ or 9-$j$ symbols in their expression, 32 that are proportional to a single 6-$j$ symbol, 16 that lead to a 9-$j$ symbol and 64 of the sort we last considered.  These account for all 120 permutations.  

\section{Summary}
A complete basis set of isotropic functions of unit vectors $\calY_\Lambda(\bfXhat=\bfxhat_1,\bfxhat_2,\ldots, \bfxhat_N)$ is conveniently constructed from spherical harmonics and Clebsch-Gordan coefficients. The quantity $\Lambda$ is specified by $N$ primary angular momenta $\Lambda_i$ together with $N-3$ intermediate angular momenta $\Lambda_{12}, \Lambda_{123},\ldots$ if $N>3$. The $\calY_\Lambda$ are eigenfunctions of each ${\bf L}_i^2$ and of $({\bf L}_1+{\bf L}_2)^2, ({\bf L}_1+{\bf L}_2+{\bf L}_3)^2,$ etc. Moreover, the $\calY_\Lambda$ are orthonormal (\S\ref{sec:ortho}).  This basis set provides a convenient means of representing any isotropic function as
\begeqar
F_{\rm{iso}}(\bfX)&=&\sum_\Lambda \calZ_\Lambda(R)\calY_\Lambda(\bfXhat),\non
\calZ_\Lambda(R)&=&\int d\bfXhat\, F_{\rm{iso}}(\bfX)\calY^*_\Lambda(\bfXhat)
\endeqar
with $R\equiv (r_1, r_2,\ldots, r_N) = (|\bfx_1|,|\bfx_2|,\ldots, |\bfx_N|)$.

Because the $\calY_\Lambda$ functions are complete it is always possible to express a product of two $\calY_\Lambda$ functions as a sum over single $\calY_\Lambda$ functions (\S\ref{sec:products}):
\begeq
\calY_\Lambda(\bfXhat)\calY_{\Lambda'}(\bfXhat)=\sum_{\Lambda''}\calOp \calG^{\Lambda\Lambda'\Lambda''}\calY_{\Lambda''}(\bfXhat)
\endeq
where $\calOp=\sum_i\Lambda_i$ and 
$\calG^{\Lambda\Lambda'\Lambda''}$ can be computed in terms of 6-$j$ and 9-$j$ symbols.  This is done conveniently with the diagrams developed by \citet{Yutsis_1962}.  The quantity $\calG^{\Lambda\Lambda'\Lambda''}$ is a generalization of the Gaunt integral to which it corresponds for $N=2$.

Permuting the arguments $\bfxhat_1,\ldots,\bfxhat_N\to \bfxhat_{F1},\ldots,\bfxhat_{FN}$ of a $\calY_\Lambda$ function produces, in general, a new function that can be expressed in terms of the canonical basis (\S\ref{sec:reordering}):
\begeq
\calY_{\Lambda}(\bfxhat_{F1},\ldots,\bfxhat_{FN})=\sum_{\Lambda'}\calB^F_{\Lambda,\Lambda'}\calY_{\Lambda'}(\bfxhat_1,\ldots,\bfxhat_N).
\endeq
The coefficients $\calB^F_{\Lambda,\Lambda'}$ (see Eq. \ref{eqn:calB_result}) can be computed again in terms of 6-$j$ and 9-$j$ symbols using Yutsis diagrams.

These relations provide a powerful means of expressing multibody correlations and for calculating covariances between the correlations as explained in Slepian, Cahn \& Eisenstein (in prep.).
\section*{Acknowledgments}
ZS thanks Daniel J. Eisenstein, Alex Krolewski, Chung-Pei Ma, and Chris Hirata for useful discussions. ZS is grateful for hospitality at Laboratoire de Physique Nucl\'eaire et des Hautes Energies in Paris for part of the period of this work, as well as at Lawrence Berkeley National Laboratory. Support for this work was provided by the National Aeronautics and Space Administration through Einstein Postdoctoral Fellowship Award Number PF7-180167
issued by the Chandra X-ray Observatory Center, which is
operated by the Smithsonian Astrophysical Observatory for
and on behalf of the National Aeronautics Space Administration under contract NAS8-03060. ZS also acknowledges
support from a Chamberlain Fellowship at Lawrence Berkeley National Laboratory (held prior to the Einstein) and
from the Berkeley Center for Cosmological Physics. RNC also thanks Laboratoire de Physique Nucl\'eaire et des Hautes Energies in Paris for their recurring hospitality. RNC's work is supported in part by the U.S. Department of Energy, Office of Science, Office of High Energy Physics  under Contract No. DE-AC02-05CH11231.

\clearpage
\bibliographystyle{mnras}
\bibliography{isotropic_bib.bib}
\clearpage

\appendix
\renewcommand\thefigure{\thesection.\arabic{figure}}  
\renewcommand\theequation{\thesection.\arabic{equation}}

\setcounter{equation}{0}
\setcounter{figure}{0}  
\section{Cartesian representations of some rotationally invariant functions }\label{ap:examples}
\setcounter{equation}{0}
It is instructive to examine explicit representations of some of the simplest $\calY_\Lambda$ functions.  These can be used to confirm some of the general relations obtained above.
We indicate specific values of $\Lambda$ by the sequence $[\Lambda_1,\Lambda_2,(\Lambda_{12}),\Lambda_3,(\Lambda_{123}),\ldots\Lambda_{N-1},\Lambda_N ]$ since $\Lambda_{12\ldots N-1}=\Lambda_N$. Explicitly, we have
\subsection{$N = 2$}
\label{subsec:A_N2}
\begeqar
\calY_{00}(\bfxhat_1,\bfxhat_2)&=&\frac 1{4\pi}\non
\calY_{11}(\bfxhat_1,\bfxhat_2)&=&-\frac{\sqrt 3}{4\pi}(\bfxhat_1\cdot\bfxhat_2)\non
\calY_{22}(\bfxhat_1,\bfxhat_2)&=&\frac 32\sqrt{\frac{5}{(4\pi)^2}}\left[(\bfxhat_1\cdot\bfxhat_2)^2-\frac 13\right].
\endeqar
\subsection{$N = 3$}
\label{subsec:A_N3}
\begeqar
\calY_{110}(\bfxhat_1,\bfxhat_2,\bfxhat_3)&=&-\frac 1{4\pi}\sqrt{\frac 3{4\pi}}(\bfxhat_1\cdot\bfxhat_2)\non 
\calY_{111}(\bfxhat_1,\bfxhat_2,\bfxhat_3)&=&-\frac {3i}{\sqrt 2(4\pi)^{3/2}}\bfxhat_1\cdot(\bfxhat_2\times\bfxhat_3)\non
\calY_{112}(\bfxhat_1,\bfxhat_2,\bfxhat_3)&=&\sqrt{\frac{27}{2(4\pi)^3}}\left[(\bfxhat_1\cdot\bfxhat_3)(\bfxhat_2\cdot\bfxhat_3)-\frac 13(\bfxhat_1\cdot\bfxhat_2)\right]\non
\calY_{222}(\bfxhat_1,\bfxhat_2,\bfxhat_3)&=&-\frac{45}{\sqrt{14(4\pi)^3}}\left[\dotprod 12\dotprod 13\dotprod23\right.\non
&&\left.-\frac 13\dotprod12^2-\frac 13\dotprod 13^2-\frac13\dotprod 23^2+\frac 29\right].
\endeqar
\subsection{$N = 4$}
\label{subsec:A_N4}
\begeqar
\calY_{11(0)11}(\bfxhat_1,\bfxhat_2,\bfxhat_3,\bfxhat_4)&=&\frac 3{(4\pi)^2}(\bfxhat_1\cdot\bfxhat_2)(\bfxhat_3\cdot\bfxhat_4)\non
\calY_{10(1)01}(\bfxhat_1,\bfxhat_2,\bfxhat_3,\bfxhat_4)&=&-\frac{\sqrt 3}{(4\pi)^2}(\bfxhat_1\cdot\bfxhat_4)\nonumber
\endeqar
\begeqar
\calY_{21(1)12}(\bfxhat_1,\bfxhat_2,\bfxhat_3,\bfxhat_4)&=&-\frac{9\sqrt 3}{2(4\pi)^2}\left[\dotprod 12\dotprod 14\dotprod 34-\frac 13\dotprod 24\dotprod 34\right.\non
&&\qquad\qquad\qquad\left.-\frac 13\dotprod 12\dotprod 13+\frac 19 \dotprod 23\right]\non
\calY_{21(2)12}(\bfxhat_1,\bfxhat_2,\bfxhat_3,\bfxhat_4)
&=&\frac{3\sqrt5}{2(4\pi)^2}\left[2\dotprod 14[\dotprod 14\dotprod23-\dotprod 13\dotprod 24]\right.\non
&&\qquad\qquad\qquad-\dotprod 14\dotprod 12\dotprod 34+\dotprod 24\dotprod 34\non
&&\qquad\qquad\qquad\qquad+\left.\dotprod12\dotprod 13-\dotprod 23\right]\non
\calY_{11(2)11}(\bfxhat_1,\bfxhat_2,\bfxhat_3,\bfxhat_4)&=&\frac{9}{2\sqrt 5(4\pi)^2}\bigg[\dotprod 13\dotprod 24+\dotprod14\dotprod 23 \non
&&-\frac 23\dotprod 12\dotprod 34\bigg]\non
\calY_{11(1)11}(\bfxhat_1,\bfxhat_2,\bfxhat_3,\bfxhat_4)&=&\frac{3\sqrt 3}{2(4\pi)^2}\left[\dotprod13\dotprod 24-\dotprod 14\dotprod 23\right]\non
\calY_{21(1)10}(\bfxhat_1,\bfxhat_2,\bfxhat_3,\bfxhat_4)&=&\frac{3\sqrt 3}{\sqrt{2}(4\pi)^2}\left[\dotprod 12\dotprod13-\frac 13\dotprod 23\right]\non
\calY_{21(3)12}(\bfxhat_1,\bfxhat_2,\bfxhat_3,\bfxhat_4)&=&-\frac{15}{2\sqrt 7(4\pi)^2}\left[\dotprod 14 ^2\dotprod 23 -\frac 45 \dotprod 12\dotprod14\dotprod 34\right.\non
&&\qquad +2\dotprod 13\dotprod14\dotprod24 
 -\frac 25 \dotprod 12\dotprod13 \non
 &&\qquad \left.-\frac 25\dotprod 24\dotprod 34-\frac 15\dotprod23\right]\non\label{eq:cartesian}
\calY_{21(2)11}(\bfxhat_1,\bfxhat_2,\bfxhat_3,\bfxhat_4)&=&i\frac{3\sqrt 3}{2(4\pi)^2}\left[
(\bfxhat_1\cdot\bfxhat_3)\ \bfxhat_4\cdot(\bfxhat_1\times\bfxhat_2)+(\bfxhat_1\cdot\bfxhat_4)\ \bfxhat_3\cdot(\bfxhat_1\times\bfxhat_2)\right]\non
\calY_{21(1)11}(\bfxhat_1,\bfxhat_2,\bfxhat_3,\bfxhat_4)&=&\frac{9i}{2(4\pi)^2}\left[ (\bfxhat_1\cdot\bfxhat_2)\bfxhat_1\cdot(\bfxhat_3\times\bfxhat_4)-\frac 13\bfxhat_2\cdot(\bfxhat_3\times\bfxhat_4)\right].
\endeqar 
\section{Properties of Wigner Angular Momentum Addition symbols}
\label{ap:symbols}
\subsection{3-$j$ symbols}
The 3-$j$ symbol or Wigner coefficient is defined in terms of the Clebsch-Gordan coefficient via 
\begeq
\six{j_1}{j_2}{j_3}{m_1}{m_2}{m_3}=\frac{(-1)^{j_1-j_2-m_3}}{\sqrt{2j_3+1}}\left<j_1m_1j_2m_2|j_3-m_3\right>.
\endeq
The 3-$j$ symbols have the symmetries
\begeqar
\six{j_1}{j_2}{j_3}{m_1}{m_2}{m_3}&=&\six{j_2}{j_3}{j_1}{m_2}{m_3}{m_1}\non
&=&(-1)^{j_1+j_2+j_3}\six{j_2}{j_1}{j_3}{m_2}{m_1}{m_3}\non
&=&(-1)^{j_1+j_2+j_3}\six{j_1}{j_2}{j_3}{-m_1}{-m_2}{-m_3}.
\label{eq:3j_perms}
\endeqar
The orthogonality relation for 3-$j$ symbols is 
\begeq
\sum_{m_2,m_3}
\six{j_1}{j_2}{j_3}{m_1}{m_2}{m_3}\six{j_4}{j_2}{j_3}{m_4}{m_2}{m_3}=\frac{\deltaK_{j_1,j_4}\deltaK_{m_1,m_4}}{2j_1+1}.
\label{eqn:3j_orth}
\endeq
Further summing the above result over $m_1$ yields
\begeq
\sum_{m_1, m_2,m_3}
\six{j_1}{j_2}{j_3}{m_1}{m_2}{m_3}\six{j_1}{j_2}{j_3}{m_1}{m_2}{m_3}=1.
\label{eq:3j_orth_two}
\endeq

\subsection{6-$j$ symbols}
The 6-$j$ symbol is
\begeqar
&&\sixj{j_1}{j_2}{j_3}{j_4}{j_5}{j_6}=\sum_{m_1\ldots m_6}(-1)^{\sum_k(j_k-m_k)}\six{j_1}{j_2}{j_3}{-m_1}{-m_2}{-m_3}\non
&&\quad\times\six{j_1}{j_5}{j_6}{m_1}{-m_5}{m_6}\six{j_4}{j_2}{j_6}{m_4}{m_2}{-m_6}\non
&&\quad\times\six{j_4}{j_5}{j_3}{-m_4}{m_5}{m_3}.
\label{eq:sixj}\endeqar
The 6-$j$ symbol is invariant under permutations of the columns and also under interchanging the top and bottom elements for any pair of columns. 

If an element of a 6-$j$ symbol is zero there is a simplification. Using the rules for permuting rows and columns, a zero can always be brought to the lower right-hand position. The 6-$j$ symbol then becomes
\begeq 
\sixj{j_1}{j_2}{j_3}{j_4}{j_5}0=\frac{\deltaK_{j_2j_4}\deltaK_{j_1j_5}}{
    \sqrt{(2j_1+1)(2j_2+1)}}(-1)^{j_1+j_2+j_3}\deltaK\left\{j_1, j_2, j_3\right\}.\label{eq:6j-one-zero}
\endeq 
where the final factor, $\deltaK\left\{j_1, j_2, j_3\right\}$, is zero unless the three elements satisfy the triangular inequality.

It follows from the symmetries of the 6-$j$ symbol that
\begeqar 
     &&\sixj{j_1}{j_2}{j_3}{j_4}{j_5}0=
     \sixj{j_1}{j_3}{j_2}{j_4}0{j_5}=
     \sixj{j_3}{j_1}{j_2}0{j_4}{j_5}\non 
     &&\sixj{j_1}{j_5}0{j_4}{j_2}{j_3}=
     \sixj{j_1}0{j_5}{j_4}{j_3}{j_5}=
     \sixj0{j_1}{j_5}{j_3}{j_4}{j_2}. 
\endeqar

Finally, we note that there is also a relationship between a 9-$j$ symbol where one element is zero and a 6-$j$ symbol (Eq. \ref{eq:9j_to_6j}).

\subsection{9-$j$ symbols}
\label{subsec:9j}
The 9-$j$ symbol can be defined in a variety of ways.
The expression in \cite{Edmonds_1996}, 6.4.4,  is the remarkably symmetric
\begeqar
&&\nine{j_{11}}{j_{12}}{j_{13}}{j_{21}}{j_{22}}{j_{23}}{j_{31}}{j_{32}}{j_{33}}\\
&&\quad =\sum_{ms} \six{j_{11}}{j_{12}}{j_{13}}{m_{11}}{m_{12}}{m_{13}}\six{j_{21}}{j_{22}}{j_{23}}{m_{21}}{m_{22}}{m_{23}}\six{j_{31}}{j_{32}}{j_{33}}{m_{31}}{m_{32}}{m_{33}}\non
&&\quad \times\six{j_{11}}{j_{21}}{j_{31}}{m_{11}}{m_{21}}{m_{31}}\six{j_{12}}{j_{22}}{j_{32}}{m_{12}}{m_{22}}{m_{32}}\six{j_{13}}{j_{23}}{j_{33}}{m_{13}}{m_{23}}{m_{33}}.\nonumber
\endeqar
The 9-$j$ symbol is invariant under even permutations of its columns or rows. Under odd permutations it acquires a phase
$(-1)^J$ where $J=\sum_k j_k$. It is also invariant under reflections about either diagonal.

Equivalently we have
\begeqar
\nine{j_1}{j_2}{j_3}{k_1}{k_2}{k_3}{\ell_1}{\ell_2}{\ell_3}&=&\sum_{m_i,n_i,p_i}
\six{j_1}{j_2}{j_3}{m_1}{m_2}{m_3}\six{k_1}{k_2}{k_3}{n_1}{n_2}{n_3}\six{\ell_1}{\ell_2}{\ell_3}{p_1}{p_2}{p_3}\non
&&\times\six{j_1}{k_1}{\ell_1}{m_1}{n_1}{p_1}\six{j_2}{k_2}{\ell_2}{m_2}{n_2}{p_2}\six{j_3}{k_3}{\ell_3}{m_3}{n_3}{p_3}\non
&=&\sum_{m_i,n_i,p_i}(-1)^{J}
\six{j_1}{j_2}{j_3}{m_1}{m_2}{m_3}\six{k_1}{k_2}{k_3}{n_1}{n_2}{n_3}\non
&&\times \six{\ell_1}{\ell_2}{\ell_3}{p_1}{p_2}{p_3}\six{j_1}{k_1}{\ell_1}{-m_1}{-n_1}{-p_1}\non
&&\times\six{j_2}{k_2}{\ell_2}{-m_2}{-n_2}{-p_2}\six{j_3}{k_3}{\ell_3}{-m_3}{-n_3}{-p_3}.
\label{eq:9pt}
\endeqar
The 9-$j$ symbol with a lower right-hand zero becomes    
\begeq 
    \nine{j_1}{j_2}{j_3}{j_4}{j_5}{j_3}{j_6}{j_6}0=
    \frac{(-1)^{j_2+j_4+j_3+j_6}}{\sqrt{(2j_3+1)(2j_6+1)}}
    \sixj{j_1}{j_2}{j_3}{j_5}{j_4}{j_6}.
    \label{eq:9j_to_6j}
\endeq 
It follows from the symmetries of the 9-$j$ symbol that
 \begeqar 
    &&\nine{j_1}{j_2}{j_3}{j_4}{j_5}{j_3}{j_6}{j_6}0=
    (-1)^J\nine{j_1}{j_3}{j_2}{j_4}{j_3}{j_5}{j_6}0{j_6}=
    \nine{j_3}{j_1}{j_2}{j_3}{j_4}{j_5}0{j_6}{j_6}\non
    &&=(-1)^J\nine{j_1}{j_2}{j_3}{j_6}{j_6}0{j_4}{j_5}{j_3}=
     \nine{j_6}{j_6}0{j_1}{j_2}{j_3}{j_4}{j_5}{j_3}=
    (-1)^J\nine{j_6}0{j_6}{j_1}{j_3}{j_2}{j_4}{j_3}{j_5}\non
    &&=\nine 0{j_6}{j_6}{j_3}{j_1}{j_2}{j_3}{j_4}{j_5}=
    (-1)^J\nine{j_3}{j_1}{j_2}0{j_6}{j_6}{j_3}{j_4}{j_5}= 
     \nine{j_1}{j_3}{j_2}{j_6}0{j_6}{j_4}{j_3}{j_5}. 
     \label{eq:9j-onezero}
\endeqar

\section{Yutsis diagrams}\label{ap:yutsis}
\setcounter{equation}{0}
A Yutsis diagram \citep{Yutsis_1962} is defined by a product of Wigner coefficients, i.e. 3-$j$ symbols, in which every repeated $j$ occurs once with $m$ and once with $-m$.    Every Wigner coefficient is assigned a point and two points are joined if they have a $j$ in common.   At each point a $+$ or $-$ is assigned.  If the $j$'s in the Wigner symbol, taken in left-to-right order,  occur counter-clockwise in the diagram, one assigns a $+$ sign at the point.  Otherwise one assigns a $-$ sign. We note that, in order for the initial expression to be described by a Yutsis diagram, the expression must have the phase
\begeq
(-1)^{\sum_i (j_i -m_i)}
\endeq
for every $(j_i,m_i),(j_i,-m_i)$ that appears in the collection,  where $m_i$ is summed over. 

\begin{figure}
\begin{center}
\includegraphics[width=3. in]{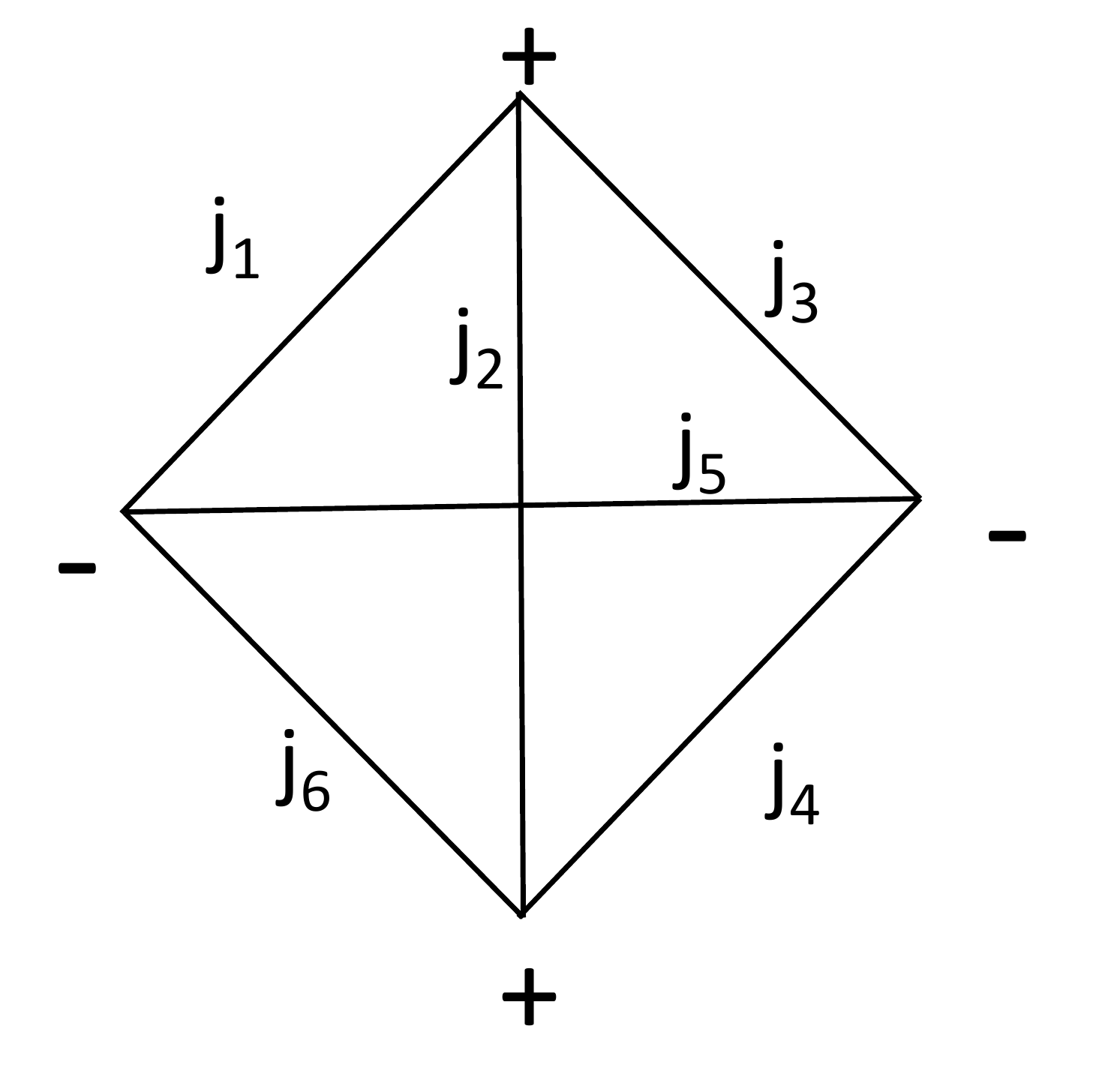}
\caption{The Yutsis diagram for a 6-$j$ symbol in canonical form.}
\label{fig:6j}
\end{center}
\end{figure}

Applying the rules for Yutsis diagrams to Eq. (\ref{eq:sixj}) we obtain Fig. \ref{fig:6j} representing
\begeq
\sixj{j_1}{j_2}{j_3}{j_4}{j_5}{j_6}.
\endeq

\begin{figure}
\begin{center}
\includegraphics[width=2.4 in]{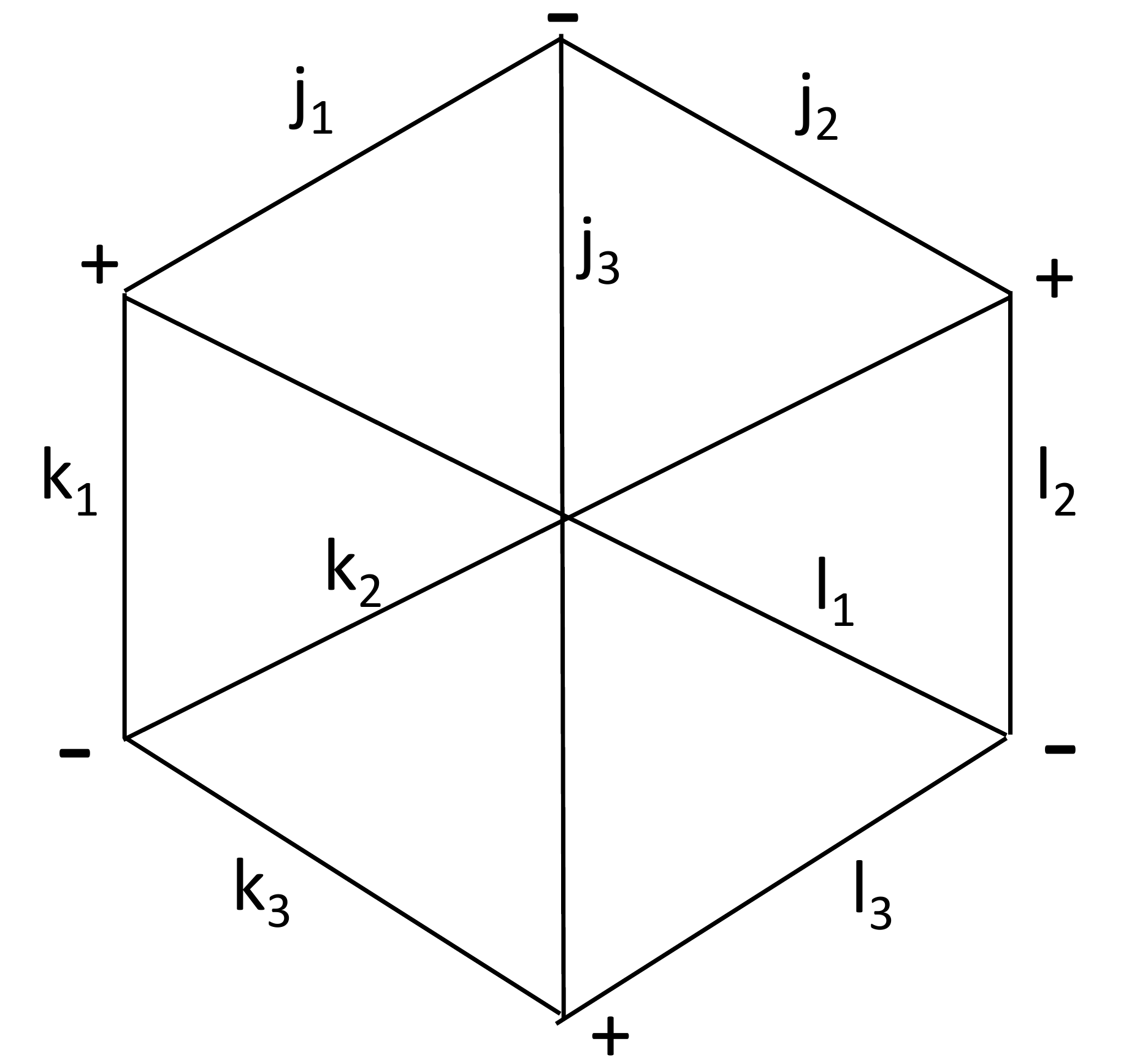}
\caption{Yutsis diagram showing the canonical 9-$j$ diagram; see Eq. (\ref{eq:9j_appendix_C}). Note that changing sign at all vertices leaves the value unchanged because it introduces a factor $(-1)^{2\sum_i(j_i+k_i+\ell_i)}=1$.  Similarly, the mirror image diagram stands for the same 9-$j$ symbol.}
\label{fig:9j}
\end{center}
\end{figure}

Applying these rules to the 9-$j$ symbol in Eq. (\ref{eq:9pt}),
\begeq
\label{eq:9j_appendix_C}
\nine{j_1}{j_2}{j_3}{k_1}{k_2}{k_3}{\ell_1}{\ell_2}{\ell_3}
\endeq
we obtain Fig. \ref{fig:9j}.
\setcounter{equation}{0}
\setcounter{figure}{0}
\section{Yutsis Theorem on Factorization}\label{ap:theorem}
\cite{Yutsis_1962} prove a general theorem on the factorization of Yutsis diagrams that can be separated into two pieces by cutting through three or fewer lines.  We need only a restricted version of the theorem, that which states that a product $R^{j_1j_2j_3}_{m_1m_2m_3}$ of 3-$j$ symbols summed over some indices but not summed over $m_1, m_2, m_3$  associated with $j_1,j_2, j_3$ can be written
\begeq
R^{j_1j_2j_3}_{m_1m_2m_3}=\six{j_1}{j_2}{j_3}{m_1}{m_2}{m_3}R
\endeq
where the value of $R$ is given by

\begin{align}
R&=\sum_{m_i}R^{j_1j_2j_3}_{m_1m_2m_2}
\six{j_1}{j_2}{j_3}{-m_1}{-m_2}{-m_3}(-1)^{\sum_i(j_i-mi)}\non
&=\sum_{m_i}R^{j_1j_2j_3}_{m_1m_2m_2}
\six{j_1}{j_2}{j_3}{m_1}{m_2}{m_3}\non
\end{align}
where we recall that $\sum_i m_i=0$ and use Eqs. (\ref{eq:3j_perms}) and (\ref{eq:3j_orth_two}).

\begin{figure}
\begin{center}
\includegraphics[width=3.4 in]{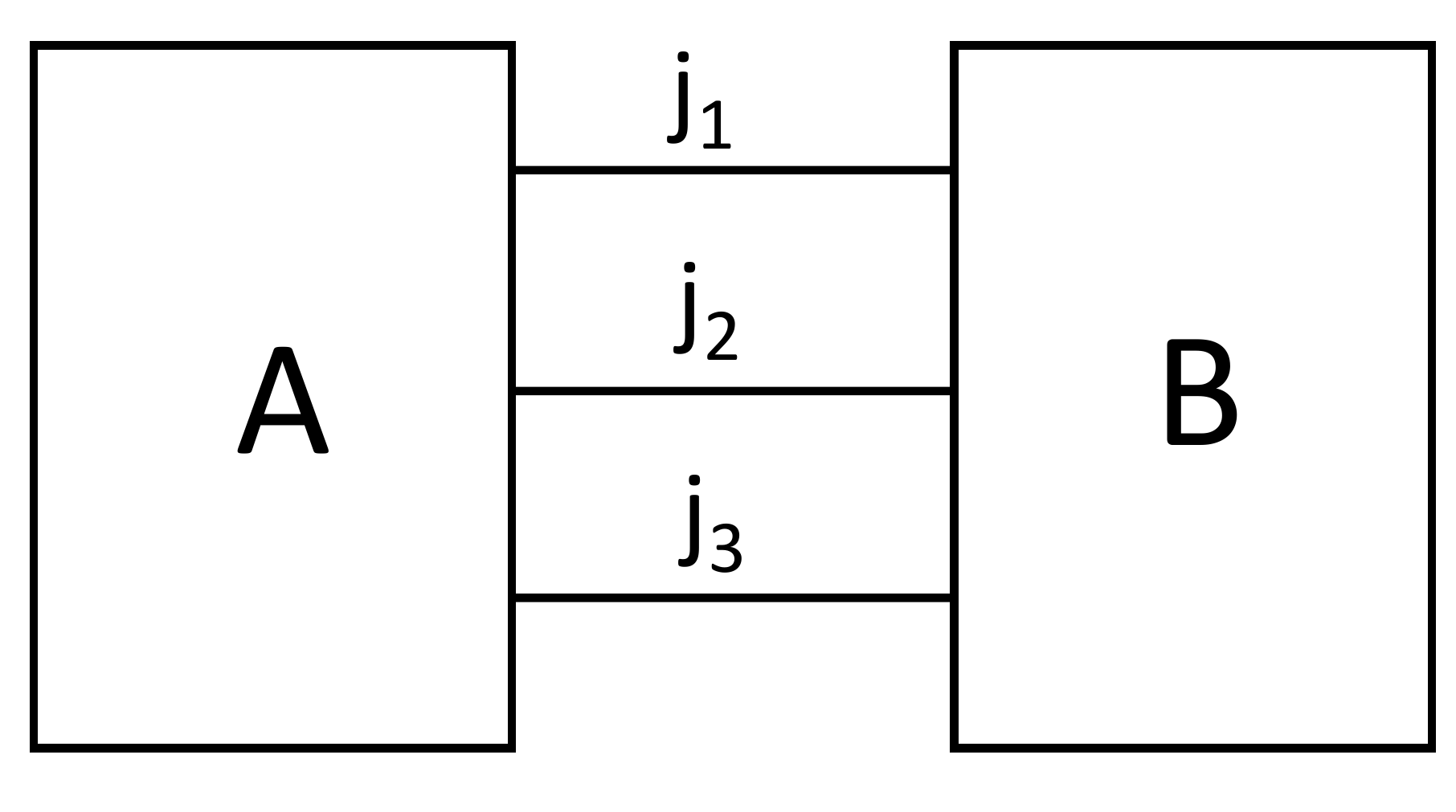} 

\includegraphics[width=3.4 in]{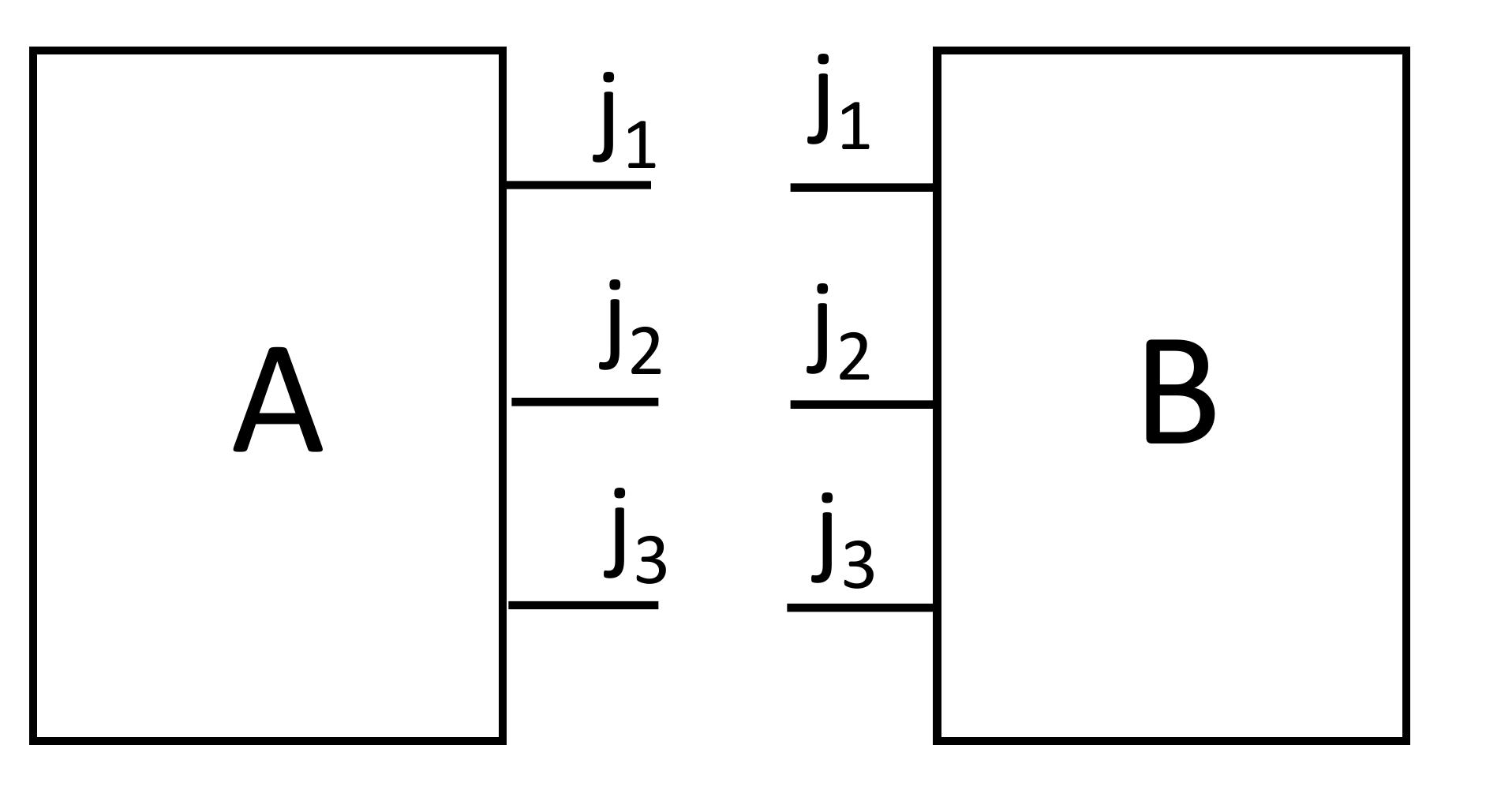}

\includegraphics[width=4.4 in]{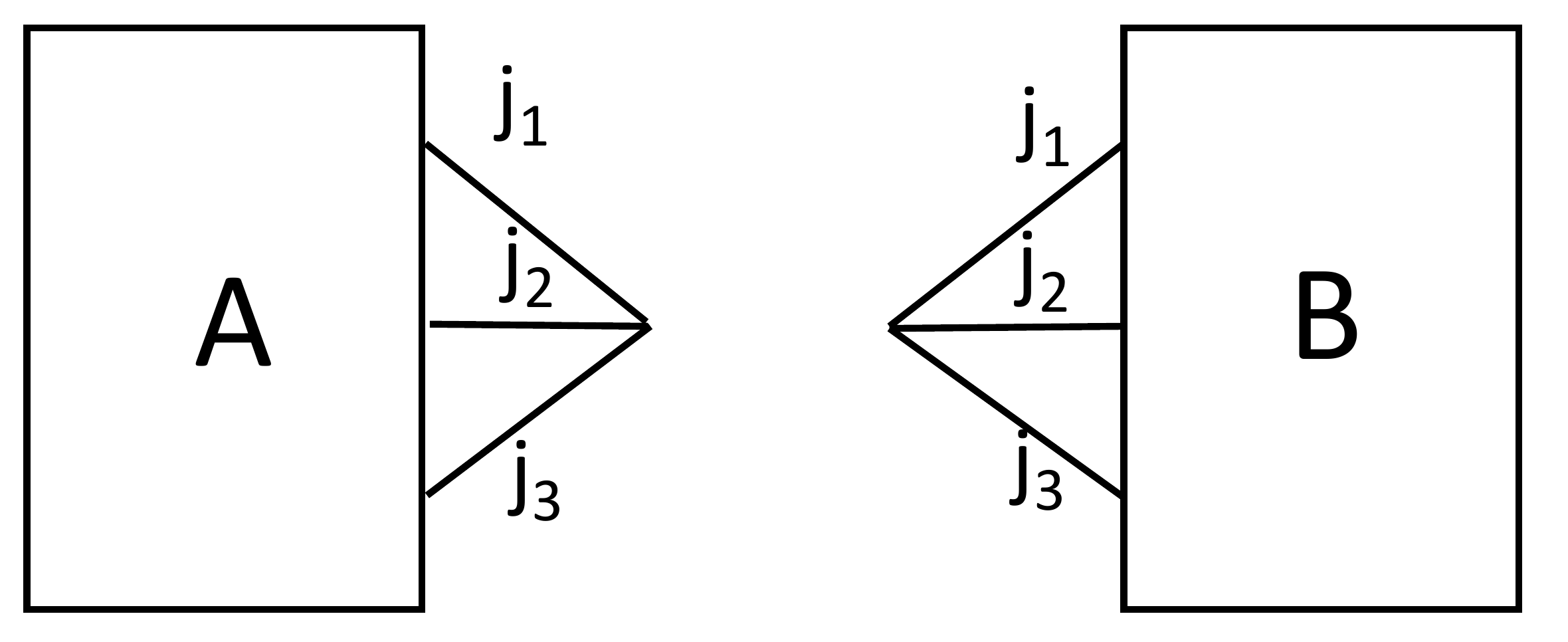}
\caption{A Yutsis diagram that can be separated into two separate diagrams by cutting just three lines can be written as the product of two simpler diagrams.   }

\label{fig:AB}
\end{center}
\end{figure}
Now consider a diagram that can be separated by a single cut through three lines. Such a diagram can generically be represented by
\begeqar
\sum_{m_i}
&&A^{j_1j_2j_3}_{m_1m_2m_3}B^{j_1j_2j_3}_{-m_1-m_2-m_3}(-1)^{\sum_i(j_i-m_i)}\non
&&=\sum_{m_i}\six{j_1}{j_2}{j_3}{m_1}{m_2}{m_3}\six{j_1}{j_2}{j_3}{-m_1}{-m_2}{-m_3}(-1)^{\sum_i(j_i-m_i)}AB\non
&&=AB.
\endeqar
The last line comes from the orthogonality relation for 3-$j$ symbols, Eq. (\ref{eqn:3j_orth}).
The signs to be attached to the diagrams for $A$ and $B$ will depend on the orientations of the lines for the $j_i$.
In Fig. \ref{fig:AB} we see that the two separate diagrams will be identical to the two split pieces, but with a new vertex.  In this example, the vertex attached to diagram A will have the opposite sign from the vertex attached to B.  The phases introduced thereby will cancel.

\end{document}